\documentclass[a4paper,11pt]{article}
\pdfoutput=1 % if your are submitting a pdflatex (i.e. if you have
\usepackage{jheppub} % for details on the use of the package, please
\usepackage[T1]{fontenc} % if needed

\usepackage[svgnames]{xcolor}
\usepackage{parskip}
\usepackage{breqn}
\allowdisplaybreaks
\usepackage{slashed}
\usepackage{bm}
\usepackage{comment}

\newcommand{\nn}{\nonumber}
\renewcommand{\t}{\tilde}
\newcommand{\h}{\hat}

\newcommand{\rma}[1]{\textcolor{red}{#1}}

\preprint{USTC-ICTS/PCFT-25-59}

\title{\boldmath Time-dependent flux backgrounds in type IIB supergravity}

\author[a,b]{Ahmed Rakin Kamal}
\author[c,d]{and Ratul Mahanta}

\affiliation[a]{Department of Theoretical Physics and Astrophysics, Faculty of Science,
Masaryk University, Kotl\'a\v{r}sk\'a 2, CS-61137 Brno, Czechia}
\affiliation[b]{\small Department of Mathematics and Natural Sciences,
BRAC University, 66 Mohakhali, Dhaka 1212, Bangladesh}
\affiliation[c]{Interdisciplinary Center for Theoretical Study, University of Science and Technology of China, Hefei, Anhui 230026, China}
\affiliation[d]{Peng Huanwu Center for Fundamental Theory, Hefei, Anhui 230026, China}

\emailAdd{ahmedrakinkamaltunok@gmail.com}
\emailAdd{ratul.mahanta@ustc.edu.cn}

\abstract{We analytically construct families of type IIB supergravity backgrounds in ten dimensions in which the four-dimensional metric is time dependent, while the six-dimensional internal space is an arbitrary compact Calabi-Yau manifold (with no restriction on holonomy) up to an overall time-dependent scale factor. Our solutions include cases with all fluxes (three-form and self-dual five-form) switched on, as well as cases with subsets of these fluxes, together with a time-dependent axiodilaton in most cases. These constructions require no local sources. We show that the associated energy-momentum tensors (both 10D and the resulting 4D effective) satisfy the null, weak, strong, and dominant energy conditions. In our explicit constructions, the Ricci scalar of the four-dimensional Einstein frame metric is negative; such backgrounds may find applications to anisotropic or FLRW cosmologies in the early universe. We also revisit the Maldacena--Nu\~nez no-go analysis, incorporating new elements that appear in our constructions, namely an overall noncompact spacetime-dependent scale factor multiplying the internal metric, and field strengths with components partially covering the noncompact directions. We argue that, with these generalizations, a four-dimensional Einstein-frame metric with positive Ricci scalar cannot be ruled out by such an analysis.}

\begin{document} 
\maketitle
\flushbottom

\section{Introduction and summary}
\label{sec:intro}

The type IIB supergravity appears in the low energy limit of the type IIB superstring theory in ten dimensions \cite{Polchinski:1998rr}. Its bosonic sector consists of the 10D metric field, a complex zero-form axiodilaton field, along with a self-dual five-form and a complex three-form field strengths. Finding its classical backgrounds amounts to solving the 10D Einstein field equations and the equations of motion and Bianchi identities for the form fields, collectively referred to as the type IIB supergravity equations. Over the years various classes of such backgrounds have been constructed, including solutions with an $AdS_n$ component (see e.g. \cite{Macpherson:2014eza,Figueroa-OFarrill:2012whx}), black hole solutions (see e.g. \cite{Duff:1999rk,Mandal:2000rp,Emparan:2006mm}), pp-wave solutions (see e.g. \cite{Blau:2001ne, Blau:2002dy, Bena:2002kq,Gran:2018ijr}), etc. Our interests are classical backgrounds that allow for a noncompact time-dependent 4D spacetime component. In this regard there are no-go theorems \cite{deWit:1986mwo, Maldacena:2000mw, Gibbons:2003gb, HariDass:2002si} (see \cite{Russo:2019fnk, Das:2019vnx, Bernardo:2022ony, Faruk:2024usu, Shukla:2019dqd, Niedermann:2025tuo} for recent explorations) concerning the existence of a 4D component with nonnegative Ricci scalar, such as Minkowski, Kasner or de Sitter spacetimes, based on specific ansatzes for the metric and other field content.\footnote{Allowing local sources, especially objects with negative tension, can loosen such no-go theorems and allow for a 4D spacetime component with zero Ricci scalar such as Minkowski or Kasner spacetimes \cite{Giddings:2001yu}. In the presence of such branes, whether a 4D de Sitter spacetime can be realized has been discussed in \cite{Dasgupta:2014pma}, adopting similar types of 10D metric and fluxes.} Notably, some of the above discussions on no-go theorems employ various energy conditions; for review on the consequences and known violations of such energy conditions, see \cite{Kontou:2020bta, Curiel:2014zba, Dain:2013vsa, Malament:2005rc}.

A noncompact time-dependent 4D spacetime component is of particular interest in cosmology. For instance, one with a negative Ricci scalar potentially could support Kination epoch in the early universe \cite{Morgante:2025lav, Apers:2022cyl, Cicoli:2023opf}. Another recent development is the work \cite{Andriot:2025cyi} on achieving scale separation between the KK length scale and the 4D curvature radius as time evolves. Notably, both works \cite{Apers:2022cyl,Andriot:2025cyi} employ higher-dimensional metrics with a time-dependent scale factor multiplying an internal space.\footnote{Interestingly time-dependent 6D scale factors also appear in the ongoing discussion of realizing 4D de Sitter spacetime as a Glauber-Sudarshan state in string theory (e.g., \cite{Dasgupta:2025ypg}).}

In general, a systematic tabulation of time-dependent 4D components allowed in supergravity and string theory, together with an analysis of their cosmological implications, will be of considerable value. Attempts have been made to construct such backgrounds, either in toy supergravity models or by switching off all fluxes \cite{Chen:2002yq, Bhattacharya:2003sh, Neupane:2009jn, Mueller:1989in, Townsend:2003fx, Ohta:2003uw, Ohta:2003pu, Roy:2003nd, Roy:2006du, Roy:2007fu, Nayek:2014iza, Nayek:2017yqx, Sabra:2020gio, Fakhredin:2025ngo}. In these discussions, models of higher-dimensional gravity coupled to a dilaton scalar field and a higher-form potential field have been considered, thus incorporating some ingredients found in various supergravity theories. However, they do not have, for example, contributions from Chern-Simons terms, as well as the full intricacy of the supergravity equations arising from multiple form fields (which couple to the dilaton differently). See \cite{Buchel:2002kj, SenGupta:2007lbr} for time-dependent examples constructed using different methods, such as gauge/gravity correspondence and analytical continuation of known static solutions. While our focus in this paper is on type IIB supergravity, see \cite{Billyard:1999dg, Cavaglia:2000rx, Cavaglia:2000uu, Cavaglia:2001sb} for a discussion of 4D cosmologies ignoring only the R--R one-form potential in 10D type IIA supergravity, while for more general constructions see \cite{Marconnet:2022fmx}.\footnote{Type IIA supergravity, which lacks self-dual field strengths, arises from compactification of 11D M-theory on an internal circle.} Moreover, cosmological solutions in supergravity systems coupled to brane sources have been constructed previously (see, e.g., \cite{Chen:2005ae}), while the present work does not incorporate branes.

In this article, we aim to construct 10D type IIB supergravity backgrounds with a time-dependent 4D component, while keeping all fluxes nontrivial. In doing so, we rely on a metric and flux ansatz which also allows us to explore subcases where some fluxes are turned off. Based on this specific ansatz, we complete the tabulation of solvable subcases with time-dependent 4D components. Subsequently, we revisit the analysis of the Maldacena--Nu\~nez (MN) no-go theorem \cite{Maldacena:2000mw}, incorporating the elements required for our constructions. Below, we summarize the key points of our methodology and results:
\begin{itemize}
  \item In our analysis we keep a time-dependent scale factor multiplying the 6D internal space ($\t g_{mn}$), and field strengths with components only partially covering the 4D noncompact spacetime ($\h g_{\mu\nu}$) -- such fluxes are referred to as non-MN type.\footnote{The fluxes used are time-dependent and not purely internal, and thus standard flux quantization conditions do not apply.} We identify the 4D Einstein frame through a Weyl rescaling of the 4D metric, which involves the aforementioned time-dependent scale factor. Notably, we do not consider any local sources in constructing backgrounds.
  
  \item We construct backgrounds by directly solving the type IIB supergravity equations. First, we satisfy the equations of motion and Bianchi identities for the form fields consistently by maintaining the internal components (if any) of the fluxes on any 6D Ricci-flat compact K\"ahler manifold.\footnote{Confining our analysis to a Ricci-flat compact K\"ahler internal manifold is motivated by the fact that the constraints on the internal components of the fluxes, arising from the supergravity equations, can be satisfied by known forms on such manifolds. For our purposes, we only employ the harmonic K\"ahler form $J_2$ and its Hodge dual. Extending the discussion to more general internal manifolds is left for future work.} No further restrictions are imposed on the 6D internal manifold, nor is any specific choice required for it. Second, we tackle the Einstein field equations, allowing only time-dependence in the independent components of the metric $\hat{g}_{\mu\nu}(t)$ and in various noncompact components of the fluxes. This reduces to solving a system of nonlinear coupled ODEs w.r.t. $t$.
  
  \item Solving the system of nonlinear coupled ODEs analytically is managed by systematically studying linear combinations of the ODEs, which result in a simpler system.
  
  \item More generally, our constructions are 4D Bianchi type I universes with negative Ricci scalar; however, special cases yield spatially flat FLRW universes (thereby demonstrating that the isometries of FLRW spacetime can be retained even in the presence of non-MN type fluxes). In some examples, the 4D scale factors exhibit power-law growth at late times.
  
  \item We show that the associated energy-momentum tensors (both 10D and the resulting 4D effective) satisfy the null, weak, strong, and dominant energy conditions.
  
  \item Next, we revisit the Maldacena--Nu\~nez no-go analysis, incorporating new elements that appear in our constructions, namely an overall noncompact spacetime-dependent scale factor multiplying the internal metric, and field strengths with components partially covering the noncompact directions. For this analysis (which is largely independent from above), we also include a warp factor dependent on the internal coordinates, which multiplies the noncompact spacetime. We argue that, with these generalizations, a 4D Einstein-frame metric with positive Ricci scalar cannot be ruled out solely by such an analysis. In this regard, even the contribution of the 6D scale factor is individually notable.
  
  \item At the end, we briefly discuss a range of topics, including the moduli space of the constructed solutions, supergravity constraints on the 6D scale factor, late-time behaviour of the 4D scale factors, potential cosmological applications, and a tabulation of the solvable flux combinations.
\end{itemize}

The rest of the paper is organized as follows. In section \ref{sec:TDepBackgIIBFlux}, we present our constructions of analytical time-dependent backgrounds in 10D type IIB supergravity with nontrivial fluxes. In section \ref{sec:EnergyConds}, we check energy conditions for the associated energy momentum tensors. In section \ref{sec:MNextension}, we extend the Maldacena--Nu\~nez analysis. In section \ref{sec:Discuss}, we provide several remarks on our constructions along with potential future directions. Appendix \ref{app:Conven} outlines all our conventions. Appendix \ref{app:4DgHatEinProf} presents some representative 4D scale factor profiles for examples we have constructed in this work.

\section{Construction of time-dependent backgrounds}
\label{sec:TDepBackgIIBFlux}

In this section, we solve the type IIB supergravity equations in 10D, as listed in appendix \ref{app:SUGRAEOMs}, based on a specific ansatz for the 10D Einstein frame metric, the 3- and 5-form fluxes, and the axiodilaton. In the first subsection, we provide the ansatz, which is general enough to capture all our explicit constructions, and set up the corresponding equations. In the second subsection, we analyze various subcases depending on which fields are kept nontrivial, and outline the general conditions on the forms entering the ansatz, along with potential solutions, under which the supergravity equations can be consistently satisfied. In the third subsection, we construct concrete examples that satisfy these conditions and explicitly realize these potential solutions.

\subsection{Ansatz, resulting equations and 4D Einstein frame}

We consider the following ansatz, which captures all our explicit constructions:\footnote{For our convention on the Hodge dual and the notations of the hat and tilde, see appendix \ref{app:Conven}.}
\begin{align}
  &ds^2_{10}= \h g_{\mu\nu}(x) dx^\mu dx^\nu+e^{2\beta(x)}\t g_{mn}(y)dy^m dy^n\,, \nn\\
  &\t F_5=\h\star \gamma_1(x)\wedge \lambda_2(y)+e^{2\beta(x)}\gamma_1(x)\wedge\t\star\lambda_2(y)\,, \nn\\
  &G_3=\zeta_1(x)\wedge \eta_2(y) +q\ \h\star\vartheta_1(x)\,. \label{eq:CommonAnsatz}
\end{align}
In the above, $e^{2\beta}$ (where  $\beta$ is a real function) acts as an external $x$-dependent scale factor on the 6D internal metric $\t g_{mn}$.\footnote{This makes the 10D geometry a warped product spacetime.} $\t F_5$ is self-dual. $\lambda_2$ and $\eta_2$ are forms on the 6D internal manifold, while $\gamma_1,\zeta_1$ and $\vartheta_1$ are forms on the noncompact 4D spacetime. The forms $\gamma_1, \zeta_1,\vartheta_1$, and $\lambda_2$ are real, while $\eta_2$ and the constant parameter $q$ can be complex. At this stage, we keep the axiodilaton $\tau(x,y)$ with dependence on both $x$ and $y$ coordinates. Here, we keep the 6D internal metric $\tilde{g}_{mn}$ general, and it will be specialized when constructing explicit backgrounds in the next subsection. 

Next, we provide the equations derived from the equations of motion and Bianchi identities \eqref{eq:SUGRAEOMForms}--\eqref{eq:SUGRAEinsteinEq} by plugging in the above ansatz.\footnote{For a detailed breakdown of the terms, see appendix \ref{app:EOMTermDetail}.}

\paragraph{$\t F_5$ Bianchi identity:} Equating the various components on both sides of the $\tilde{F}_5$ Bianchi identity yield the following equations:
\begin{align}
  [4,2]\ \textrm{components}:&\quad \h d\h\star \gamma_1\wedge \lambda_2=\frac{i}{2\textrm{Im}\tau} \zeta_1 \wedge\h\star\vartheta_1\wedge \left(\bar{q}\eta_2 - q\bar\eta_2\right)\,, \nn\\
  [3,3]\ \textrm{components}:&\quad \h\star \gamma_1\wedge \t d\lambda_2=0\,, \nn\\
  [2,4]\ \textrm{components}:&\quad \h d \left(e^{2\beta}\gamma_1\right)\wedge\t\star\lambda_2=0\,, \nn\\
  [1,5]\ \textrm{components}:&\quad e^{2\beta}\gamma_1\wedge\t d\t\star\lambda_2=0\,. \label{eq:F5BianchiAnsatz}
\end{align}

\paragraph{$G_3$ EOM:} Equating the various components on both sides of the $G_3$ equation of motion gives:
\begin{align}
  [4,4]\ \textrm{components}:&\quad \h d\left(e^{2\beta}\h\star\zeta_1\right)\wedge \t\star\eta_2 = i\h\star \gamma_1\wedge\zeta_1\wedge \lambda_2\wedge\eta_2 -\frac{i}{\textrm{Im}\tau} e^{2\beta}\h d\tau\wedge\h\star\zeta_1\wedge \textrm{Re}\t\star\eta_2 \nn\\
  &\qquad\qquad\qquad\qquad\qquad+i q e^{2\beta}\gamma_1\wedge\h\star\vartheta_1\wedge\t\star\lambda_2\,, \nn\\
  [3,5]\ \textrm{components}:&\quad e^{2\beta}\h\star\zeta_1\wedge\t d\t\star\eta_2 = -\frac{i}{\textrm{Im}\tau}e^{2\beta} \h\star\zeta_1\wedge\t d\tau\wedge \textrm{Re}\t\star\eta_2\,, \nn\\
  [2,6]\ \textrm{components}:&\quad q\h d\left(e^{6\beta}\vartheta_1\right)\wedge\t\epsilon=ie^{2\beta}\gamma_1\wedge\zeta_1\wedge\t\star\lambda_2\wedge\eta_2-\frac{i\textrm{Re}q}{\textrm{Im}\tau} e^{6\beta}\h d\tau\wedge\vartheta_1\wedge\t\epsilon\,. \label{eq:G3EOMAnsatz}
\end{align}

\paragraph{$G_3$ Bianchi identity:} Equating the various components on both sides of the $G_3$ Bianchi identity gives:
\begin{align}
  [4,0]\ \textrm{components}:&\quad q\h d\h\star\vartheta_1=\frac{\textrm{Im}q}{\textrm{Im}\tau}\h d\tau\wedge\h\star\vartheta_1\,, \nn\\
  [3,1]\ \textrm{components}:&\quad \frac{\textrm{Im}q}{\textrm{Im}\tau}\h\star\vartheta_1\wedge\t d\tau=0\,, \nn\\
  [2,2]\ \textrm{components}:&\quad \h d\zeta_1\wedge\eta_2=\h d\tau\wedge\zeta_1\wedge\frac{\textrm{Im}\eta_2}{\textrm{Im}\tau}\,, \nn\\
  [1,3]\ \textrm{components}:&\quad \zeta_1\wedge\t d\eta_2=\zeta_1\wedge\t d\tau\wedge\frac{\textrm{Im}\eta_2}{\textrm{Im}\tau}\,. \label{eq:G3BianchiAnsatz}
\end{align}

\paragraph{$\tau$ EOM:} The $\tau$ equation of motion (comprises of $[4,6]$ components) reads as:
\begin{align}
  \h d\left(e^{6\beta}\h\star\h d\tau\right)\wedge \t\epsilon + e^{4\beta}\h\epsilon\wedge\t d\t\star\t d\tau =& -\frac{i}{\textrm{Im}\tau}\left(e^{6\beta}\h d\tau\wedge\h\star\h d\tau\wedge\t\epsilon + e^{4\beta}\h\epsilon\wedge\t d\tau\wedge\t\star\t d\tau\right) \nn\\
  &-\frac{i}{2}e^{2\beta}\zeta_1\wedge\h\star\zeta_1\wedge \eta_2\wedge\t\star\eta_2 +\frac{iq^2}{2} e^{6\beta}\vartheta_1\wedge\h\star\vartheta_1\wedge\t\epsilon\,. \label{eq:TauEOMAnsatz}
\end{align}

\paragraph{Einstein field equations:} The $\mu\nu$-components of the Einstein field equations gives:
\begin{align}
  &\hat{G}_{\mu \nu}-6\left(\hat{\nabla}_\mu \hat{\nabla}_\nu \beta+\partial_\mu \beta \partial_\nu \beta\right) +3\h g_{\mu\nu}\left(2\hat{\nabla}^{\h 2} \beta+7\partial_\sigma \beta \partial^{\h\sigma} \beta\right) -\frac{1}{2}\h g_{\mu\nu}e^{-2\beta}\tilde{R} \nn\\
  &=\frac{e^{-4\beta}}{4}\lambda_2\cdot\lambda_2\left[2\gamma_\mu\gamma_\nu-\h g_{\mu\nu}\gamma_\sigma\gamma^{\h\sigma}\right] + \frac{e^{-4\beta}}{4\textrm{Im}\tau}\eta_2\cdot\bar\eta_2 \left[2\zeta_{\mu}\zeta_{\nu} - \h g_{\mu\nu} \zeta_\sigma\zeta^{\h\sigma}\right] +\frac{q\bar q}{4\textrm{Im}\tau} \left[2\vartheta_{\mu}\vartheta_{\nu} - \h g_{\mu\nu} \vartheta_\sigma\vartheta^{\h\sigma}\right] \nn\\
  &\quad +\frac1{2(\textrm{Im}\tau)^2}\partial_{(\mu}\tau\partial_{\nu)}\bar{\tau}-\frac1{4(\textrm{Im}\tau)^2} \h g_{\mu\nu}\partial_\sigma\tau\partial^{\h \sigma}\bar\tau -\frac1{4(\textrm{Im}\tau)^2} \h g_{\mu\nu} e^{-2\beta}\partial_m\tau\partial^{\t m}\bar\tau\,. \label{eq:munuEinsteinEqAnsatz}
\end{align}
The $mn$-components of the Einstein field equations gives:
\begin{align}
  &\t G_{mn} + 5e^{2\beta}\t g_{mn}\left(\hat{\nabla}^{\h 2} \beta+3\partial_\mu \beta \partial^{\h\mu} \beta\right)-\frac{1}{2}\t g_{mn}e^{2\beta}\hat{R} \nn\\
  &=\frac{e^{-2\beta}}{4}\gamma_\mu\gamma^{\h\mu}\left[\t g_{mn}\lambda_2\cdot\lambda_2-2\lambda_m{}^{\t p}\lambda_{np}\right] + \frac{e^{-2\beta}}{4\textrm{Im}\tau} \zeta_\mu\zeta^{\h\mu} \left[2\eta{}_{(m}{}^{\t p} \bar\eta{}_{n)p} - \t g_{mn} \eta_2\cdot\bar\eta_2\right] +\frac{q\bar q}{4\textrm{Im}\tau}e^{2\beta}\vartheta_\sigma\vartheta^{\h\sigma}\t g_{mn} \nn\\
  &\quad +\frac1{2(\textrm{Im}\tau)^2}\partial_{(m}\tau\partial_{n)}\bar{\tau}-\frac1{4(\textrm{Im}\tau)^2} \t g_{mn} \partial_l\tau\partial^{\t l}\bar\tau -\frac1{4(\textrm{Im}\tau)^2} \t g_{mn} e^{2\beta}\partial_\mu\tau\partial^{\h \mu}\bar\tau\,. \label{eq:mnEinsteinEqAnsatz}
\end{align}
The $\mu m$-components of the Einstein field equations gives:
\begin{align}
  \frac1{2(\textrm{Im}\tau)^2}\partial_{(\mu}\tau\partial_{m)}\bar{\tau}=0\,. \label{eq:mumEinsteinEqAnsatz}
\end{align}

\paragraph{4D Einstein frame:} Before closing this subsection, we note for later use that one may pass to the 4D Einstein frame via a Weyl rescaling of the metric $\h g_{\mu\nu}$, as follows. The Ricci scalar corresponding to the 10D Einstein frame metric of \eqref{eq:CommonAnsatz} is given by
\begin{align}
  R_g=\hat{R} + e^{-2\beta}\tilde{R} -6\left(2\hat{\nabla}^{\h 2} \beta+7\partial_\mu \beta \partial^{\h\mu} \beta\right)\,.
\end{align}
Substituting this into the 10D supergravity action \eqref{eq:10DIIBSUGRAAction} and integrating over the 6D internal space coordinates yields an effective 4D action. Focusing on the Einstein-Hilbert term, we obtain:
\begin{align}
  S_{\textrm{IIB}}=\frac1{2\kappa_{10}^2} \int d^6y \sqrt{\t g}\ \int d^4x \sqrt{-\h g^E}\ e^{\Omega}e^{6\beta}\h R^E +\cdots \,,\quad \h g_{\mu\nu}=e^{\Omega(x)} \h g^{E}_{\mu\nu}\,.
\end{align}
For the 4D Einstein frame, where the 4D gravitational constant is not spacetime dependent, the Weyl factor $\Omega$ must be chosen as
\begin{align}
  e^{\Omega}=e^{-2\kappa_4}e^{-6\beta}\,,\quad \h g^{E}_{\mu\nu}=e^{2\kappa_4}e^{6\beta} \h g_{\mu\nu}\,,\quad \h R^E=e^{-2\kappa_4}e^{-6\beta} \left[\h R-18 \left(\hat{\nabla}^{\h 2}\beta +3 \partial_\mu \beta \partial^{\h\mu} \beta\right)\right]\,, \label{eq:4DEinsteinFrameWeyl}
\end{align}
where $\kappa_4$ is a constant parameter that we set to zero hereafter for notational simplicity.

\subsection{Subcases and general solvability}
\label{subsec:SubCaseGenSolv}

Here we subsequently consider the following scenarios: where $G_3$ is turned off, where $\t F_5$ is turned off, and where neither of them is turned off. We ignore the $y$-dependence of $\tau$. Our strategy for solving the supergravity equations is as follows. First, we focus on the equations of motion and Bianchi identities for the form fields, determining the constraints on both the internal and noncompact components of the form fields. The factorization of the $x$-dependence from the $y$-dependence in the internal components of the 10D Einstein field equations also imposes additional constraints on the internal components of the form fields. We maintain these constraints on the internal components of the form fields using known forms on any 6D Ricci-flat compact K\"ahler manifold. We then discuss how the constraints on the noncompact components of the form fields can be maintained. Finally, we argue the solvability of the remaining equations -- i.e., the noncompact components of the 10D Einstein field equations and a compatibility condition on $\h R$ coming from the internal components of the 10D Einstein field equations -- by counting the numbers of equations and unknowns.

Note that for the cases we have $G_3$ turned on, $F_3,H_3$ fluxes are nontrivial and are given by: $F_3=\textrm{Re}G_3-\textrm{Im}G_3\ \textrm{Re}\tau/\textrm{Im}\tau\,,\ H_3=-\textrm{Im}G_3/\textrm{Im}\tau$. Notably, $\textrm{Re} \tau$ is either kept constant or allowed to vary in time.

\subsubsection{\texorpdfstring{$G_3= 0,\ \textrm{self-dual}\ \t F_5,\ \tau=\tau_0$}{z}}
\label{sssec:Zero3formConsT}

We first consider the case where the 3-form flux is turned off, i.e., $G_3 = 0$ (which can be done by setting $\zeta_1,\eta_2=0$), and the axiodilaton $\tau$ is kept constant at any $\tau_0$ (while the case with time-dependent $\tau$ will be treated subsequently). In this setup, the $G_3$ equation of motion \eqref{eq:G3EOMAnsatz}, the $G_3$ Bianchi identity \eqref{eq:G3BianchiAnsatz}, the $\tau$ equation of motion \eqref{eq:TauEOMAnsatz}, and the $\mu m$-components of the Einstein field equations \eqref{eq:mumEinsteinEqAnsatz} are all trivially satisfied.

Now, the $\t F_5$ Bianchi identity \eqref{eq:F5BianchiAnsatz} can be satisfied with nonzero $\t F_5$ only if we require
\begin{align}
  \h d\h\star\gamma_1=0\,,\quad \h d\left(e^{2\beta}\gamma_1\right)=0\,,\quad \t d\lambda_2=0\,,\quad \t d\t\star\lambda_2=0\,, \label{eq:GammaLambdaCondTauC}
\end{align}
i.e., $\lambda_2$ is harmonic.

We now turn to the $mn$-components of the Einstein field equations. If we demand the internal manifold to be Ricci flat, i.e., $\t R_{mn}=0$, we have $\lambda_m{}^{\t p}\lambda_{np}\propto \t g_{mn}$, which follows from the factorization of the $x$-dependence from the $y$-dependence in \eqref{eq:mnEinsteinEqAnsatz}. Consistent with the harmonicity of $\lambda_2$, on a compact K\"ahler internal manifold we can simply take\footnote{For the properties of $J_2$, see \cite{Candelas:1987is,huybrechts2005complex}. Its dual is given by: $\t\star J_2=J_2\wedge J_2/2\,.$ \label{fn:CandelasJ2prop}}
\begin{align}
  \lambda_2= k J_2\,,\quad J_m{}^{\t p}J_{pn}=-\t g_{mn}\,, \label{eq:LambdaJ2relTauC}
\end{align}
where $k$ is a real constant and $J_2$ is the K\"ahler form. Note that $J_2$ is a real form, which ensures the reality of $\t F_5$. With this choice, the $mn$-components of the Einstein equations yield the following condition on $\h R$:
\begin{align}
  \hat{R}=10\left(\hat{\nabla}^{\h 2} \beta+3\partial_\mu \beta \partial^{\h\mu} \beta\right)-\frac{k^2}{2}e^{-4\beta}\gamma_\mu\gamma^{\h\mu}\,. \label{eq:mnEinsteinRHatTauC}
\end{align}

Finally, the $\mu\nu$-components of the Einstein field equations \eqref{eq:munuEinsteinEqAnsatz} lead to
\begin{align}
  \hat{G}_{\mu \nu}-6\left(\hat{\nabla}_\mu \hat{\nabla}_\nu \beta+\partial_\mu \beta \partial_\nu \beta\right) +3\h g_{\mu\nu}\left(2\hat{\nabla}^{\h 2} \beta+7\partial_\sigma \beta \partial^{\h\sigma} \beta\right)=\frac{3k^2}{4}e^{-4\beta}\left(2\gamma_\mu\gamma_\nu-\h g_{\mu\nu}\gamma_\sigma\gamma^{\h\sigma}\right)\,. \label{eq:GHatmunuCondTauC}
\end{align}
From above the resulting expression for $\h R$ when compared with \eqref{eq:mnEinsteinRHatTauC} yields:
\begin{align}
  \left(\hat{\nabla}^{\h 2} \beta+6\partial_\mu \beta \partial^{\h\mu} \beta\right) +\frac{k^2}{4}e^{-4\beta}\gamma_\mu\gamma^{\h\mu}=0\,, \label{eq:RHatCompatTauC}
\end{align}
which provides a simpler condition to satisfy.

We now argue that the system of equations above admits solutions. In general, one could consider solutions with dependence on all the 4D spacetime coordinates; however, since we are interested in a time-dependent metric $\h g_{\mu\nu}(t)$, consider $\beta = \beta(t)$. In \eqref{eq:GammaLambdaCondTauC}, the closedness of $e^{2\beta}\gamma_1$ can be satisfied by choosing $\gamma_1 = \gamma_t(t)\,dt$, while the coclosedness of $\gamma_1$ then determines the component $\gamma_t$ in terms of $\beta$ and the independent components of the metric $\hat{g}_{\mu\nu}$. The Einstein equations \eqref{eq:GHatmunuCondTauC} together with the $\hat{R}$ compatibility condition \eqref{eq:RHatCompatTauC} yield a system of equations whose number matches that of the unknowns, and can therefore, in general, be solved to determine the metric components and the function $\beta(t)$. Explicit constructions of the functions $\h g_{\mu\nu},\beta,\gamma_1$ are provided in section \ref{subsec:ExplicitSols}. Note that no particular choice of the 6D internal manifold is required. The solutions constructed by our procedure are valid for any Ricci-flat compact K\"ahler internal manifold. For the torus $T^6$, the components of $\lambda_2$ will be constant, whereas for other compatible manifolds, such as Calabi-Yau threefolds, $\lambda_2$ (and consequently $\tilde{F}_5$) will exhibit nontrivial $y$ dependence.

\subsubsection{\texorpdfstring{$G_3= 0,\ \textrm{self-dual}\ \t F_5,\ \tau(t)$}{z}}
\label{sssec:Zero3formTauT}

In the previous setup with zero 3-form flux, now we allow $\tau$ to depend on the 4D spacetime coordinates, it remains decoupled from the equations for $G_3$ and $\t F_5$. Hence the conditions in \eqref{eq:GammaLambdaCondTauC} still apply, and we consider \eqref{eq:LambdaJ2relTauC} as in the previous discussion. However, $\tau$ is now governed by the equation of motion:
\begin{align}
  \h d\left(e^{6\beta}\h\star\h d\tau\right)= -\frac{i}{\textrm{Im}\tau}e^{6\beta}\h d\tau\wedge\h\star\h d\tau\,. \label{eq:TauEOMTauT}
\end{align}
The $\h R$ condition resulting from the $mn$-components of 10D Einstein field equations is modified to:
\begin{align}
  \h R=10\left(\hat{\nabla}^{\h 2} \beta+3\partial_\mu \beta \partial^{\h\mu} \beta\right) -\frac{k^2}{2}e^{-4\beta}\gamma_\mu\gamma^{\h\mu} +\frac1{2(\textrm{Im}\tau)^2} \partial_\mu\tau\partial^{\h \mu}\bar\tau\,.
\end{align}
While, the $\mu\nu$-components of 10D Einstein field equations are modified as:
\begin{align}
  &\hat{G}_{\mu \nu}-6\left(\hat{\nabla}_\mu \hat{\nabla}_\nu \beta+\partial_\mu \beta \partial_\nu \beta\right) +3\h g_{\mu\nu}\left(2\hat{\nabla}^{\h 2} \beta+7\partial_\sigma \beta \partial^{\h\sigma} \beta\right) \nn\\
  &=\frac{3k^2}{4}e^{-4\beta} \left[2\gamma_\mu\gamma_\nu-\h g_{\mu\nu}\gamma_\sigma\gamma^{\h\sigma}\right] +\frac1{2(\textrm{Im}\tau)^2}\partial_{(\mu}\tau\partial_{\nu)}\bar{\tau}-\frac1{4(\textrm{Im}\tau)^2} \h g_{\mu\nu}\partial_\sigma\tau\partial^{\h \sigma}\bar\tau\,. \label{eq:GHatmunuCondTauT}
\end{align}
However, it is straightforward to check the resulting $\h R$ compatibility condition remains the same as in the previous case, i.e., \eqref{eq:RHatCompatTauC} holds with the underlying metric $\h g_{\mu\nu}$ satisfying Einstein equations \eqref{eq:GHatmunuCondTauT} above.

Again considering only time dependences, the real and the imaginary parts of $\tau(t)$ are determined by \eqref{eq:TauEOMTauT} in terms of $\beta(t)$ and the independent metric components $\h g_{\mu\nu}(t)$. As in the previous case, considering $\gamma_1=\gamma_t(t)dt$, \eqref{eq:GammaLambdaCondTauC} determines $\gamma_t$. The remaining equations, \eqref{eq:GHatmunuCondTauT} and \eqref{eq:RHatCompatTauC}, form a system in which the number of equations matches the number of unknowns, $\h g_{\mu\nu}$ and $\beta$. Explicit constructions of the functions $\h g_{\mu\nu},\beta,\gamma_1,\tau$ are provided in section \ref{subsec:ExplicitSols}.

\subsubsection{\texorpdfstring{$\t F_5=0,\ G_3,\tau\neq 0,\ \tau(t)$}{x}}
\label{sssec:ZeroF5TauT}

We now consider the case where the 5-form flux is turned off, i.e., $\tilde{F}_5 = 0$ (which can be done by setting $\gamma_1, \lambda_2=0$), and allow the axiodilaton $\tau$ to depend only on the 4D spacetime coordinates, and include only the $[1,2]$ components of $G_3$ (which can be done by setting $q, \vartheta_1=0$). In this setup, the $\t F_5$ Bianchi identity \eqref{eq:F5BianchiAnsatz}, the $[2,6]$ components of the $G_3$ equation of motion in \eqref{eq:G3EOMAnsatz}, the $\mu m$-components of the Einstein field equations \eqref{eq:mumEinsteinEqAnsatz} are all trivially satisfied.

The $[3,5]$ components of the $G_3$ equation of motion in \eqref{eq:G3EOMAnsatz} and the $[1,3]$ components of the $G_3$ Bianchi identity in \eqref{eq:G3BianchiAnsatz} can be satisfied with nonzero $G_3$ only if we respectively require:
\begin{align}
  \t d\t\star\eta_2=0\,,\quad \t d\eta_2=0\,,
\end{align}
i.e., $\eta_2$ is harmonic.

Now in order to satisfy the $[4,4]$ components of the $G_3$ equation of motion in \eqref{eq:G3EOMAnsatz} and the $[2,2]$ components of the $G_3$ Bianchi identity in \eqref{eq:G3BianchiAnsatz}, we respectively assume:
\begin{align}
  \textrm{Re}\eta_2=0\,,\quad \h d\tau\wedge\zeta_1=0\,. \label{eq:EtaZetaAssum0F5}
\end{align}
Consequently, we are left with the following conditions:
\begin{align}
  \h d\left(e^{2\beta}\h\star\zeta_1\right)=0\,,\quad \h d\zeta_1=0\,. \label{eq:ZetaCond0F5}
\end{align}

$\tau$ is governed by the equation of motion \eqref{eq:TauEOMAnsatz}:\footnote{Simplified by using the identity: $\eta_2\wedge\t\star\eta_2=\eta_2\cdot\eta_2\ \t\epsilon$.}
\begin{align}
  \h d\left(e^{6\beta}\h\star\h d\tau\right) = -\frac{i}{\textrm{Im}\tau} e^{6\beta}\h d\tau\wedge\h\star\h d\tau -\frac{i}{2}e^{2\beta} \eta_2\cdot\eta_2\ \zeta_1\wedge\h\star\zeta_1\,. \label{eq:TauEOM0F5}
\end{align}
Note that $\eta_2\cdot\eta_2$ needs to be $y$-independent.\footnote{Alternatively, one could require $\zeta_\mu\zeta^{\h\mu}=0$, in which case the term involving $\eta_2\cdot\eta_2$ drops since $\zeta_1\wedge\h\star\zeta_1=-\zeta_\mu\zeta^{\h\mu}\h\epsilon$. But this choice leads to an overdetermined system of equations.} This condition is satisfied in the setup below, where $\eta_2\cdot\eta_2=-3k^2$ for some real parameter $k$.

We now turn to the $mn$-components of the Einstein field equations. If we demand the internal manifold to be Ricci flat, i.e., $\t R_{mn}=0$, we have $\eta_m{}^{\t p}\bar{\eta}_{np}\propto \t g_{mn}$, which follows from the factorization of the $x$-dependence from the $y$-dependence in \eqref{eq:mnEinsteinEqAnsatz}. Consistent with the harmonicity of $\eta_2$, on a compact K\"ahler internal manifold we can simply take\footnote{See footnote \ref{fn:CandelasJ2prop}.}
\begin{align}
  \eta_2= ik J_2\,,\quad J_m{}^{\t p}J_{pn}=-\t g_{mn}\,, \label{eq:EtaJ2rel0F5}
\end{align}
where $k$ is a real constant and $J_2$ is the K\"ahler form. Since $J_2$ is a real form, $\eta_2$ is ensured to be purely imaginary as demanded in \eqref{eq:EtaZetaAssum0F5}. With this choice, the $mn$-components of the Einstein equations yield the following condition on $\h R$:
\begin{align}
  \hat{R}=10\left(\hat{\nabla}^{\h 2} \beta+3\partial_\mu \beta \partial^{\h\mu} \beta\right) +\frac{k^2}{2\textrm{Im}\tau} e^{-4\beta} \zeta_\mu\zeta^{\h\mu} +\frac1{2(\textrm{Im}\tau)^2} \partial_\mu\tau\partial^{\h \mu}\bar\tau\,. \label{eq:mnEinsteinRHat0F5}
\end{align}

Finally, the $\mu\nu$-components of the Einstein field equations \eqref{eq:munuEinsteinEqAnsatz} lead to
\begin{align}
  &\hat{G}_{\mu \nu}-6\left(\hat{\nabla}_\mu \hat{\nabla}_\nu \beta+\partial_\mu \beta \partial_\nu \beta\right) +3\h g_{\mu\nu}\left(2\hat{\nabla}^{\h 2} \beta+7\partial_\sigma \beta \partial^{\h\sigma} \beta\right) \nn\\
  &=\frac{3k^2}{4\textrm{Im}\tau}e^{-4\beta} \left[2\zeta_{\mu}\zeta_{\nu} - \h g_{\mu\nu} \zeta_\sigma\zeta^{\h\sigma}\right] +\frac1{2(\textrm{Im}\tau)^2}\partial_{(\mu}\tau\partial_{\nu)}\bar{\tau}-\frac1{4(\textrm{Im}\tau)^2} \h g_{\mu\nu}\partial_\sigma\tau\partial^{\h \sigma}\bar\tau\,. \label{eq:GHatmunuCond0F5}
\end{align}
From above the resulting expression for $\h R$ when compared with \eqref{eq:mnEinsteinRHat0F5} yields:
\begin{align}
  \left(\hat{\nabla}^{\h 2} \beta+6\partial_\mu \beta \partial^{\h\mu} \beta\right) +\frac{k^2}{8\textrm{Im}\tau} e^{-4\beta} \zeta_\mu\zeta^{\h\mu}=0\,, \label{eq:RHatCompat0F5}
\end{align}
which provides a simpler condition to satisfy.

We now argue that the system of equations above admits solutions. Considering only time dependences as in the previous flux setups, the real and the imaginary parts of $\tau(t)$ are determined by \eqref{eq:TauEOM0F5} in terms of $\beta(t)$ and the independent metric components $\h g_{\mu\nu}(t)$. By choosing $\zeta_1=\zeta_t(t)dt$, we can satisfy $\h d\zeta_1=0$ and $\h d\tau\wedge\zeta_1=0$ as required. Then the first condition given in \eqref{eq:ZetaCond0F5} determines the component $\zeta_t$. The remaining equations, \eqref{eq:GHatmunuCond0F5} and \eqref{eq:RHatCompat0F5}, form a system in which the number of equations matches the number of unknowns, $\h g_{\mu\nu}$ and $\beta$. Explicit constructions of the functions $\h g_{\mu\nu},\beta,\zeta_1,\tau$ are provided in section \ref{subsec:ExplicitSols}.

\subsubsection{\texorpdfstring{$\textrm{Self-dual}\ \t F_5,\ G_3,\tau\neq 0,\ \tau(t)$}{x}}
\label{sssec:SDF5NZG3TauT}

We now turn to the case where both the 3-form and 5-form fluxes are non-vanishing, following \eqref{eq:CommonAnsatz}, and, we allow the axiodilaton $\tau$ to depend only on the 4D spacetime coordinates. This constitutes the most general flux configuration considered in this article; however, solving the supergravity equations has been managed by turning the $[1,2]$ components of the $G_3$ flux off (i.e., by setting $\zeta_1,\eta_2=0$).\footnote{Note that instead keeping the $[1,2]$ components of $G_3$ while turning off the $[3,0]$ components leads to an overdetermined system of equations.}

In this flux setup, the $[3,5]$ components of the $G_3$ equation of motion in\eqref{eq:G3EOMAnsatz}, the $[3,1]$, $[2,2]$ and $[1,3]$ components of the $G_3$ Bianchi identity in \eqref{eq:G3BianchiAnsatz}, and the $\mu m$-components of the Einstein field equations \eqref{eq:mumEinsteinEqAnsatz} are all trivially satisfied.

Now, the $\t F_5$ Bianchi identity \eqref{eq:F5BianchiAnsatz} can be satisfied with nonzero $\t F_5$ only if we require
\begin{align}
  \h d\h\star\gamma_1=0\,,\quad \h d\left(e^{2\beta}\gamma_1\right)=0\,,\quad \t d\lambda_2=0\,,\quad \t d\t\star\lambda_2=0\,, \label{eq:GammaLambdaCondAllFlux}
\end{align}
i.e., $\lambda_2$ is harmonic.

The $[4,4]$ components of the $G_3$ equation of motion in\eqref{eq:G3EOMAnsatz} yield:
\begin{align}
  \gamma_1\wedge\h\star\vartheta_1=0\,,
\end{align}
while the $[2,6]$ components give:
\begin{align}
  q\h d\left(e^{6\beta}\vartheta_1\right)=-\frac{i\textrm{Re}q}{\textrm{Im}\tau} e^{6\beta}\h d\tau\wedge\vartheta_1\,. \label{eq:G3EOM26AllFlux}
\end{align}
Alongside these, we have the $[4,0]$ components of the $G_3$ Bianchi identity in \eqref{eq:G3BianchiAnsatz}:
\begin{align}
  q\h d\h\star\vartheta_1=\frac{\textrm{Im}q}{\textrm{Im}\tau}\h d\tau\wedge\h\star\vartheta_1\,.  \label{eq:G3Bianchi40AllFlux}
\end{align}

$\tau$ is governed by the equation of motion \eqref{eq:TauEOMAnsatz}:
\begin{align}
  \h d\left(e^{6\beta}\h\star\h d\tau\right) = -\frac{i}{\textrm{Im}\tau} e^{6\beta}\h d\tau\wedge\h\star\h d\tau +\frac{iq^2}{2} e^{6\beta}\vartheta_1\wedge\h\star\vartheta_1\,. \label{eq:TauEOMAllFlux}
\end{align}

We now turn to the $mn$-components of the Einstein field equations. If we demand the internal manifold to be Ricci flat, i.e., $\t R_{mn}=0$, we have $\lambda_m{}^{\t p}\lambda_{np}\propto \t g_{mn}$, which follows from the factorization of the $x$-dependence from the $y$-dependence in \eqref{eq:mnEinsteinEqAnsatz}. Consistent with the harmonicity of $\lambda_2$, on a compact K\"ahler internal manifold we can simply take\footnote{See footnote \ref{fn:CandelasJ2prop}.}
\begin{align}
  \lambda_2= k J_2\,,\quad J_m{}^{\t p}J_{pn}=-\t g_{mn}\,, \label{eq:LambdaJ2relAllFlux}
\end{align}
where $k$ is a real constant and $J_2$ is the K\"ahler form. Note that $J_2$ is a real form, which ensures the reality of $\t F_5$. With this choice, the $mn$-components of the Einstein equations yield the following condition on $\h R$:
\begin{align}
  \hat{R}=10\left(\hat{\nabla}^{\h 2} \beta+3\partial_\mu \beta \partial^{\h\mu} \beta\right) -\frac{k^2 e^{-4\beta}}{2}\gamma_\mu\gamma^{\h\mu} -\frac{q\bar q}{2\textrm{Im}\tau}\vartheta_\sigma\vartheta^{\h\sigma} +\frac1{2(\textrm{Im}\tau)^2} \partial_\mu\tau\partial^{\h \mu}\bar\tau\,. \label{eq:mnEinsteinRHatAllFlux}
\end{align}

Finally, the $\mu\nu$-components of the Einstein field equations \eqref{eq:munuEinsteinEqAnsatz} lead to
\begin{align}
  &\hat{G}_{\mu \nu}-6\left(\hat{\nabla}_\mu \hat{\nabla}_\nu \beta+\partial_\mu \beta \partial_\nu \beta\right) +3\h g_{\mu\nu}\left(2\hat{\nabla}^{\h 2} \beta+7\partial_\sigma \beta \partial^{\h\sigma} \beta\right) \nn\\
  &=\frac{3k^2}{4}e^{-4\beta}\left[2\gamma_\mu\gamma_\nu-\h g_{\mu\nu}\gamma_\sigma\gamma^{\h\sigma}\right] +\frac{q\bar q}{4\textrm{Im}\tau} \left[2\vartheta_{\mu}\vartheta_{\nu} - \h g_{\mu\nu} \vartheta_\sigma\vartheta^{\h\sigma}\right] \nn\\
  &\quad +\frac1{2(\textrm{Im}\tau)^2}\partial_{(\mu}\tau\partial_{\nu)}\bar{\tau}-\frac1{4(\textrm{Im}\tau)^2} \h g_{\mu\nu}\partial_\sigma\tau\partial^{\h \sigma}\bar\tau\,. \label{eq:GHatmunuCondAllFlux}
\end{align}
From above the resulting expression for $\h R$ when compared with \eqref{eq:mnEinsteinRHatAllFlux} yields:
\begin{align}
  \left(\hat{\nabla}^{\h 2} \beta+6\partial_\mu \beta \partial^{\h\mu} \beta\right) +\frac{k^2}{4}e^{-4\beta}\gamma_\mu\gamma^{\h\mu}+\frac{q\bar q}{8\textrm{Im}\tau}\vartheta_\mu\vartheta^{\h\mu}=0\,, \label{eq:RHatCompatAllFlux}
\end{align}
which provides a simpler condition to satisfy.

We now argue that the system of equations above admits solutions. Considering only time dependences as in the previous flux setups, the real and the imaginary parts of $\tau(t)$ are determined by \eqref{eq:TauEOMAllFlux} in terms of $\beta(t)$ and the independent metric components $\h g_{\mu\nu}(t)$. By choosing $\gamma_1=\gamma_t(t)dt$ and letting $\vartheta_1$ have a single time-dependent spatial component in any chosen direction, say $\vartheta_3(t)$ along the third spatial direction, we can satisfy both $\h d (e^{2\beta}\gamma_1)=0$ and $\gamma_1\wedge\h\star\vartheta_1=0$ as required.\footnote{This holds provided there are no off-diagonal metric components involving the differential of the chosen spatial coordinate ($dx^3$ in the example above) in $\h g_{\mu\nu}$.} Notably \eqref{eq:G3Bianchi40AllFlux} is also satisfied under this ansatz. Then the first condition given in \eqref{eq:GammaLambdaCondAllFlux} determines the component $\gamma_t$, while \eqref{eq:G3EOM26AllFlux} determines the component $\vartheta_3(t)$. The remaining equations, \eqref{eq:GHatmunuCondAllFlux} and \eqref{eq:RHatCompatAllFlux}, form a system in which the number of equations matches the number of unknowns, $\h g_{\mu\nu}$ and $\beta$. Explicit constructions of the functions $\h g_{\mu\nu},\beta,\gamma_1,\vartheta_1,\tau$ are provided in section \ref{subsec:ExplicitSols}.

Note that \eqref{eq:G3EOM26AllFlux} yields two equations when $q$ is a generic complex number, which can potentially make the system of equations overdetermined. However, in the following two cases only one equation needs to be solved: $\textrm{Re}q=0$, or $\textrm{Im}q=0$ with $\textrm{Re}\tau=\textrm{constant}$. In our explicit constructions we focus on the first case, since this allows the possibility of a time-dependent $\textrm{Re}\tau(t)$. In this case, \eqref{eq:G3EOM26AllFlux} simply leads to $\h d(e^{6\beta}\vartheta_1)=0$.

\subsection{Explicit realizations}
\label{subsec:ExplicitSols}

For the explicit construction of the time-dependent backgrounds discussed in the previous subsection, we focus on the following ansatz for the 4D metric $\h g_{\mu\nu}$:
\begin{align}
  \h ds^2=-e^{\sigma(t)}dt^2+\sum_{i=1}^3e^{\alpha_i(t)}dx^idx^i\,, \label{eq:4DmetricAnsatz}
\end{align}
where the reality of the functions $\alpha_i(t),\sigma(t)$ ensures the Lorentzian signature of the metric.

Note that, the 4D Einstein frame metric is given by $\h g^{E}_{\mu\nu}=e^{6\beta} \h g_{\mu\nu}$. While passing to the Einstein frame corresponds to a Weyl rescaling of the metric, one can further perform a coordinate transformation $t \to t'$, with $t'=\int_0^t e^{3\beta(s)+\sigma(s)/2}ds$, to absorb the scale factor in the time component, transforming $-e^{6\beta+\sigma} dt^2$ into $-{dt'}^{2}$. However, the scale factors multiplying the spatial directions, $e^{6\beta+\alpha_i}$, cannot be removed through such a reparametrization, as although their functional forms in $t'$ change via the inverse map $t(t')$, this does not alter their functional values.\footnote{In terms of cosmic time, the 4D Einstein frame metric reads: $\h ds_E^2=-{dt'}^{2}+\sum_i e^{6\beta(t(t'))+\alpha_i(t(t'))}dx^idx^i$.} Therefore, for the cases where we obtain all $\alpha_i$ equal, the Einstein-frame metric takes a spatially flat FLRW form in the coordinates $(t',x^i)$; otherwise, it is spatially homogeneous yet anisotropic, a Bianchi type I universe.

The remainder of this subsection is organized according to the subcases discussed earlier.

\subsubsection{\texorpdfstring{$G_3= 0,\ \textrm{self-dual}\ \t F_5,\ \tau=\tau_0$}{z}}

For the case with vanishing $G_3$, nontrivial self-dual $\tilde{F}_5$ featuring $[3,2]$ and $[1,4]$ components, and constant $\tau$, here we provide the explicit forms of the functions and differential forms appearing in the ansatzes \eqref{eq:CommonAnsatz} and \eqref{eq:4DmetricAnsatz} that solve the relevant supergravity equations discussed in section \ref{sssec:Zero3formConsT}:
\begin{align}
  &\lambda_2=kJ_2\,,\quad \gamma_1=\gamma_tdt\,,\quad \gamma_t=4 c_1 e^{6 \beta} \sqrt{\frac{\dot\beta^2}{c_1^2 k^2 \left(8 c_2^2-e^{8 \beta }\right)}}\,, \nn\\
  &\sigma=c_4+c_6+c_8-\ln \left[\frac{c_1^2 k^2 \left(8 c_2^2-e^{8 \beta }\right)}{16\dot\beta^2}\right]-\frac{c_3+c_5+c_7}{8 \sqrt{2} c_2} \coth ^{-1}\left[\frac{2 \sqrt{2} c_2}{\sqrt{8 c_2^2-e^{8 \beta }}}\right]-6 \beta\,, \nn\\
  &\alpha_1=c_4-\frac{c_3}{8 \sqrt{2} c_2} \coth ^{-1}\left[\frac{2 \sqrt{2} c_2}{\sqrt{8 c_2^2-e^{8 \beta }}}\right]-6 \beta\,,\quad \alpha_2=c_6-\frac{c_5}{8 \sqrt{2} c_2} \coth ^{-1}\left[\frac{2 \sqrt{2} c_2}{\sqrt{8 c_2^2-e^{8 \beta }}}\right]-6 \beta\,, \nn\\
  &\alpha_3=c_8-\frac{c_7}{8 \sqrt{2} c_2} \coth ^{-1}\left[\frac{2 \sqrt{2} c_2}{\sqrt{8 c_2^2-e^{8 \beta }}}\right]-6 \beta\,, \nn\\
  &384 c_2^2-c_3 c_5-c_5 c_7-c_7 c_3=0\,,\quad c_2>0\,, \label{eq:0G3F5TauCEx}
\end{align}
where the function $\beta(t)$ remains unfixed due to the aforesaid relation among the integration constants $c_2,c_3,c_5,c_7$, as explained below. The reality of the above expressions requires all $c_i$ to be real, and the condition $e^{8\beta} < 8 c_2^2$ must hold.\footnote{\label{fn:AcothProp}Note that $\coth^{-1}[z]$ is complex for $-1 < z < 1$. As $c_2^{-1} \coth^{-1}[\frac{2 \sqrt{2} c_2}{\sqrt{8 c_2^2-e^{8 \beta }}}]$ does not depend on the sign of $c_2$, consider $c_2 > 0$. Hence, with the choice $0 < e^{8\beta} < 8 c_2^2$, we have $\frac{2 \sqrt{2} c_2}{\sqrt{8 c_2^2-e^{8 \beta }}} > 1$.} In addition, $\dot{\beta} \neq 0$ is necessary to ensure that the 4D metric is nonsingular. Therefore, for any given value of $c_2$, one can choose various forms of $\beta(t)$ to generate interesting 4D metrics. Note that, in general, the scale factors are distinct. However, by an appropriate choice of the constants $c_3,\dots,c_8$, any two or all of the $\alpha_i$ can be made equal. For example, choosing $c_3 = c_5=c_7,\ c_4 = c_6=c_8$, leads to $\alpha_1(t) = \alpha_2(t) = \alpha_3(t)$, while $\sigma(t)$ remains different nontrivially due to the logarithmic term in it.

Note also that $\gamma_t$ depends only on the sign of $c_1$, while $\sigma$ depends on its magnitude.

We now turn to the origin of the freedom of $\beta(t)$. Solving the $\hat{R}$ compatibility condition \eqref{eq:RHatCompatTauC} gives $\sigma$ in terms of $\beta$ and $\alpha_i$. Next, we organize the four independent Einstein equations \eqref{eq:GHatmunuCondTauC} into four simpler combinations, solving three of which determines $\alpha_i$ in terms of $\beta$. Substituting these into the fourth equation then imposes the aforesaid relation among the integration constants $c_i$ (while avoiding the vanishing of $\dot\beta$), leaving $\beta$ unfixed.

Notably, while the Ricci scalar associated with the metric $\hat{g}_{\mu\nu}$ is positive, the Ricci scalar associated with the 4D Einstein frame metric $\h g^{E}_{\mu\nu}=e^{6\beta} \h g_{\mu\nu}$ as in \eqref{eq:4DEinsteinFrameWeyl} is strictly negative:
\begin{align}
  &\h R(t)=\frac{3}{8} c_1^2 k^2 \left(40 c_2^2+3 e^{8 \beta}\right) e^{\frac{384 c_2^2+c_3^2+c_5^2+c_3 c_5}{8 \sqrt{2} c_2 \left(c_3+c_5\right)} \coth ^{-1}\left[\frac{2 \sqrt{2} c_2}{\sqrt{8 c_2^2-e^{8 \beta }}}\right]+6 \beta-c_4-c_6-c_8}>0\,, \nn\\
  &\h R^E(t)=-12 c_1^2 c_2^2 k^2 e^{\frac{384 c_2^2+c_3^2+c_5^2+c_3 c_5}{8 \sqrt{2} c_2 \left(c_3+c_5\right)} \coth ^{-1}\left[\frac{2 \sqrt{2} c_2}{\sqrt{8 c_2^2-e^{8 \beta }}}\right]-c_4-c_6-c_8}<0\,.
\end{align}

See appendix \ref{sapp:Zero3formConsT} (and figure \ref{fig:Zero3formConsT}), where profiles of the time-dependent scale factors and the Ricci scalar associated with the 4D Einstein frame metric are presented for representative choices of the constant parameters and $\beta(t)$. These solutions feature scale factors that grow with time from positive initial values, with some cases exhibiting power-law growth at late times.

\subsubsection{\texorpdfstring{$G_3= 0,\ \textrm{self-dual}\ \t F_5,\ \tau(t)$}{z}}

In the previous flux setup, allowing time-dependence in the axiodilaton $\tau$ contributes to the 10D energy-momentum tensor, which enters the nonlinear 4D Einstein equations \eqref{eq:GHatmunuCondTauT}. The time evolution of the real and imaginary parts of $\tau$ is governed by \eqref{eq:TauEOMTauT}. We consider two classes of solutions: one in which the real part of $\tau$ remains constant, and another in which both the real and imaginary parts vary over time.\footnote{Another possibility is to keep the imaginary part of $\tau$ constant; however, the imaginary part of equation \eqref{eq:TauEOMTauT} then forces the real part of $\tau$ to be time-independent.}

\textbf{For the first class (i.e., $\textrm{Re}\tau=\textrm{constant}$)}, we provide below the explicit forms of the functions and differential forms appearing in the ansatzes \eqref{eq:CommonAnsatz} and \eqref{eq:4DmetricAnsatz} that solve the relevant supergravity equations discussed in section \ref{sssec:Zero3formTauT}:
\begin{align}
  &\lambda_2=kJ_2\,,\quad \gamma_1=\gamma_tdt\,,\quad \gamma_t=4 c_1 e^{6 \beta} \sqrt{\frac{\dot\beta^2}{c_1^2 k^2 \left(8 c_2^2-e^{8 \beta }\right)}}\,, \nn\\
  &\textrm{Im}\tau=c_4\ e^{-\frac{c_3}{8 \sqrt{2} c_2} \coth ^{-1}\left[\frac{2 \sqrt{2} c_2}{\sqrt{8 c_2^2-e^{8 \beta } }}\right]}\,,\quad \textrm{Re}\tau=c_0\,, \nn\\
  &\sigma=c_6+c_8+c_{10}-\ln \left[\frac{c_1^2 k^2 \left(8 c_2^2-e^{8 \beta }\right)}{16\dot\beta^2}\right]-\frac{\left(c_5+c_7+c_9\right)}{8 \sqrt{2} c_2} \coth ^{-1}\left[\frac{2 \sqrt{2} c_2}{\sqrt{8 c_2^2-e^{8 \beta }}}\right]-6 \beta\,, \nn\\
  &\alpha_1=c_6-\frac{c_5}{8 \sqrt{2} c_2} \coth ^{-1}\left[\frac{2 \sqrt{2} c_2}{\sqrt{8 c_2^2-e^{8 \beta }}}\right]-6 \beta\,,\quad \alpha_2=c_8-\frac{c_7}{8 \sqrt{2} c_2} \coth ^{-1}\left[\frac{2 \sqrt{2} c_2}{\sqrt{8 c_2^2-e^{8 \beta }}}\right]-6 \beta\,, \nn\\
  &\alpha_3=c_{10}-\frac{c_9}{8 \sqrt{2} c_2} \coth ^{-1}\left[\frac{2 \sqrt{2} c_2}{\sqrt{8 c_2^2-e^{8 \beta }}}\right]-6 \beta\,, \nn\\
  &384 c_2^2+c_3^2-c_5 c_7-c_7 c_9-c_9 c_5=0\,,\quad c_2,c_4>0\,, \label{eq:0G3F5ImTauTEx}
\end{align}
where $c_0$ can take arbitrary values since the real part of \eqref{eq:TauEOMTauT} is trivially satisfied, and also the function $\beta(t)$ remains unfixed due to the aforesaid relation among the integration constants $c_2,c_3,c_5,c_7,c_9$. The reality of the above expressions requires all $c_i$ to be real, and the condition $e^{8\beta} < 8 c_2^2$ must hold. In addition, $\dot{\beta} \neq 0$. Therefore, for any given value of $c_2$, one can choose various forms of $\beta(t)$ to generate interesting 4D metrics. By an appropriate choice of the constants $c_5,\dots,c_{10}$, any two or all of the $\alpha_i$ can be made equal. For example, choosing $c_5 = c_7=c_9,\ c_6 = c_8=c_{10}$, leads to $\alpha_1(t) = \alpha_2(t) = \alpha_3(t)$, while $\sigma(t)$ remains different nontrivially due to the logarithmic term in it.

Note also that $\gamma_t$ depends only on the sign of $c_1$, while $\sigma$ depends on its magnitude.

Notably, while the Ricci scalar associated with the metric $\hat g_{\mu\nu}$ can take positive values depending on $c_2,c_3$, and $\beta$\footnote{If $c_3^2>768\,c_2^2$, then $\hat R<0$ for any admissible $\beta$.}, the Ricci scalar corresponding to the 4D Einstein-frame metric $\hat g^{E}_{\mu\nu}=e^{6\beta}\hat g_{\mu\nu}$ as in \eqref{eq:4DEinsteinFrameWeyl} is strictly negative:
\begin{align}
  &\h R(t)=\frac{1}{32} c_1^2 k^2 \left(36 e^{8 \beta }+480 c_2^2-c_3^2 \right) e^{\frac{384 c_2^2+c_3^2+c_5^2+c_7^2+c_5 c_7}{8 \sqrt{2} c_2 \left(c_5+c_7\right)} \coth ^{-1}\left[\frac{2 \sqrt{2} c_2}{\sqrt{8 c_2^2-e^{8 \beta } }}\right]+6 \beta-c_6-c_8-c_{10}}\,, \nn\\
  &\h R^E(t)=-\frac{1}{32} c_1^2 k^2 \left(384 c_2^2+c_3^2\right) e^{\frac{384 c_2^2+c_3^2+c_5^2+c_7^2+c_5 c_7}{8 \sqrt{2} c_2 \left(c_5+c_7\right)} \coth ^{-1}\left[\frac{2 \sqrt{2} c_2}{\sqrt{8 c_2^2-e^{8 \beta } }}\right]-c_6-c_8-c_{10}}<0\,.
\end{align}

See appendix \ref{sapp:Zero3formTauT} (and figure \ref{fig:Zero3formImTauT}), where profiles of the time-dependent scale factors and the Ricci scalar associated with the 4D Einstein frame metric are presented for representative choices of the constant parameters and $\beta(t)$. These solutions feature scale factors and the imaginary part of axiodilaton that grow with time from positive initial values, with some cases exhibiting power-law growth at late times.

\textbf{For the second class (i.e., both $\textrm{Re}\tau,\textrm{Im}\tau$ are time-dependent)}, we provide below the explicit forms of the functions and differential forms appearing in the ansatzes \eqref{eq:CommonAnsatz} and \eqref{eq:4DmetricAnsatz} that solve the relevant supergravity equations discussed in section \ref{sssec:Zero3formTauT}:
\begin{align}
  &\lambda_2=kJ_2\,,\quad \gamma_1=\gamma_tdt\,,\quad \gamma_t=4 c_1 e^{6 \beta} \sqrt{\frac{\dot\beta^2}{c_1^2 k^2 \left(8 c_2^2-e^{8 \beta }\right)}}\,, \nn\\
  &\textrm{Re}\tau=\pm\ c_3 \sqrt{\tanh ^2\left[\frac{c_3 c_5}{8 \sqrt{2} c_2} \coth ^{-1}\left(\frac{2 \sqrt{2} c_2}{\sqrt{8 c_2^2-e^{8 \beta } }}\right)+c_3 c_6\right]}+c_4 \equiv \textrm{Re}\tau_{\pm}\,,  \nn\\
  &\textrm{Im}\tau=c_3\ \textrm{sech}\left[\frac{c_3 c_5}{8 \sqrt{2} c_2} \coth ^{-1}\left(\frac{2 \sqrt{2} c_2}{\sqrt{8 c_2^2-e^{8 \beta } }}\right)+c_3 c_6\right]\,, \nn\\
  &\sigma=c_8+c_{10}+c_{12}-\ln \left[\frac{c_1^2 k^2 \left(8 c_2^2-e^{8 \beta }\right)}{16\dot\beta^2}\right]-\frac{c_7+c_9+c_{11}}{8 \sqrt{2} c_2} \coth ^{-1}\left[\frac{2 \sqrt{2} c_2}{\sqrt{8 c_2^2-e^{8 \beta }}}\right]-6 \beta\,, \nn\\
  &\alpha_1=c_8-\frac{c_7}{8 \sqrt{2} c_2} \coth ^{-1}\left[\frac{2 \sqrt{2} c_2}{\sqrt{8 c_2^2-e^{8 \beta }}}\right]-6 \beta\,,\quad \alpha_2=c_{10}-\frac{c_9}{8 \sqrt{2} c_2} \coth ^{-1}\left[\frac{2 \sqrt{2} c_2}{\sqrt{8 c_2^2-e^{8 \beta }}}\right]-6 \beta\,, \nn\\
  &\alpha_3=c_{12}-\frac{c_{11}}{8 \sqrt{2} c_2} \coth ^{-1}\left[\frac{2 \sqrt{2} c_2}{\sqrt{8 c_2^2-e^{8 \beta }}}\right]-6 \beta\,, \nn\\
  &384 c_2^2+c_3^2 c_5^2-c_7 c_9-c_9 c_{11}-c_{11} c_7=0\,,\quad c_2,c_3>0\,, \label{eq:0G3F5ReImTauTEx}
\end{align}
where the function $\beta(t)$ remains unfixed due to the aforesaid relation among the integration constants $c_2,c_3,c_5,c_7,c_9,c_{11}$. The reality of the above expressions requires all $c_i$ to be real, and the condition $e^{8\beta} < 8 c_2^2$ must hold. In addition, $\dot{\beta} \neq 0$. Therefore, for any given value of $c_2$, one can choose various forms of $\beta(t)$ to generate interesting 4D metrics. By an appropriate choice of the constants $c_7,\dots,c_{12}$, any two or all of the $\alpha_i$ can be made equal. For example, choosing $c_7 = c_9=c_{11},\ c_8 = c_{10}=c_{12}$, leads to $\alpha_1(t) = \alpha_2(t) = \alpha_3(t)$, while $\sigma(t)$ remains different nontrivially due to the logarithmic term in it.

Note also that $\gamma_t$ depends only on the sign of $c_1$, while $\sigma$ depends on its magnitude. The sign of the first term in $\textrm{Re}\tau$ does not influence the other functions, since all relevant supergravity equations depend solely on $(\textrm{Re}\dot\tau)^2$.

Notably, while the Ricci scalar associated with the metric $\hat g_{\mu\nu}$ can take positive values depending on $c_2,c_3,c_5$, and $\beta$\footnote{If $c_3^2c_5^2>768\,c_2^2$, then $\hat R<0$ for any admissible $\beta$.}, the Ricci scalar corresponding to the 4D Einstein-frame metric $\hat g^{E}_{\mu\nu}=e^{6\beta}\hat g_{\mu\nu}$ as in \eqref{eq:4DEinsteinFrameWeyl} is strictly negative:
\begin{align}
  &\h R(t)=\frac{1}{32} c_1^2 k^2 \left(36 e^{8 \beta }+480 c_2^2-c_3^2 c_5^2 \right) e^{\frac{384 c_2^2+c_3^2 c_5^2+c_7^2+c_9^2+c_7 c_9}{8 \sqrt{2} c_2 \left(c_7+c_9\right)} \coth ^{-1}\left[\frac{2 \sqrt{2} c_2}{\sqrt{8 c_2^2-e^{8 \beta } }}\right]+6 \beta-c_8-c_{10}-c_{12}}\,, \nn\\
  &\h R^E(t)=-\frac{1}{32} c_1^2 k^2 \left(384 c_2^2+c_3^2 c_5^2\right) e^{\frac{384 c_2^2+c_3^2 c_5^2+c_7^2+c_9^2+c_7 c_9}{8 \sqrt{2} c_2 \left(c_7+c_9\right)} \coth ^{-1}\left[\frac{2 \sqrt{2} c_2}{\sqrt{8 c_2^2-e^{8 \beta } }}\right]-c_8-c_{10}-c_{12}}<0\,.
\end{align}

See appendix \ref{sapp:Zero3formTauT} (and figure \ref{fig:Zero3formReImTauT}), where profiles of the time-dependent scale factors and the Ricci scalar associated with the 4D Einstein frame metric are presented for representative choices of the constant parameters and $\beta(t)$. The solution features scale factors that grow with time from positive initial values, while the real part of the axiodilaton approaches a constant and the imaginary part decays to zero exponentially.

\subsubsection{\texorpdfstring{$\t F_5=0,\ G_3,\tau\neq 0,\ \tau(t)$}{x}}

For the case with vanishing $\t F_5$, nontrivial 3-form flux $G_3$ featuring $[1,2]$ components, and time-dependent axiodilaton $\tau$, in the following we consider two classes of solutions: one in which the real part of $\tau$ remains constant, and another in which both the real and imaginary parts vary over time.\footnote{Another possibility is to keep $\textrm{Im}\tau$ constant while allowing $\textrm{Re}\tau$ to vary in time; however, equation \eqref{eq:TauEOM0F5} then forces $\beta$ to be time-independent.}

\textbf{For the first class (i.e., $\textrm{Re}\tau=\textrm{constant}$)}, we provide below the explicit forms of the functions and differential forms appearing in the ansatzes \eqref{eq:CommonAnsatz} and \eqref{eq:4DmetricAnsatz} that solve the relevant supergravity equations discussed in section \ref{sssec:ZeroF5TauT}:
\begin{align}
  &\eta_2=ikJ_2\,,\quad \zeta_1=\zeta_tdt\,,\quad \zeta_t=8 c_1 e^{4 \beta}\sqrt{\frac{c_2 \dot\beta^2}{c_1^2 k^2 \left(16 c_3^2-e^{16 \beta}\right)}}\,,\quad \textrm{Im}\tau=c_2 e^{-12 \beta}\,,\quad \textrm{Re}\tau=c_0\,, \nn\\
  &\sigma=c_5+c_7+c_9-\ln \left[\frac{c_1^2 k^2 \left(16 c_3^2-e^{16 \beta}\right)}{64c_2 \dot\beta^2}\right]-\frac{c_4+c_6+c_8}{32 c_3} \coth ^{-1}\left[\frac{4 c_3}{\sqrt{16 c_3^2-e^{16 \beta} }}\right]-6 \beta\,, \nn\\
  &\alpha_1=c_5-\frac{c_4}{32 c_3} \coth ^{-1}\left[\frac{4 c_3}{\sqrt{16 c_3^2-e^{16 \beta} }}\right]-6 \beta\,,\quad \alpha_2=c_7-\frac{c_6}{32 c_3} \coth ^{-1}\left[\frac{4 c_3}{\sqrt{16 c_3^2-e^{16 \beta} }}\right]-6 \beta\,, \nn\\
  &\alpha_3=c_9-\frac{c_8}{32 c_3} \coth ^{-1}\left[\frac{4 c_3}{\sqrt{16 c_3^2-e^{16 \beta} }}\right]-6 \beta\,, \nn\\
  &3072 c_3^2-c_4 c_6-c_6 c_8-c_8 c_4=0\,,\quad c_2,c_3>0\,, \label{eq:0F5G3ImTauTEx}
\end{align}
where $c_0$ can take arbitrary values since the real part of \eqref{eq:TauEOM0F5} is trivially satisfied, and also the function $\beta(t)$ remains unfixed due to the aforesaid relation among the integration constants $c_3,c_4,c_6,c_8$. The reality of the above expressions requires all $c_i$ to be real, and the condition $e^{16\beta} < 16 c_3^2$ must hold. In addition, $\dot{\beta} \neq 0$. Therefore, for any given value of $c_3$, one can choose various forms of $\beta(t)$ to generate interesting 4D metrics. By an appropriate choice of the constants $c_4,\dots,c_9$, any two or all of the $\alpha_i$ can be made equal. For example, choosing $c_4 = c_6=c_8,\ c_5 = c_7=c_9$, leads to $\alpha_1(t) = \alpha_2(t) = \alpha_3(t)$, while $\sigma(t)$ remains different nontrivially due to the logarithmic term in it.

Note also that $\zeta_t$ depends only on the sign of $c_1$, while $\sigma$ depends on its magnitude.

Solving the supergravity equations in the present class requires some explanation. Due to the 3-form flux contribution, the equation governing $\textrm{Im}\tau$ \eqref{eq:TauEOM0F5} differs qualitatively from the cases considered previously.\footnote{With the substitution $\textrm{Im}\,\tau = e^{\sigma - \alpha_1 - \alpha_2 - \alpha_3 - 8\beta + f}$, the equation can be recast in the form $\ddot f + h_1 \dot f + p\, e^{-f} + h_2 = 0$, where $p$ is a constant and the functions $h_i$ are determined by derivatives of $\sigma, \alpha_i,$ and $\beta$. The resulting exponential nonlinearity prevents direct integration.} Rather than attempting to solve this equation in isolation, it is more effective to consider it together with the $\hat{R}$ compatibility condition \eqref{eq:RHatCompat0F5}. The two equations admit a common solution only if
\begin{align}
  \textrm{Im}\tau=c_2e^{-12\beta(t) + c_{10} \int_1^t ds\ e^{\frac{1}{2} \left[\sigma(s)-\alpha_1(s)-\alpha_2(s)-\alpha_3-12 \beta(s)\right]}}\,,
\end{align}
for arbitrary constants $c_2$ and $c_{10}$. This expression solves an appropriate linear combination of the two equations. For simplicity, we set $c_{10}=0$ and then use a second independent linear combination to determine $\sigma$ in terms of $\beta$ and $\alpha_i$, thereby completing the solution of both equations. Next, we organize the four independent Einstein equations \eqref{eq:GHatmunuCond0F5} into four simpler combinations, solving three of which determines $\alpha_i$ in terms of $\beta$. Substituting these into the fourth equation then imposes the aforesaid relation among the integration constants $c_i$ (while avoiding the vanishing of $\dot\beta$), leaving $\beta$ unfixed.

Notably, while the Ricci scalar associated with the metric $\hat g_{\mu\nu}$ can take positive values depending on $c_3$ and $\beta$\footnote{For any admissible $\beta$, we have: $-112 c_3^2<15 e^{16 \beta}-112 c_3^2<128c_3^2$.}, the Ricci scalar corresponding to the 4D Einstein-frame metric $\hat g^{E}_{\mu\nu}=e^{6\beta}\hat g_{\mu\nu}$ as in \eqref{eq:4DEinsteinFrameWeyl} is strictly negative:
\begin{align}
  &\h R(t)=\frac{3 c_1^2 k^2}{32 c_2} \left(15 e^{16 \beta}-112 c_3^2 \right) e^{\frac{3072 c_3^2+c_4^2+c_6^2+c_4 c_6}{32 c_3 \left(c_4+c_6\right)} \coth ^{-1}\left[\frac{4 c_3}{\sqrt{16 c_3^2-e^{16 \beta} }}\right]+6 \beta-c_5-c_7-c_9}\,, \nn\\
  &\h R^E(t)=-\frac{24 c_1^2 c_3^2 k^2}{c_2} e^{\frac{3072 c_3^2+c_4^2+c_6^2+c_4 c_6}{32 c_3 \left(c_4+c_6\right)} \coth ^{-1}\left[\frac{4 c_3}{\sqrt{16 c_3^2-e^{16 \beta} }}\right]-c_5-c_7-c_9}<0\,.
\end{align}

See appendix \ref{sapp:ZeroF5TauT} (and figure \ref{fig:ZeroF5ImTauT}), where profiles of the time-dependent scale factors and the Ricci scalar associated with the 4D Einstein frame metric are presented for representative choices of the constant parameters and $\beta(t)$. These solutions feature scale factors and the imaginary part of axiodilaton that grow with time from positive initial values, with some cases exhibiting power-law growth at late times.

\textbf{For the second class (i.e., both $\textrm{Re}\tau,\textrm{Im}\tau$ are time-dependent)}, we provide below the explicit forms of the functions and differential forms appearing in the ansatzes \eqref{eq:CommonAnsatz} and \eqref{eq:4DmetricAnsatz} that solve the relevant supergravity equations discussed in section \ref{sssec:ZeroF5TauT}:
\begin{align}
  &\eta_2=ikJ_2\,,\quad \zeta_1=\zeta_tdt\,,\quad \zeta_t= 12 \sqrt{\frac{2}{5}} \frac{e^{4 \beta}}{c_1  k^2 }\sqrt{\frac{c_2^2 \dot\beta^2}{8 c_4^2-9 e^{\frac{8 \beta}{3}}}}\,, \nn\\
  &\textrm{Re}\tau= c_3 +12 \sqrt{\frac{2}{5}}\ \frac{c_2}{k^2 c_1^2} \int _1^t ds\ e^{\frac{8}{3} \beta (s)}\sqrt{\frac{c_2^2 \dot\beta (s)^2}{8 c_4^2-9 e^{\frac{8}{3} \beta (s)}}}\,,\quad \textrm{Im}\tau= \frac{3 c_2^2}{5 c_1^2 k^2} e^{\frac{4 \beta}{3}}\,, \nn\\
  &\sigma=c_6+c_8+c_{10} -\ln \left[\frac{5 c_1^4  k^4 \left(8 c_4^2-9 e^{\frac{8 \beta}{3}}\right)}{288 c_2^2 \dot\beta^2}\right] -\frac{3 \left(c_5+c_7+c_9\right)}{8 \sqrt{2} c_4} \coth ^{-1}\left[\frac{2 \sqrt{2} c_4}{\sqrt{8 c_4^2-9 e^{\frac{8 \beta}{3}} }}\right]-6 \beta\,, \nn\\
  &\alpha_1= c_6-\frac{3 c_5}{8 \sqrt{2} c_4} \coth ^{-1}\left[\frac{2 \sqrt{2} c_4}{\sqrt{8 c_4^2-9 e^{\frac{8 \beta}{3}} }}\right]-6 \beta\,,\quad \alpha_2= c_8 -\frac{3 c_7}{8 \sqrt{2} c_4} \coth ^{-1}\left[\frac{2 \sqrt{2} c_4}{\sqrt{8 c_4^2-9 e^{\frac{8 \beta}{3}} }}\right]-6 \beta\,, \nn\\
  &\alpha_3= c_{10} -\frac{3 c_9}{8 \sqrt{2} c_4} \coth ^{-1}\left[\frac{2 \sqrt{2} c_4}{\sqrt{8 c_4^2-9 e^{\frac{8 \beta}{3}} }}\right]-6 \beta\,, \nn\\
  &3584 c_4^2-9 c_5 c_7-9 c_7 c_9-9 c_9c_5=0\,,\quad c_4>0\,, \label{eq:0F5G3ReImTauTEx}
\end{align}
where the function $\beta(t)$ remains unfixed due to the aforesaid relation among the integration constants $c_4,c_5,c_7,c_9$. The reality of the above expressions requires all $c_i$ to be real, and the condition $9 e^{\frac{8 \beta}{3}} < 8 c_4^2$ must hold. In addition, $\dot{\beta} \neq 0$. Therefore, for any given value of $c_4$, one can choose various forms of $\beta(t)$ to generate interesting 4D metrics. By an appropriate choice of the constants $c_5,\dots,c_{10}$, any two or all of the $\alpha_i$ can be made equal. For example, choosing $c_5 = c_7=c_9,\ c_6 = c_8=c_{10}$, leads to $\alpha_1(t) = \alpha_2(t) = \alpha_3(t)$, while $\sigma(t)$ remains different nontrivially due to the logarithmic term in it.

Note also that $c_3$ and the sign of the second term in $\textrm{Re}\tau$ does not influence the other functions, since all relevant supergravity equations depend solely on $(\textrm{Re}\dot\tau)^2$. And, $\textrm{Re}\dot\tau(t)$ is simply given by the integrand in the second term evaluated at $t$.

Notably, while the Ricci scalar associated with the metric $\hat g_{\mu\nu}$ can take positive values depending on $c_4$ and $\beta$\footnote{For any admissible $\beta$, we have: $-32 c_4^2<1048 c_4^2-1215 e^{\frac{8 \beta (t)}{3}}<1048 c_4^2$.}, the Ricci scalar corresponding to the 4D Einstein-frame metric $\hat g^{E}_{\mu\nu}=e^{6\beta}\hat g_{\mu\nu}$ as in \eqref{eq:4DEinsteinFrameWeyl} is strictly negative:
\begin{align}
  &\h R(t)=\frac{5 c_1^4 k^4 }{1296 c_2^2}\left(1048 c_4^2-1215 e^{\frac{8 \beta}{3}}\right) e^{\frac{3584 c_4^2+9 \left(c_5^2+c_7 c_5+c_7^2\right)}{24 \sqrt{2} c_4 \left(c_5+c_7\right)} \coth ^{-1}\left[\frac{2 \sqrt{2} c_4}{\sqrt{8 c_4^2-9 e^{\frac{8 \beta}{3}} }}\right]+6 \beta-c_6-c_8-c_{10}}\,, \nn\\
  &\h R^E(t)=-\frac{280 c_1^4 c_4^2 k^4}{81 c_2^2} e^{\frac{3584 c_4^2+9 \left(c_5^2+c_7 c_5+c_7^2\right)}{24 \sqrt{2} c_4 \left(c_5+c_7\right)} \coth ^{-1}\left[\frac{2 \sqrt{2} c_4}{\sqrt{8 c_4^2-9 e^{\frac{8 \beta}{3}} }}\right]-c_6-c_8-c_{10}} <0\,.
\end{align}

See appendix \ref{sapp:ZeroF5TauT} (and figure \ref{fig:ZeroF5ReImTauT}), where profiles of the time-dependent scale factors and the Ricci scalar associated with the 4D Einstein frame metric are presented for representative choices of the constant parameters and $\beta(t)$. The solution features scale factors that grow with time from positive initial values, while the real part of the axiodilaton approaches a constant and the imaginary part decays to zero exponentially.

\subsubsection{\texorpdfstring{$\textrm{Self-dual}\ \t F_5,\ G_3,\tau\neq 0,\ \tau(t)$}{x}}

For the case with nontrivial self-dual 5-from flux $\t F_5$ featuring $[3,2]$ and $[1,4]$ components, 3-form flux $G_3$ featuring $[3,0]$ components, and time-dependent axiodilaton $\tau$, in the following we consider two classes of solutions: one in which the real part of $\tau$ remains constant, and another in which both the real and imaginary parts vary over time.\footnote{Another possibility is to keep $\textrm{Im}\tau$ constant while allowing $\textrm{Re}\tau$ to vary in time; however, the supergravity equations (though not immediately from \eqref{eq:TauEOMAllFlux}) eventually force $\beta$ to be time-independent.}

\textbf{For the first class (i.e., $\textrm{Re}\tau=\textrm{constant}$)}, we provide below the explicit forms of the functions and differential forms appearing in the ansatzes \eqref{eq:CommonAnsatz} and \eqref{eq:4DmetricAnsatz} that solve the relevant supergravity equations discussed in section \ref{sssec:SDF5NZG3TauT}:\footnote{Note that $k,r$ are not constants of integration.}
\begin{align}
  &\lambda_2=kJ_2\,,\quad \gamma_1=\gamma_tdt\,,\quad \gamma_t=-\frac{8 \sqrt{7} c_1 e^{\frac{1}{2} \left(c_4-c_5\right)+6 \beta} \dot\beta}{\sqrt{c_6^2-c_8^2 e^{8 \beta} }}\,, \nn\\
  &q=ir\,,\quad r\in\mathbb{R}\setminus\{0\}\,,\quad \vartheta_1=\vartheta_3dx^3\,,\quad \vartheta_3=c_2 e^{-6 \beta}\,, \nn\\
  &\textrm{Im}\tau=c_3 e^{c_9 \sqrt{c_6^2+1} +\frac{c_7}{2} +\frac{2 \sqrt{7}}{c_6}\sqrt{c_6^2+1} \coth ^{-1}\left[\frac{c_6}{\sqrt{c_6^2-c_8^2 e^{8 \beta} }}\right]-20 \beta}\,,\quad \textrm{Re}\tau=c_0\,, \nn\\
  &\sigma=c_9 (3 \sqrt{c_6^2+1}-1)+c_4+c_7+ \ln \left[\frac{448 \dot\beta^2}{c_6^2-c_8^2 e^{8 \beta} }\right] +\frac{2 \sqrt{7}}{c_6} (3 \sqrt{c_6^2+1}-1) \coth ^{-1}\left[\frac{c_6}{\sqrt{c_6^2-c_8^2 e^{8 \beta} }}\right] -46 \beta\,, \nn\\
  &\alpha_1=c_9 (\sqrt{c_6^2+1}-1)+c_5+\frac{2 \sqrt{7}}{c_6} (\sqrt{c_6^2+1}-1) \coth ^{-1}\left[\frac{c_6}{\sqrt{c_6^2-c_8^2 e^{8 \beta} }}\right] -6 \beta\,, \nn\\
  &\alpha_2=c_9+\frac{2 \sqrt{7}}{c_6} \coth ^{-1}\left[\frac{c_6}{\sqrt{c_6^2-c_8^2 e^{8 \beta} }}\right] -6 \beta\,, \nn\\
  &\alpha_3=c_9 (2 \sqrt{c_6^2+1}-1)+c_7+\frac{2 \sqrt{7}}{c_6} (2 \sqrt{c_6^2+1}-1) \coth ^{-1}\left[\frac{c_6}{\sqrt{c_6^2-c_8^2 e^{8 \beta} }}\right] -46 \beta\,, \nn\\
  &c_3=\frac{3 c_2^2 r^2}{5 c_1^2 k^2} e^{c_5-\frac{c_7}{2}}>0\,,\quad c_8^2=\frac{14}{3} e^{c_4-c_5} c_1^2 k^2\,,\quad c_6,c_8>0\,, \label{eq:AllFluxImTauTEx}
\end{align}
where $c_0$ can take arbitrary values since the real part of \eqref{eq:TauEOMAllFlux} is trivially satisfied, and also the function $\beta(t)$ remains unfixed due to the aforesaid two relations among the integration constants $c_1,c_3,c_4,c_5,c_7,c_8$ given a pair $(k,r)$. The reality of the above expressions requires all $c_i$ to be real, and the condition $e^{8\beta} < c_6^2/c_8^2$ must hold. In addition, $\dot{\beta} \neq 0$ to ensure that the 4D metric is nonsingular. Therefore, for any given values of $c_6,c_8$, one can choose various forms of $\beta(t)$ to generate interesting 4D metrics.

By choosing $c_5=0, c_6=\sqrt{3}$, $\alpha_1$ and $\alpha_2$ can be made equal. However, unlike the previous flux setups, there is no choice of the constants $c_i$ that can make $\alpha_3$ equal to $\alpha_1$ or $\alpha_2$, or make all three $\alpha_i$ equal. $\sigma(t)$ remains different nontrivially from the $\alpha_i$ due to the logarithmic term in it.

Here we give some crucial details involved in solving the supergravity equations in the present class. The 3-from flux contribution to the equation governing $\textrm{Im}\tau$ \eqref{eq:TauEOMAllFlux} can be eliminated by taking an appropriate linear combination with the $(t,t)$ and $(1,1)$ components of the Einstein equations \eqref{eq:GHatmunuCondAllFlux}. The resulting equation can then be integrated to obtain:
\begin{align}
  \textrm{Im}\tau=c_3 e^{\frac{1}{2} \left[\alpha _2(t)+\alpha _3(t)+12 \beta(t) \right]}\ e^{c_{10} \int_1^t ds\ e^{\frac{1}{2} \left[\sigma(s)-\alpha_1(s)-\alpha_2(s)-\alpha_3-12 \beta(s)\right]}}\,,
\end{align}
for arbitrary constants $c_3$ and $c_{10}$. For simplicity, we proceed by setting $c_{10}=0$. Next, a linear combination of the $(t,t)$ and $(3,3)$ components of the Einstein equations \eqref{eq:GHatmunuCondAllFlux}, together with a separate combination of the $(1,1)$ and $(2,2)$ components, can be solved to obtain $\sigma$ and $\alpha_1$. Substituting these solutions into the remaining three independent equations, namely the $(t,t)$ and $(1,1)$ components of the Einstein equations \eqref{eq:GHatmunuCondAllFlux}, and the $\h R$ compatibility condition \eqref{eq:RHatCompatAllFlux}, we analyze their structure. We then arrive at an ansatz for $\alpha_3$ in terms of $\alpha_2$ and $\beta$ that makes some of the terms in each equation proportional, thereby giving a common structure. This step does not solve any of the three equations by itself; rather, it ensures that a common solution to them can exist.\footnote{As direct integration of the three equations is not straightforward, we adopt this trick as a practical method.} Plugging this back in, an appropriate linear combination of the three equations then determines $\alpha_2$ in terms of $\beta$, and the remaining two equations can then be solved by imposing the two aforesaid relations among the integration constants $c_i$ (while avoiding the vanishing of $\dot\beta$), leaving $\beta$ unfixed. All resulting expressions have been checked directly to satisfy the full system.

Notably, the Ricci scalars associated with both the 4D metric $\hat g_{\mu\nu}$ and the Einstein-frame metric $\hat g^{E}_{\mu\nu}=e^{6\beta}\hat g_{\mu\nu}$ as in \eqref{eq:4DEinsteinFrameWeyl} are strictly negative:
\begin{align}
  &\h R(t)=-\frac{1}{224} \left[80 \sqrt{7} \sqrt{(c_6^2+1) (c_6^2-c_8^2 e^{8 \beta}) }+108 c_6^2+89 (c_6^2-c_8^2 e^{8 \beta}) +112\right] \nn\\
  &\quad\quad\quad \times e^{c_9-c_4-c_7-3 c_9 \sqrt{c_6^2+1} +\frac{2 \sqrt{7}}{c_6} (1-3 \sqrt{c_6^2+1}) \coth ^{-1}\left[\frac{c_6}{\sqrt{c_6^2-c_8^2 e^{8 \beta} }}\right]+46 \beta}<0\,, \nn\\
  &\h R^E(t)=-\frac{1}{14} \left[5 \sqrt{7} \sqrt{\left(c_6^2+1\right) (c_6^2-c_8^2 e^{8 \beta}) }+9 c_6^2+5 (c_6^2-c_8^2 e^{8 \beta}) +7\right] \nn\\
  &\quad\quad\quad \times e^{c_9-c_4-c_7-3 c_9 \sqrt{c_6^2+1}+ \frac{2 \sqrt{7}}{c_6} (1-3 \sqrt{c_6^2+1}) \coth ^{-1}\left[\frac{c_6}{\sqrt{c_6^2-c_8^2 e^{8 \beta} }}\right]+40 \beta}<0\,,
\end{align}
since $c_6^2-c_8^2 e^{8 \beta}>0$.

See appendix \ref{sapp:AllFluxTauT} (and figure \ref{fig:AllFluxImTauT}), where profiles of the time-dependent scale factors and the Ricci scalar associated with the 4D Einstein frame metric are presented for representative choices of the constant parameters and $\beta(t)$. These solutions feature scale factors and the imaginary part of axiodilaton that grow with time from positive initial values, with some cases exhibiting power-law growth at late times.

\textbf{For the second class (i.e., both $\textrm{Re}\tau,\textrm{Im}\tau$ are time-dependent)}, we provide below the explicit forms of the functions and differential forms appearing in the ansatzes \eqref{eq:CommonAnsatz} and \eqref{eq:4DmetricAnsatz} that solve the relevant supergravity equations discussed in section \ref{sssec:SDF5NZG3TauT}:
\begin{align}
  &\lambda_2=kJ_2\,,\quad \gamma_1=\gamma_tdt\,,\quad \gamma_t=\frac{24 c_1}{c_5+2} e^{\frac{1}{2} \left(c_3-c_4\right)+6 \beta} \dot\beta\,, \nn\\
  &q=ir\,,\quad r\in\mathbb{R}\setminus\{0\}\,,\quad \vartheta_1=\vartheta_3dx^3\,,\quad \vartheta_3=c_2 e^{-6 \beta (t)}\,, \nn\\
  &\textrm{Re}\tau=c_7+ \frac{24 k^2 c_1^2 e^{c_3-c_4}}{\left(c_5+2\right) c_6} \int _1^t ds\ e^{8 \beta (s)} \dot\beta(s)\,,\quad \textrm{Im}\tau=\frac{c_1 k}{c_6}e^{\frac{1}{2} \left(c_3-c_4\right)+4 \beta}\,, \nn\\
  &\sigma=c_3+\left(c_5+2\right) c_8+c_{10} +\frac{18 c_1^2 k^2 e^{c_3-c_4}}{\left(c_5+2\right)^2}e^{8 \beta}+2 \ln \left[\frac{24 \dot\beta}{c_5+2}\right]+\left(c_9+24\right) \beta\,, \nn\\
  &\alpha_1=c_4+\left(c_5+1\right) c_8+\frac{6 \left(3 c_5+2\right)}{c_5+2} \beta\,,\quad \alpha_2=c_8-\frac{6 \left(c_5-2\right)}{c_5+2} \beta\,, \nn\\
  &\alpha_3=c_{10}+\frac{18 c_1^2 k^2 e^{c_3-c_4}}{\left(c_5+2\right)^2}e^{8 \beta}+c_9 \beta\,,\quad c_6= \frac{2 c_1^3 k^3}{c_2^2 r^2} e^{\frac{1}{2} \left(c_3-3 c_4\right)- \left(c_5+2\right) c_8}\,, \nn\\
  &3 c_9 \left(c_5+2\right){}^2+2 c_5 \left(5 c_5+56\right)+112=0\,, \label{eq:AllFluxReImTauTEx}
\end{align}
where the function $\beta(t)$ remains unfixed due to the aforesaid two relations among the integration constants $c_1,\dots,c_6,c_8,c_9$ given a pair $(k,r)$. The reality of the above expressions requires all $c_i$ to be real. In addition, $\dot{\beta} \neq 0$ to ensure that the 4D metric is nonsingular. Therefore, one can choose various forms of $\beta(t)$ to generate interesting 4D metrics. By choosing $c_4=0, c_5=0$, $\alpha_1$ and $\alpha_2$ can be made equal. However, there is no choice of the constants $c_i$ that can make $\alpha_3$ equal to $\alpha_1$ or $\alpha_2$, or make all three $\alpha_i$ equal. $\sigma(t)$ remains different nontrivially from the $\alpha_i$ due to the logarithmic term in it.

Note also that $c_7$ and the sign of the second term in $\textrm{Re}\tau$ does not influence the other functions, since all relevant supergravity equations depend solely on $(\textrm{Re}\dot\tau)^2$. And, $\textrm{Re}\dot\tau(t)$ is simply given by the integrand in the second term evaluated at $t$.

Notably, while the Ricci scalar associated with the metric $\hat g_{\mu\nu}$ can take positive values depending on $k,c_1,c_3,c_4,c_5$ and $\beta$, the Ricci scalar corresponding to the 4D Einstein-frame metric $\hat g^{E}_{\mu\nu}=e^{6\beta}\hat g_{\mu\nu}$ as in \eqref{eq:4DEinsteinFrameWeyl} is strictly negative:
\begin{align}
  &\h R(t)=\frac{1}{288} \left[11 e^{c_4} \left(c_5+2\right)^2-288 c_1^2 k^2 e^{c_3+8 \beta}\right] \nn\\
  &\quad\quad\quad \times  e^{-\frac{54 c_1^2 k^2 e^{c_3-c_4}}{3 \left(c_5+2\right)^2}e^{8 \beta} -\frac{2 \left(c_5 \left(31 c_5+88\right)+88\right)}{3 \left(c_5+2\right)^2} \beta -c_3 - c_4-\left(c_5+2\right) c_8+c_{10}}\,, \nn\\
  &\h R^E(t)=-\frac{1}{18} \left[18 c_1^2 k^2 e^{c_3+8 \beta}+e^{c_4} \left(c_5+2\right)^2\right] \nn\\
  &\quad\quad\quad \times  e^{-\frac{54 c_1^2 k^2 e^{c_3-c_4}}{3 \left(c_5+2\right)^2}e^{8 \beta} -\frac{8 \left(c_5 \left(10 c_5+31\right)+31\right)}{3 \left(c_5+2\right)^2} \beta-c_3 - c_4-\left(c_5+2\right) c_8-c_{10}} <0\,.
\end{align}

See appendix \ref{sapp:AllFluxTauT} (and figure \ref{fig:AllFluxReImTauT}), where profiles of the time-dependent scale factors and the Ricci scalar associated with the 4D Einstein frame metric are presented for representative choices of the constant parameters and $\beta(t)$. The solutions feature scale factors, and real and imaginary part of axiodilaton, that grow with time from positive initial values, with some cases exhibiting power-law growth at late times for certain (but not all) scale factors.

\section{Checks for energy conditions}
\label{sec:EnergyConds}

In this section, we examine the null, weak, strong, and dominant energy conditions (referred to as NEC, WEC, SEC, and DEC, respectively) for the energy-momentum tensors that arise in this work. Our analysis is performed for both the 10D and 4D energy-momentum tensors appearing in the corresponding Einstein field equations,
\begin{align}
  G_{MN}=T_{MN}\,,\quad M,N=0,\dots,9\,,\quad \h G_{\mu\nu}=\h T^{\textrm{eff}}_{\mu\nu}\,,\quad \mu,\nu=0,\dots,3\,.
\end{align}
The various components of the 10D energy-momentum tensor $T_{MN}$ are given in appendix \ref{app:EOMTermDetail}, while the various 4D effective energy-momentum tensors $\h T^{\mathrm{eff}}_{\mu\nu}$ for different flux configurations can be read from the equations \eqref{eq:GHatmunuCondTauC}, \eqref{eq:GHatmunuCondTauT}, \eqref{eq:GHatmunuCond0F5}, and \eqref{eq:GHatmunuCondAllFlux}.

To state the energy conditions, let $\mathcal{T}_{\mathfrak{a}\mathfrak{b}}$ denotes either the 10D or the 4D energy-momentum tensor. The indices $\mathfrak{a},\mathfrak{b}$ refer to the relevant spacetime (10D or 4D), $n$ denotes the dimension of the spacetime, and raising or lowering is performed with the corresponding metric (denoted by $g_{\mathfrak{a}\mathfrak{b}}$). The energy conditions are \cite{Kontou:2020bta, Curiel:2014zba, Dain:2013vsa, Malament:2005rc}:
\begin{align}
  \textrm{WEC}:\quad &\mathcal{T}_{\mathfrak{a}\mathfrak{b}}l^{\mathfrak{a}}l^{\mathfrak{b}}\geq 0\quad \forall\ l^{\mathfrak{a}}:\quad l_{\mathfrak{a}}l^{\mathfrak{a}}\leq 0\,, \nn\\
  \textrm{SEC}:\quad &\left[\mathcal{T}_{\mathfrak{a}\mathfrak{b}}-\frac{\mathcal{T}}{n-2}g_{\mathfrak{a}\mathfrak{b}}\right] l^{\mathfrak{a}}l^{\mathfrak{b}}\geq 0\quad \forall\ l^{\mathfrak{a}}:\quad l_{\mathfrak{a}}l^{\mathfrak{a}}< 0\,,\quad [\mathcal{T}\equiv\mathcal{T}_{\mathfrak{a}}^{\mathfrak{a}}]\,,\nn\\
  \textrm{DEC}:\quad &\mathcal{T}_{\mathfrak{a}\mathfrak{b}}l^{\mathfrak{a}}l^{\mathfrak{b}}\geq 0\,,\quad \mathcal{T}_{\mathfrak{a}\mathfrak{b}}\mathcal{T}^{\mathfrak{a}}{}_{\mathfrak{c}}l^{\mathfrak{b}}l^{\mathfrak{c}}\leq 0\quad \forall\ l^{\mathfrak{a}}:\quad l_{\mathfrak{a}}l^{\mathfrak{a}} < 0\,.
\end{align}
Note that in the WEC above, we have allowed $l^{\mathfrak{a}}$ to be null. Therefore WEC as stated includes the null energy condition (NEC).

\subsection{Checks in 10D}
\label{subsec:EnergyConds10D}

Let $l^M$ be any causal (i.e., non-spacelike) vector in the 10D spacetime ($g_{MN}$) of \eqref{eq:CommonAnsatz}:
\begin{align}
  l_{\mu}l^{\h \mu} +e^{-2\beta} l_ml^{\t m}=-a\,,\quad a\geq 0\,,\quad b\equiv e^{-2\beta}l_ml^{\t m}\geq 0\,. \label{eq:CausalVec10D}
\end{align}
Thereby $l^{\h \mu}$ is causal in the 4D spacetime ($\h g_{\mu\nu}$), since $a+b\geq 0$.

\textbf{Regarding the WEC}, for the 5-form contributions we have\footnote{In our constructions in section \ref{subsec:ExplicitSols}, $\lambda_m{}^{\t p}\lambda_{np}=k^2\t g_{mn},\ \lambda_2\cdot\lambda_2=3k^2$, and $\gamma^{\h \mu}$ is timelike.}
\begin{align}
  T^5_{\mu\nu}l^{\h\mu}l^{\h\nu}+e^{-4\beta} T^5_{mn} l^{\t m}l^{\t n}&=k^2e^{-4\beta}\left[\frac{3}{2}(\gamma_\mu l^{\h\mu})^2 -\gamma_\sigma\gamma^{\h\sigma}l_{\mu}l^{\h\mu}\right] -\frac{a k^2}{4}e^{-4\beta}\gamma_\mu\gamma^{\h\mu} \nn\\
  &\geq \frac{k^2}{2}e^{-4\beta}(\gamma_\mu l^{\h\mu})^2 -\frac{a k^2}{4}e^{-4\beta}\gamma_\mu\gamma^{\h\mu}\geq 0\,,
\end{align}
where we have used $\gamma_\mu\gamma^{\h\mu}<0$, and the reverse Cauchy-Schwarz inequality for causal vectors in Lorentzian manifolds \cite{ONeill:1983semi, Malament:2009Notes,Malament:2012topics}: $\gamma_\sigma\gamma^{\h\sigma}l_{\mu}l^{\h\mu}\leq (\gamma_\mu l^{\h\mu})^2$.

Now for the 3-form contributions we have\footnote{In our constructions in section \ref{subsec:ExplicitSols}, $\eta_m{}^{\t p}\bar\eta_{np}=k^2\t g_{mn},\ \eta_2\cdot\bar\eta_2=3k^2$, and $\zeta^{\h \mu}$ is timelike, $\textrm{Im}\tau>0$, while $\vartheta^{\h \mu}$ is spacelike.}
\begin{align}
  T^3_{\mu\nu}l^{\h\mu}l^{\h\nu}+e^{-4\beta} T^3_{mn} l^{\t m}l^{\t n}&=\frac{k^2}{2\textrm{Im}\tau}e^{-4\beta} \underbrace{\left[3(\zeta_{\mu}l^{\h\mu})^2 - \zeta_\sigma\zeta^{\h\sigma}l_{\mu}l^{\h\mu}\right]}_{\geq\ 2(\zeta_{\mu}l^{\h\mu})^2 \geq 0} +\frac{ak^2}{4\textrm{Im}\tau} e^{-4\beta} \zeta_\mu\zeta^{\h\mu} \nn\\
  &\quad +\frac{q\bar q}{2\textrm{Im}\tau} \underbrace{\left[(\vartheta_{\mu}l^{\h\mu})^2 - \vartheta_\sigma\vartheta^{\h\sigma}l_{\mu}l^{\h\mu}\right]}_{\geq 0} -\frac{a q\bar q}{4\textrm{Im}\tau} \vartheta_\sigma\vartheta^{\h\sigma}\,.
\end{align}
For the NEC, one sets $a=0$, in which case both the contributions above are non-negative as indicated by the underbraces. Here, for causal vectors $\zeta_\mu,l_\mu$, we have applied reverse Cauchy-Schwarz inequality, while for the pair of vectors $\vartheta_\mu,l_\mu$ we have used $\vartheta_\mu\vartheta^{\h\mu}>0, l_{\mu}l^{\h\mu}\leq 0$. For the WEC, we have $0<a\leq -l_{\mu}l^{\h\mu}$, and the contributions from the timelike $\zeta_\mu$ become:
\begin{align}
  &\frac{k^2}{2\textrm{Im}\tau}e^{-4\beta} \left[3(\zeta_{\mu}l^{\h\mu})^2 - \zeta_\sigma\zeta^{\h\sigma}l_{\mu}l^{\h\mu}\right] +\frac{ak^2}{4\textrm{Im}\tau} e^{-4\beta} \zeta_\sigma\zeta^{\h\sigma} \nn\\
  &\geq \frac{k^2}{\textrm{Im}\tau}e^{-4\beta} \left[(\zeta_{\mu}l^{\h\mu})^2+\frac{a}{4}\zeta_\sigma\zeta^{\h\sigma}\right]\geq \frac{k^2}{\textrm{Im}\tau}e^{-4\beta} \left[(\zeta_{\mu}l^{\h\mu})^2-\frac{1}{4}\zeta_\sigma\zeta^{\h\sigma}l_{\mu}l^{\h\mu}\right]\geq \frac{3k^2}{4\textrm{Im}\tau}e^{-4\beta} (\zeta_{\mu}l^{\h\mu})^2\,,
\end{align}
while the contributions from the spacelike $\vartheta_\mu$ can be rearranged as:
\begin{align}
  \frac{q\bar q}{2\textrm{Im}\tau} (\vartheta_{\mu}l^{\h\mu})^2 + \frac{q\bar q}{4\textrm{Im}\tau} \vartheta_\sigma\vartheta^{\h\sigma} \underbrace{\left[-2l_{\mu}l^{\h\mu}-a\right]}_{\geq a}\geq 0\,.
\end{align}

Now for the axiodilaton contributions we have\footnote{Axiodilaton does not depend on the internal $y$-coordinates in our constructions.}
\begin{align}
  T^1_{\mu\nu}l^{\h\mu}l^{\h\nu}+e^{-4\beta} T^1_{mn} l^{\t m}l^{\t n}=\frac1{2(\textrm{Im}\tau)^2} (l^{\h\mu}\partial_{\mu}\tau) (l^{\h\sigma}\partial_{\sigma}\bar{\tau})  +\frac{a}{4(\textrm{Im}\tau)^2} \partial_\mu\tau\partial^{\h \mu}\bar\tau \geq 0\,.
\end{align}
Therefore, our 10D energy-momentum tensor $T_{MN}$ satisfies both the NEC and WEC.

\textbf{Regarding the SEC}, we compute below the trace of $T_{MN}$. The contributions from the 5-from, 3-from fluxes and the axiodilaton are respectively:
\begin{align}
  &T^5_{\mu\nu}\h g^{\mu\nu}+e^{-2\beta} T^5_{mn} \t g^{mn}= 0\,, \nn\\
  &T^3_{\mu\nu}\h g^{\mu\nu}+e^{-2\beta} T^3_{mn} \t g^{mn}= -\frac{3k^2}{\textrm{Im}\tau}e^{-4\beta} \zeta_\sigma \zeta^{\h\sigma} +\frac{q\bar q}{\textrm{Im}\tau} \vartheta_\sigma\vartheta^{\h\sigma} > 0\,, \nn\\
  &T^1_{\mu\nu}\h g^{\mu\nu}+e^{-2\beta} T^1_{mn} \t g^{mn}= -\frac{2}{(\textrm{Im}\tau)^2}\partial_\sigma\tau\partial^{\h\sigma}\bar\tau > 0\,.
\end{align}
Given that WEC holds, as shown previously, and the total trace $T=g^{MN}T_{MN}$ is positive, it follows that the 10D energy-momentum tensor $T_{MN}$ satisfies the SEC.

\textbf{Regarding the DEC}, given that WEC holds, it remains only to verify that $T_{MN}l^N$ is causal for strictly $a>0$. Below we compute the individual contributions whose vector sum yields $T_{MN}l^N$:
\begin{align}
  T^5_{MN}l^N=&\frac{k^2}{4}e^{-4\beta} \left(6\gamma_\mu\gamma_\nu l^{\h\nu} - 3l_{\mu}\gamma_\nu\gamma^{\h\nu},\ l_{m}\gamma_\nu\gamma^{\h\nu}\right)\,, \nn\\
  T^3_{MN}l^N=&\frac{k^2e^{-4\beta}}{4\textrm{Im}\tau}\left(6\zeta_{\mu}\zeta_{\nu}l^{\h\nu} - 3l_{\mu} \zeta_\nu\zeta^{\h\nu},\ -l_m \zeta_\nu\zeta^{\h\nu} \right) +\frac{q\bar q}{4\textrm{Im}\tau}\left(2\vartheta_{\mu}\vartheta_{\nu}l^{\h\nu} - l_{\mu} \vartheta_\nu\vartheta^{\h\nu},\ l_m \vartheta_\nu\vartheta^{\h\nu} \right)\,, \nn\\
  T^1_{MN}l^N=&\frac1{4(\textrm{Im}\tau)^2} \left(2\partial_{(\mu}\tau\partial_{\nu)}\bar{\tau} l^{\h\nu}- l_{\mu}\partial_\nu\tau\partial^{\h\nu}\bar\tau,\ - l_m \partial_\nu\tau\partial^{\h \nu}\bar\tau\right)\,.
\end{align}
Each of the four parenthesized 10D vector appearing above is individually causal. Indeed, their squared norms w.r.t. $g_{MN}$ are, respectively,\footnote{Here, we use that $\tau$ depends only on time in our constructions in section \ref{subsec:ExplicitSols}. Accordingly, the following combination vanishes: $\partial_\mu\tau\partial^{\h \mu}\tau (l^{\h\nu}\partial_\nu\bar\tau)^2+\partial_\mu\bar\tau\partial^{\h \mu}\bar\tau (l^{\h\nu}\partial_\nu\tau)^2 -2\partial_\mu\tau\partial^{\h \mu}\bar\tau l^{\h\nu}\partial_\nu\tau l^{\h\sigma}\partial_\sigma\bar\tau=0$.}
\begin{align}
  &(\gamma_\nu\gamma^{\h\nu})^2\left(8l_{\mu}l^{\h\mu} -a\right)<0\,,\quad (\zeta_\nu\zeta^{\h\nu})^2\left(8l_{\mu}l^{\h\mu} -a\right)<0\,,\quad -a(\vartheta_\nu\vartheta^{\h\nu})^2<0\,,\quad -a(\partial_\nu\tau\partial^{\h \nu}\bar\tau)^2<0\,.
\end{align}
Note that, for the solutions constructed in section \ref{subsec:ExplicitSols} in which multiple vectors from the set $\{\gamma_\mu,\zeta_\mu,\vartheta_\mu,\partial_\mu\tau\}$ contribute, only the following subsets appear within a single case: $\{\gamma_\mu,\partial_\mu\tau\}$, $\{\zeta_\mu,\partial_\mu\tau\}$, or $\{\gamma_\mu,\vartheta_\mu,\partial_\mu\tau\}$.\footnote{The vectors $\gamma_\mu$ and $\zeta_\mu$ never appear in the same case. Accordingly, the constants labeled $k$ in the respective cases are independent, and this reuse of the symbol does not affect the analysis.} Therefore, to determine whether the total $T_{MN}l^N$ is causal or not, we need to check the cross terms in its squared norm.

For a general $\h g_{\mu\nu}$ the signs of these cross terms are difficult to assess. However, for our metric \eqref{eq:4DmetricAnsatz} they take particularly simple forms. Using that $\gamma_\mu$, $\zeta_\mu$, and $\partial_\mu\tau$ are purely temporal, while $\vartheta_\mu$ has only a $\vartheta_3$ component in our constructions, the $\gamma_\mu$–$\partial_\mu\tau$, $\zeta_\mu$–$\partial_\mu\tau$, $\vartheta_\mu$–$\partial_\mu\tau$, and $\gamma_\mu$–$\vartheta_\mu$ cross terms are, respectively,\footnote{Up to overall positive proportionality constants.}
\begin{align}
  &-e^{-2 \sigma } (3 a+4 b) \gamma _t^2 \left[(\textrm{Re}\dot\tau)^2+(\textrm{Im}\dot\tau)^2\right]\,,\quad -e^{-2 \sigma } (3 a+2 b) \zeta _t^2 \left[(\textrm{Re}\dot\tau)^2+(\textrm{Im}\dot\tau)^2\right]\,, \nn\\
  &-e^{-\alpha _3-\sigma } \left(a+2 e^{-\alpha _1} l_1^2+2 e^{-\alpha _2} l_2^2\right) \vartheta_3^2 \left[(\textrm{Re}\dot\tau)^2+(\textrm{Im}\dot\tau)^2\right]\,, \nn\\
  &-e^{-\alpha _3-\sigma } \left(3 a+4 b+6 e^{-\alpha _1} l_1^2+6 e^{-\alpha _2} l_2^2\right) \vartheta _3^2 \gamma _t^2\,,
\end{align}
all of which are strictly negative.

Combining, our 10D energy-momentum tensor $T_{MN}$ satisfies DEC.

\subsection{Checks in 4D}
\label{subsec:EnergyConds4D}

Following \eqref{eq:munuEinsteinEqAnsatz}, we write here the effective energy-momentum tensor that sources our 4D Einstein frame metric $\h g^E_{\mu\nu}$:\footnote{Using the transformation formula: $\h G_{\mu\nu}^E=\h G_{\mu\nu} -6\left(\h\nabla_\mu\h\nabla_\nu\beta-3\partial_\mu\beta \partial_\nu\beta\right) +3\h g_{\mu\nu} \left(2\hat{\nabla}^{\h 2}\beta + 3\partial_\sigma \beta \partial^{\h\sigma} \beta\right)\,,$ where on the r.h.s., hat denotes the use of $\h g_{\mu\nu}$. \label{fn:EinsteinTensorWeyl}}
\begin{align}
  \h T^{E,\textrm{eff}}_{\mu\nu}= &12\left[2\partial_\mu \beta \partial_\nu \beta -\h g^E_{\mu\nu}\partial_\sigma \beta \partial^{\h\sigma} \beta\right] +\frac{e^{-4\beta}}{4}\lambda_2\cdot\lambda_2\left[2\gamma_\mu\gamma_\nu-\h g^E_{\mu\nu}\gamma_\sigma\gamma^{\h\sigma}\right] \nn\\
  &+ \frac{e^{-4\beta}}{4\textrm{Im}\tau}\eta_2\cdot\bar\eta_2 \left[2\zeta_{\mu}\zeta_{\nu} - \h g^E_{\mu\nu} \zeta_\sigma\zeta^{\h\sigma}\right] +\frac{q\bar q}{4\textrm{Im}\tau} \left[2\vartheta_{\mu}\vartheta_{\nu} - \h g^E_{\mu\nu} \vartheta_\sigma\vartheta^{\h\sigma}\right] \nn\\
  &+\frac{1}{4(\textrm{Im}\tau)^2}\left[2\partial_{(\mu}\tau\partial_{\nu)}\bar{\tau} -\h g^E_{\mu\nu}\partial_\sigma\tau\partial^{\h \sigma}\bar\tau\right]\,, \label{eq:4DEffEMTensorE}
\end{align}
written generally; however, in our explicit constructions, not all vectors from $\{\gamma_\mu,\zeta_\mu,\vartheta_\mu,\partial_\mu\tau\}$ contribute simultaneously within a single case. Note that a hat on raised indices refers to the contraction performed using $\hat g^E_{\mu\nu}$, which is a convention restricted to this subsection and applies to all subsequent equations.\footnote{Aside from the transformation formula between 4D frames given in footnote \ref{fn:EinsteinTensorWeyl}.}

In order to check energy conditions for above energy-momentum tensor let us consider an arbitrary causal vector $l^{\h\mu}$ in 4D spacetime $\h g^{E}_{\mu\nu}$: $l_\mu l^{\h\mu}=-a\,,\ a\geq 0$. Note that as all prefactors multiplying the parenthesized contributions to $\h T^{E,\textrm{eff}}_{\mu\nu}$ are positive in our constructions, we will not keep track of them in the analysis below.

\textbf{Regarding the WEC}, for any of the timelike vectors, denoted by $w_\mu$ (in our constructions in section \ref{subsec:ExplicitSols}, $\partial_\mu\beta$, $\gamma_\mu$, $\zeta_\mu$, or $\partial_\mu\tau$) we have:
\begin{align}
  \left[2w_\mu w_\nu-\h g^E_{\mu\nu}w_\sigma w^{\h\sigma}\right]l^{\h\mu}l^{\h\nu} \geq l_\mu l^{\h\mu} w_\sigma w^{\h\sigma}\geq 0\,,
\end{align}
using reverse Cauchy-Schwarz inequality, while for spacelike vector $\vartheta_\mu$ we have:
\begin{align}
  \left[2\vartheta_{\mu}\vartheta_{\nu} - \h g^E_{\mu\nu} \vartheta_\sigma\vartheta^{\h\sigma}\right]l^{\h\mu}l^{\h\nu}=2(\vartheta_\mu l^{\h\mu})^2 + (-l_\mu l^{\h\mu}) \vartheta_\sigma\vartheta^{\h\sigma}\geq 0\,,
\end{align}
as both the terms are non-negative individually. Therefore, NEC and WEC are satisfied.

\textbf{Regarding the SEC}, the contribution of any of the timelike vectors $w_\mu$ in the trace $\h T^{E,\textrm{eff}}=g^E{}^{\mu\nu}\h T^{E,\textrm{eff}}_{\mu\nu}$ is positive:
\begin{align}
  \h g^E{}^{\mu\nu}\left[2w_\mu w_\nu-\h g^E_{\mu\nu}w_\sigma w^{\h\sigma}\right]=-2w_\sigma w^{\h\sigma}>0\,.
\end{align}
Therefore, energy-momentum tensor contributions from any $w_\mu$ satisfies SEC since WEC holds. However, the trace contribution from the spacelike vector $\vartheta_\mu$ is negative, given by $-2\vartheta_\sigma\vartheta^{\h\sigma}$. Thus, in this case we need to consider its contribution to the Ricci tensor, which is clearly non-negative:
\begin{align}
  \left[\h T^{E,\textrm{eff}}_{\mu\nu}-\frac{\h T^{E,\textrm{eff}}}{2}\h g^E_{\mu\nu}\right]_{\vartheta_\sigma}l^{\h\mu}l^{\h\nu}= \frac{q\bar q}{2\textrm{Im}\tau}(\vartheta_\mu l^{\h\mu})^2\geq 0.
\end{align}
Hence, $T^{E,\textrm{eff}}_{\mu\nu}$ satisfies SEC.

\textbf{Regarding the DEC}, given that WEC holds, it remains only to verify that $\h T^{E,\textrm{eff}}_{\mu\nu}l^{\h\nu}$ is causal for strictly $a>0$. The individual contributions to it from $w_\mu$ and $\vartheta_\mu$ are timelike; their squared norms are namely:
\begin{align}
  l_\mu l^{\h\mu}(w_\sigma w^{\h\sigma})^2<0\,,\quad l_\mu l^{\h\mu}(\vartheta_\sigma \vartheta^{\h\sigma})^2<0\,.
\end{align}
There are two types of crossed terms generated in the squared norm of $\h T^{E,\textrm{eff}}_{\mu\nu}l^{\h\nu}$: those between two timelike vectors $w_\mu,w'_\mu$, and those between a timelike vector $w_\mu$ and the spacelike vector $\vartheta_\mu$. Since in our constructions the timelike vectors $w_\mu,w'_\mu$ are purely temporal, while $\vartheta_\mu$ has only a $\vartheta_3$ component, for our metric $e^{6\beta}\times$\eqref{eq:4DmetricAnsatz} these cross terms reduce to, respectively:
\begin{align}
  -a e^{-2\sigma -12\beta} w_t^2 w_t^{\prime 2}\,,\quad - e^{-\sigma -\alpha _1-\alpha _2-\alpha _3-18 \beta} \left(a e^{\alpha _1+\alpha _2+6 \beta }+2 e^{\alpha _2} l_1^2+2 e^{\alpha _1} l_2^2\right) \vartheta _3^2 w_t^2\,,
\end{align}
all of which are strictly negative. Therefore, DEC holds.

\begin{comment}
%
\begin{align}
  &(2w_\mu w_\nu l^{\h\nu}-l_{\mu} w_\sigma w^{\h\sigma},\ 2\vartheta_\mu \vartheta_\nu l^{\h\nu}-l_{\mu} \vartheta_\sigma \vartheta^{\h\sigma}) \nn\\
  &=4w_\mu\vartheta^{\h\mu} w_\nu l^{\h\nu}\vartheta_\nu l^{\h\nu} -2l_{\mu}\vartheta^{\h\mu} w_\sigma w^{\h\sigma}\vartheta_\nu l^{\h\nu} -2w_\mu l^{\h\mu}w_\nu l^{\h\nu}\vartheta_\sigma \vartheta^{\h\sigma} +l_\mu l^{\h\mu}w_\nu w^{\h\nu}\vartheta_\sigma \vartheta^{\h\sigma}\,.
\end{align}
%
\end{comment}

\section{Revisiting Maldacena--Nu\~nez analysis}
\label{sec:MNextension}

Our constructions of time-dependent flux backgrounds above feature: an overall time-depen\-dent scale factor multiplying the internal metric, and a 3-form field strength with $[1,2]$ or $[3,0]$ components (in the latter case lying entirely in the noncompact 4D spacetime), while a 5-form field strength with $[3,2]$ and $[1,4]$ components. This motivates revisiting the analysis of the Maldacena--Nu\~nez (MN) no go theorem \cite{Maldacena:2000mw} incorporating these elements.\footnote{\cite{Maldacena:2000mw} focuses on $n$-form field strengths either with indices completely along the internal dimensions if $n\leq 6$, or with $4$ indices along the non-compact dimensions and rest along the internal dimensions if $n\geq 4$.}

The following analysis is not specific to type IIB supergravity. The energy momentum tensor used below may come from any underlying action of scalar and form fields coupled to gravity. The analysis uses only the Einstein field equations. We work with 4 noncompact and 6 compact dimensions, but extending the discussion to general noncompact and compact dimensionalities is straightforward.

Consider the following ansatz for the metric $g_{MN}$ on 10D spacetime:
\begin{align}
  d s^2_{10}=e^{2 A(y)} \hat{g}_{\mu \nu}(x) d x^\mu d x^\nu+ e^{2 \beta(x)} \tilde{g}_{m n}(y) d y^m d y^n\,,\label{eq:10DMetricDoubleWarp}
\end{align}
where $A,\beta$ are real valued functions, which make the 10D geometry a doubly warped product spacetime.\footnote{One may further generalize the setup by allowing off-diagonal components in the 10D metric; we leave this for future work.}

The trace-reversed Einstein field equations are:
\begin{align}
  R_{MN} = T_{MN} -\frac{1}{8}g_{MN}T\,,\quad T\equiv g^{MN}T_{MN}=e^{-2A}T_{\mu}^{\h\mu}+e^{-2\beta}T_{m}^{\t m}\,. \label{eq:TrRevEinsteinGen}
\end{align}
In particular, the noncompact $\mu\nu$-components, contracted with $\h g^{\mu\nu}$, lead to:\footnote{See appendix \ref{app:EOMTermDetail} for details of the curvature tensors and scalar associated with the metric \eqref{eq:10DMetricDoubleWarp}.} \footnote{Using following identities: $e^{-nA}\tilde{\nabla}^{\t 2}e^{nA}=n\tilde{\nabla}^{\t 2}A+n^2\partial_{m}A\partial^{\t m}A\,,\quad e^{-n\beta}\h\nabla^{\h 2}e^{n\beta}=n\h\nabla^{\h 2}\beta+n^2\partial_{\mu}\beta\partial^{\h\mu}\beta\,.$}
\begin{align}
  &e^{2A}\h R - 6e^{2A}e^{-\beta}\h\nabla^{\h 2}e^{\beta} -e^{-2\beta}\tilde{\nabla}^{\t 2}e^{4A} = \frac{1}{2}e^{4A}\left(e^{-2A}T^{\h\mu}_{\mu}-e^{-2\beta}T_{m}^{\t m}\right)\,. \label{eq:MNRhat}
\end{align}
One might examine the sign of $\h R$ above; however, the sign can change for the 4D Einstein frame metric, as we observed in our explicit constructions. At this stage, focusing on the Einstein-Hilbert term in action we obtain the 4D Einstein frame metric $\h g_{\mu\nu}^E$ via a Weyl rescaling of the metric $\h g_{\mu\nu}$, as follows.
\begin{align}
  S&=\frac1{2\kappa_{10}^2} \int d^{10}X \sqrt{-g}\ R_g + \cdots \nn\\
  &=\frac1{2\kappa_{10}^2} \int d^6y \sqrt{\t g}\ e^{2A} \int d^4x \sqrt{-\h g^E}\ e^{\Omega}e^{6\beta}\h R^E +\cdots \,,\quad \h g_{\mu\nu}=e^{\Omega(x)} \h g^{E}_{\mu\nu}\,.
\end{align}
For the 4D Einstein frame, the Weyl factor $\Omega$ must be chosen as
\begin{align}
  e^{\Omega}=e^{-2\kappa_4}e^{-6\beta}\,,\quad \h g^{E}_{\mu\nu}=e^{2\kappa_4}e^{6\beta} \h g_{\mu\nu}\,,\quad \h R^E=e^{-2\kappa_4}e^{-6\beta} \left[\h R-6e^{-3\beta}\hat{\nabla}^{\h 2}e^{3\beta}\right]\,,
\end{align}
where $\kappa_4$ is a constant parameter. In this case, the 4D gravitational constant depends on the warp factor $A(y)$.\footnote{This is given by, $(16\pi G_N)^{-1}=\left(e^{-2\kappa_4}/2\kappa_{10}^2\right) \int d^6y \sqrt{\t g}\ e^{2A}$.}

Hence, from \eqref{eq:MNRhat}, the Ricci scalar of the 4D Einstein frame metric can be expressed as:\footnote{Where on the r.h.s., hat denotes the use of $\h g_{\mu\nu}$, not $\h g^E_{\mu\nu}$.}
\begin{align}
  e^{2\kappa_4}e^{2A}e^{6\beta}\h R^E=\frac{1}{2}e^{4A}\left(e^{-2A}T^{\h\mu}_{\mu}-e^{-2\beta}T_{m}^{\t m}\right) - 3 e^{2A} e^{-4\beta}\hat{\nabla}^{\h 2}e^{4\beta} +e^{-2\beta}\tilde{\nabla}^{\t 2}e^{4A}\,. \label{eq:MNRhatE}
\end{align}
Integrating over the 6D internal space coordinates we get:
\begin{align}
  &e^{2\kappa_4}e^{6\beta}\h R^E \int d^6y\sqrt{\t g}e^{2A}=\frac{1}{2}\int d^6y\sqrt{\t g} e^{4A} \bar{T}_{\textrm{I}} - 3 e^{-4\beta}\hat{\nabla}^{\h 2}e^{4\beta} \int d^6y\sqrt{\t g}e^{2A}\,, \nn\\
  &\bar{T}_{\textrm{I}}\equiv e^{-2A}T^{\h\mu}_{\mu}-e^{-2\beta}T_{m}^{\t m}\,, \label{eq:MNRhatEInty}
\end{align}
where the $\tilde{\nabla}^{\t 2}e^{4A}$ contribution is dropped assuming a closed 6D internal space.

Similar equation has usually been used to comment on the sign of $\h R^E$. Before coming to that, we first record the consequence of contracting the internal $mn$-components of \eqref{eq:TrRevEinsteinGen} with $\t g^{mn}$. This leads to:
\begin{align}
  e^{-A} e^{-6\beta}\hat{\nabla}^{\h 2} e^{6\beta}=\frac{1}{4}e^{A}\left(3e^{-2A}T_{\mu}^{\h\mu}-e^{-2\beta}T_{m}^{\t m}\right) +e^{A-2\beta}\tilde{R} -4 e^{-2\beta}\t\nabla^{\t 2}e^{A}\,. \label{eq:MNRtilde}
\end{align}
Integrating over the 6D internal space coordinates we get:
\begin{align}
  &e^{-6\beta}\hat{\nabla}^{\h 2} e^{6\beta} \int d^6y\sqrt{\t g}e^{-A} =\frac{1}{4} \int d^6y\sqrt{\t g}e^{A}\bar{T}_{\textrm{II}} +e^{-2\beta}\int d^6y\sqrt{\t g}e^{A}\tilde{R}\,, \nn\\
  &\bar{T}_{\textrm{II}}\equiv 3e^{-2A}T_{\mu}^{\h\mu}-e^{-2\beta}T_{m}^{\t m}\,,  \label{eq:MNRtildeInty}
\end{align}
where, as before, the $\tilde{\nabla}^{\t 2}e^{A}$ contribution is dropped assuming a closed 6D internal space.

It is noteworthy that the $\hat{\nabla}^{\h 2}e^{4\beta}$ term in \eqref{eq:MNRhatEInty} is new compared to \cite{Maldacena:2000mw}. The sign of $\h R^E$ then depends on both the sign of this new term and the sign of $\bar{T}_{\textrm{I}}$. On the other hand, \eqref{eq:MNRtildeInty} can be used to determine the sign of $\hat{\nabla}^{\h 2}e^{6\beta}$ for possible signs of $\t R$ and $\bar{T}_{\textrm{II}}$. We do this when the 6D internal space is Ricci flat, i.e., $\t R=0$.

\subsection{Sign of $\bar{T}_{\textrm{I}}$ and $\bar{T}_{\textrm{II}}$}

Let us consider a generic $n$-form field strength $F_n$. Its contribution to the energy-momentum tensor, up to an overall positive factor, is given by:
\begin{align}
  T_{MN}=F_{MA_1\dots A_{n-1}}F_{N}{}^{A_1\dots A_{n-1}}-\frac{1}{2n}g_{MN}F_{A_1\dots A_{n}}F^{A_1\dots A_{n}}\,.
\end{align}
The proportionality factor may include dilaton contributions. The above expression is general across theories with a standard quadratic kinetic term for an $n$-from field strength, including all classical supergravity theories.

There are two classes of $F_n$, characterized by their $[r,n-r]$ components, that we need to analyze. The first consists of field strengths with $1\leq r \leq 3$ (only partially covering the 4D noncompact spacetime). The second consists of those with either $r=4$ (fully covering the 4D noncompact spacetime) or $r=0$ (purely internal components). The second class was studied in \cite{Maldacena:2000mw}, where they were shown to give negative contributions to $\bar{T}_{\textrm{I}}$. We will refer to these field strengths as non-MN-type and MN-type, respectively.

Here we give the general formulae that hold for any $[r,n-r]$:
\begin{align}
  &\bar{T}_{\textrm{I}}= c (2 r - n +1) e^{-2rA-2(n-r)\beta} K^2 S^2\,,\quad \bar{T}_{\textrm{II}}= c (4 r - n - 3) e^{-2rA-2(n-r)\beta} K^2 S^2\,, \\
  &F_n\equiv K_r\wedge S_{n-r}\,,\quad c\equiv(n-1)! \,,\quad S^2\equiv\frac{1}{(n-r)!}S_{l_1\cdots l_{n-r}}S^{\t l_1\cdots\t l_{n-r}}>0\,, \nn\\
  &K^2\equiv\frac{1}{r!}K_{\mu_1\cdots\mu_r}K^{\h\mu_1\cdots\h\mu_r}\,. \nn
\end{align}

\paragraph{Non-MN-type field strengths:} In this case we have $1 \leq r \leq 3$. Due to Hodge duality, in 10D we may impose an upper bound: $r\leq n \leq 5$. Depending on the pair $(r,n)$, $\bar{T}_{\textrm{I}}$ and $\bar{T}_{\textrm{II}}$ may have the same sign or opposite signs, and the corresponding sets of admissible $(r,n)$ values can be tabulated straightforwardly. Ultimately, the signs of $\bar{T}_{\textrm{I}},\bar{T}_{\textrm{II}}$ are determined by the sign of $K^2$, which can be either positive or negative since the 4D spacetime is Lorentzian.

\paragraph{MN-type field strengths:} In this case we have either $r=0$ or $r=4$. When $r=0$, $K_0$ is a scalar and therefore $K^2=K_0^2>0$. For $n\geq 1$ this implies $\bar{T}_{\textrm{I}}\leq 0$ and $\bar{T}_{\textrm{II}}<0$. When $r=4$, $K_4=f(x)\h\epsilon$, and therefore $K^2=-f(x)^2<0$. For $n\geq 4$ this implies $\bar{T}_{\textrm{I}}< 0$ and $\bar{T}_{\textrm{II}}<0$ again.

\subsection{Sign of $\h R^E$}
%\subsection{Sign of 4D Einstein frame Ricci scalar}

For the case $\t R=0$, \eqref{eq:MNRtildeInty} implies that $\hat{\nabla}^{\h 2}e^{6\beta}$ and $\bar{T}_{\textrm{II}}$ are of same sign. However as indicated above that one can cook up $\bar{T}_{\textrm{II}}$ of either signs. Note that $\bar{T}_{\textrm{I}}>0$, which can be cooked up with non-MN-type field strengths, generates positive contribution to $\h R^E$ via \eqref{eq:MNRhatEInty}. When $\bar{T}_{\textrm{II}}<0$, we have $\hat{\nabla}^{\h 2}e^{6\beta}<0$. On top of this, if $\partial_\mu\beta\partial^{\h\mu}\beta>0$, we have $0>\hat{\nabla}^{\h 2}e^{6\beta}>\hat{\nabla}^{\h 2}e^{4\beta}$.\footnote{We have: $\frac{1}{6}e^{-6\beta}\hat{\nabla}^{\h 2}e^{6\beta}-\frac{1}{4}e^{-4\beta}\hat{\nabla}^{\h 2}e^{4\beta}=2\partial_\mu\beta\partial^{\h\mu}\beta\,.$} Such a $\hat{\nabla}^{\h 2}e^{4\beta}$ generates positive contribution to $\h R^E$ via \eqref{eq:MNRhatEInty}. When $\bar{T}_{\textrm{II}}>0$, we have $\hat{\nabla}^{\h 2}e^{6\beta}>0$. On top of this, if $\partial_\mu\beta\partial^{\h\mu}\beta<0$, we have $0<\hat{\nabla}^{\h 2}e^{6\beta}<\hat{\nabla}^{\h 2}e^{4\beta}$. Such a $\hat{\nabla}^{\h 2}e^{4\beta}$ generates negative contribution to $\h R^E$ via \eqref{eq:MNRhatEInty}.

In summary, the existence of a positive $\h R^E$ cannot be ruled out by considering only the traced Einstein equations. In the pursuit of $\h R^E>0$, this suggests the need for a comprehensive search in full generality, in which all types of components in the field strengths and both warp factors $A(y)$ and $\beta(x)$ are retained, and the full set of supergravity equations is solved.

\paragraph{Specializing to our constructions:} Before closing this section, below we present the explicit $\bar{T}_{\textrm{I}}, \bar{T}_{\textrm{II}}$ obtained in our constructions in section \ref{subsec:ExplicitSols} where we work with $A(y)=0$. Recall that in all these constructions we have $\h R^E<0$. Here, we justify this result in light of \eqref{eq:MNRhatEInty}.

In section \ref{subsec:ExplicitSols}, the contributions from the 5-form, 3-form fluxes, and the axiodilaton to $\bar{T}_{\textrm{I}}$ and $\bar{T}_{\textrm{II}}$ are as follows:
\begin{align}
  &\bar{T}_{\textrm{I}}^5=-3k^2e^{-4\beta}\gamma_\mu\gamma^{\h\mu}>0\,,\quad \bar{T}_{\textrm{II}}^5=-6k^2e^{-4\beta}\gamma_\mu\gamma^{\h\mu}>0\,, \nn\\
  &\bar{T}_{\textrm{I}}^3=0 -\frac{2q\bar q}{\textrm{Im}\tau} \vartheta_\mu\vartheta^{\h\mu}<0\,,\quad \bar{T}_{\textrm{II}}^3=\underbrace{-\frac{3 k^2 e^{-4\beta}}{\textrm{Im}\tau} \zeta_\mu\zeta^{\h\mu}}_{>0} \underbrace{-\frac{3 q\bar q}{\textrm{Im}\tau} \vartheta_\mu\vartheta^{\h\mu}}_{<0}\,, \nn\\
  &\bar{T}_{\textrm{I}}^1=\frac{1}{(\textrm{Im}\tau)^2}\partial_\mu\tau\partial^{\h \mu}\bar\tau<0\,,\quad \bar{T}_{\textrm{II}}^1=0\,,
\end{align}
written generally; however not all vectors from $\{\gamma_\mu,\zeta_\mu,\vartheta_\mu,\partial_\mu\tau\}$ contribute simultaneously within a single case.

In all our constructions, the scale factor exponents $\sigma,\alpha_i,\beta$ are such that both $\h\nabla^{\h 2}e^{4\beta}$ and $\h\nabla^{\h 2}e^{6\beta}$ turn out to be positive. A positive $\h\nabla^{\h 2}e^{6\beta}$ requires the corresponding $\bar{T}_{\textrm{II}}$ to be positive, see \eqref{eq:MNRtildeInty}. Although $\vartheta_\mu$ contributes negatively to $\bar{T}_{\textrm{II}}$, in the same setup the contribution from $\gamma_\mu$ is positive, and together they yield a positive $\bar{T}_{\textrm{II}}$. Now, a positive $\h\nabla^{\h 2}e^{4\beta}$ gives a negative contribution to $\h R^E$ via \eqref{eq:MNRhatEInty}. While $\gamma_\mu$  contributes positively to $\bar{T}_{\textrm{I}}$ and hereby to $\h R^E$, in the same setup the contribution from $\h\nabla^{\h 2}e^{4\beta}$ is larger in magnitude. As a result, the total $\h R^E$ is negative.

\section{Discussions}
\label{sec:Discuss}

In section \ref{sec:TDepBackgIIBFlux}, we have constructed analytic time-dependent backgrounds in 10D type IIB supergravity with nontrivial fluxes. Specifically, we have considered the following flux configurations:
\begin{itemize}
  \item[A.] The self-dual 5-form flux $\t F_5$ featuring $[3,2]$ and $[1,4]$ components, with both time-dependent and time-independent nonvanishing axiodilaton $\tau$, while the 3-form flux $G_3$ vanishes. Explicit solutions falling under this category are provided by: \eqref{eq:0G3F5TauCEx}, \eqref{eq:0G3F5ImTauTEx}, \eqref{eq:0G3F5ReImTauTEx}.
  \item[B.] The 3-form flux $G_3$ featuring $[1,2]$ components, with a time-dependent axiodilaton $\tau$, while the 5-form flux $\t F_5$ vanishes. Explicit solutions falling under this category are provided by: \eqref{eq:0F5G3ImTauTEx}, \eqref{eq:0F5G3ReImTauTEx}.
  \item[C.] The self-dual 5-form flux $\t F_5$ featuring $[3,2]$ and $[1,4]$ components, the 3-form flux $G_3$ featuring $[3,0]$ components, and a time-dependent axiodilaton $\tau$. Explicit solutions falling under this category are provided by: \eqref{eq:AllFluxImTauTEx}, \eqref{eq:AllFluxReImTauTEx}.
\end{itemize}
The internal components of the 10D Einstein field equations have been solved keeping the 6D internal metric $\t g_{mn}$ (as introduced in \eqref{eq:CommonAnsatz}) Ricci flat on any compact K\"ahler manifold. Meanwhile, the internal components of the aforelisted fluxes, which are 2-forms and 4-forms as seen above, are maintained respectively by the K\"ahler form $J_2$ of the internal manifold and its Hodge dual $\t\star J_2$ (up to proportionality constants). Such choices have solved constraints coming from the internal components of the 10D Einstein field equations as well as the equations of motion and Bianchi identities for the fluxes. In section \ref{sec:EnergyConds}, we have verified that the energy-momentum tensors (both 10D and the resulting 4D effective) appeared in our constructions satisfy the null, weak, strong, and dominant energy conditions. In section \ref{sec:MNextension}, we have revisited the Maldacena--Nu\~nez no-go analysis, incorporating elements from our constructions -- an overall noncompact spacetime-dependent scale factor multiplying the internal metric and field strengths only partially covering the noncompact directions -- and argued that with these generalizations a 4D Einstein frame metric with positive Ricci scalar cannot be ruled out solely by such an analysis.

%\subsection{Remarks and outlook}

This article achieves our primary aim of explicitly constructing time-dependent flux backgrounds in type IIB supergravity, leaving a detailed study of their cosmological implications for future work. In the following, we present several remarks relevant to our constructions along with some potential future directions:

\subsection*{\textit{1. Moduli space of the constructed backgrounds}}

The time-dependent flux backgrounds constructed in this paper utilize a 6D internal Ricci-flat metric $\t g_{mn}$ on any compact K\"ahler manifolds with internal components of various form fluxes depending only on the respective K\"ahler form and its Hodge dual. Such a metric $\t g_{mn}$ can be continuously deformed through K\"ahler and complex structure deformations, which preserve Ricci-flatness. The deformed K\"ahler form can then still be used in our constructions, continuing to satisfy the supergravity equations. Consequently, the moduli space of our constructed backgrounds will at least include these geometric moduli. It will be important to fully characterize all the moduli including any arising from deformations of the noncompact components of the fluxes. Dimensional reduction of such moduli fields around our backgrounds would then yield their 4D EFT. This could have phenomenological implications in early universe cosmology, as discussed later. We leave a detailed characterization of the moduli space of the constructed backgrounds for future work.

\subsection*{\textit{2. SUGRA constraints on the scale factor multiplying the internal metric:}}

In all our constructions, the 6D scale factor exponent $\beta(t)$, as introduced via \eqref{eq:CommonAnsatz}, remains unfixed. However, there are several constraints on its functional dependence in order for the supergravity equations to be sensibly solved. The constraint common across all constructions is that $\beta(t)$ must be a strictly monotonic function. This is because a factor of $\dot{\beta}^2$ appears in one of the 4D scale factors, $e^\sigma$ that multiplies the time direction as introduced via \eqref{eq:4DmetricAnsatz}, and any local extrema of $\beta(t)$ would cause $e^\sigma$ to vanish at those instants of time. Moreover, for a sensible 4D spacetime \eqref{eq:4DmetricAnsatz} a strictly monotonic $\beta(t)$ that approaches a constant at late times is also not allowed, since $\dot\beta$ (thereby $e^\sigma$) would tend to zero.

To discuss further constraints, we focus on two situations that we have encountered. In all the flux configurations listed above (except for a subcase of the last one in C, where time dependence in both $\textrm{Re}\tau$ and $\textrm{Im}\tau$ is considered), we have obtained the following constraint: $\forall\ t>0,\ e^{a\beta} < b^2$ for some constants $a,b > 0$. This condition arises from the appearance of a combination $b^2 - e^{a\beta}$ in the 4D scale factor solutions, and is necessary for the reality of such scale factor exponents\footnote{In other words, for the positivity of the 4D scale factors.}. A strictly increasing $\beta(t)$ that is not allowed to approach a constant at late times cannot satisfy this, so $\beta(t)$ must be chosen a strictly decreasing function which does not approach a constant asymptotically. In contrast, in the subcase of the flux configuration C where time dependence is considered in both $\textrm{Re}\tau$ and $\textrm{Im}\tau$, we must take $\beta(t)$ to be strictly increasing which does not approach a constant asymptotically. This condition arises because, in this case (see, \eqref{eq:AllFluxReImTauTEx}), one of the 4D scale factors contains a factor $e^{c\beta}$ while another contains a factor $e^{d\beta}$ for some constants $c,d$, and these constants cannot both be negative simultaneously, but can both be positive. We require that none of the 4D scale factors approach zero for any $t > 0$, even at late times.

We have argued above that $\beta(t)$ cannot approach a constant asymptotically. Among such possible functions, interestingly, $\beta\sim\mp\ln t$ at late times is allowed even though $\dot\beta$ approaches to zero as $\mp 1/t$.\footnote{Note that $\ln t$ is bounded by $t^n$ for any real $n>0$ at late times. Thus, logarithmic behaviours shall be in the boundary of allowed space of functions $\beta(t)$.} The minus sign corresponds to the situation where $\beta$ shall be strictly decreasing, while the plus sign corresponds to the situation where $\beta$ shall be strictly increasing as discussed above. To justify this, we need to closely examine the other factors that appear in the scale factor $e^{6\beta+\sigma}$\footnote{In the 4D Einstein frame as in \eqref{eq:4DEinsteinFrameWeyl}.} aside from the factor $\dot\beta^2$.

For the first situation, generically we have:
\begin{align}
  e^{6\beta+\sigma}=\frac{\dot\beta^2 e^{k \coth ^{-1}\left[\frac{b}{\sqrt{b^2-e^{a \beta}}}\right]+l}}{b^2-e^{a \beta}} \underset{t\to+\infty}{\sim} 2^k e^l b^{k-2} t^{\frac{a k}{2}-2}\,,\quad \beta\sim-\ln t\,,
\end{align}
which does not necessarily approach zero at late times, as the constant $k>4/a$ can be chosen to avoid this, given the constant $a>0$ that appears in the relevant constructions. We now provide a proof that the factors above in $e^{6\beta+\sigma}$ do not affect the statement that when $\beta(t)$ approaches to a constant asymptotically, $e^{6\beta+\sigma}$ approaches zero -- a statement we used to discard such functions. For such a $\beta(t)$, the late-time behaviour is generically of the form: $\beta\sim\beta_0+\beta_n t^{-n}+\mathcal{O}(t^{-n-1})$ for some $n>0$ (even fractions are allowed) and constants $\beta_0,\beta_n$. Moreover, we should have $e^{b\beta_0}<b^2$, although this does not affect the leading power of $t$ in $e^{6\beta+\sigma}$ at late times. Based on the above formula, we obtain $e^{6\beta+\sigma}\sim t^{-2 (n+1)}$ (up to a positive proportionality constant).

Now for the second situation (see, \eqref{eq:AllFluxReImTauTEx}), generically we have with some $a,b>0$:
\begin{align}
  e^{6\beta+\sigma}= \dot\beta^2 e^{b^2 e^{a \beta}+k \beta+l}\underset{t\to+\infty}{\sim} t^{k-2} e^{b^2 t^a+l}\,,\quad \beta\sim +\ln t\,,
\end{align}
which does not necessarily approach zero at late times, as the constant $k>2$ can be chosen to avoid this in the relevant construction. Similar to the previous case, at late times $e^{6\beta+\sigma}\sim t^{-2 (n+1)}$ when $\beta\sim\beta_0+\beta_n t^{-n}+\mathcal{O}(t^{-n-1})$ for some $n>0$.

We arrive at the above constraints on $\beta(t)$, requiring a well-defined 4D Einstein frame metric $e^{6\beta}\times$\eqref{eq:4DmetricAnsatz}. However, in the coordinate system of cosmic time, this requires more careful analysis. The cosmic time is defined as $t'=\int_0^t e^{3\beta(s)+\sigma(s)/2}ds$. Clearly, positivity of the scale factors implies that as $t$ increases from zero, the area under the curve (thereby $t'$) increases from zero. However, for the cosmic time to span up to infinity, the area under the curve must not be finite as $t \to \infty$. As the $t$-dependence of $\sigma$ comes through $\beta$, this certainly restricts the space of allowed functions $\beta(t)$. A $\beta(t)$ which approaches a constant at late times would make $e^{3\beta(t)+\sigma(t)/2}$ approach zero as $t^{-n-1}$ for some $n>0$, thus the integral over $t$ from $0$ to $\infty$ would be finite, and hence such $\beta$ are ruled out. This will be interesting to study further implications of the well-definedness of cosmic time on the allowed space of $\beta$.

The aforesaid constraints on the 6D scale factor exponent $\beta(t)$ arise from the way it appears in the 4D scale factors as solutions to the 10D supergravity equations. Representative choices of $\beta$ and various constant parameters that appear in our constructions, while maintaining the above constraints, have been considered in appendix \ref{app:4DgHatEinProf}. The above discussion may be of particular relevance in the context of achieving scale separation between the KK length scale and the 4D curvature radius in time-dependent compactifications, as recently studied in \cite{Andriot:2025cyi}. We leave the study of such scale separation as time flows using our explicit flux backgrounds for future work.

\begin{comment}
%
Notably, by relaxing some of the constraints on $\beta(t)$ noted above we can cook up cases where all/some of the 4D scale factors decay to zero asymptotically at late times. Such 4D metrics could also be of interest in studying scenarios like the Big Crunch and related cosmological phenomena \cite{ToledoSesma:2015wxj, Apers:2022cyl, Hertog:2004rz}. In any case, the constraints those arising from the reality of the metric components should not be violated. \rma{What about Big Bang, Milne universe?}
%
\end{comment}

\subsection*{\textit{3. Late time 4D scale factors and Kasner-like backgrounds}}

In our constructions, it is not possible for all the scale factors associated with the 4D Einstein frame metric to asymptotically approach constant values at late times (i.e., for the 4D spacetime to approach Minkowski space). This is because the time dependence of these scale factors is governed by a single function $\beta(t)$, and while the scale factors $e^{6\beta + \alpha_i}$ in  $e^{6\beta}\times$\eqref{eq:4DmetricAnsatz} can become positive constants at late times only if $\beta$ approaches a constant, in that case $e^{6\beta + \sigma}$  would vanish at late times, as discussed above.

At late times, the Ricci scalar $\h R^E$ of the 4D Einstein frame metric asymptotically approaches to zero in all our constructions. Therefore, the 4D metric does not represent a vacuum solution at any large (but finite) time $t$. While our scale factors are not simple powers of $t$ at generic time, at late times we can obtain power-law growths, as constructed in appendix \ref{app:4DgHatEinProf} by considering logarithmic time dependence of $\beta$ (except for the subcase of the flux configuration C where time dependence in both $\textrm{Re}\tau$ and $\textrm{Im}\tau$ is considered). In such cases, the 4D Einstein frame metric at late times takes the form:
\begin{align}
  \h ds_E^2=-\epsilon_0t^{p_0}dt^2+\sum_{i=1}^3 \epsilon_it^{p_i}dx^idx^i\,,\quad \sum_{i=1}^3 p_i=2+p_0\,,
\end{align}
The condition on the powers of $t$ is known as first Kasner condition. Indeed, the powers of $t$ derived in appendix \ref{app:4DgHatEinProf} satisfy this Kanser condition. However, they do not satisfy the second Kasner condition, which requires $\sum_{i=1}^3 p_i^2=\left(2+p_0\right)^2$\footnote{When $p_0=0$, one recovers the well known Kasner conditions.}, thus preventing the 4D metric from being a Kasner metric (which corresponds to a vacuum solution) at late times.

Also see appendix \ref{app:4DgHatEinProf}, for the exponential growth of all the 4D scale factors at late times constructed by considering linear time dependence of $\beta$, except for the subcase of flux configuration C where time dependence in both $\textrm{Re}\tau$ and $\textrm{Im}\tau$ is considered. In the latter case, we have obtained a mix of exponential growth (for the scale factors $e^{6\beta+\alpha_1},e^{6\beta+\alpha_2}$) and double-exponential growth (for the scale factors $e^{6\beta+\sigma},e^{6\beta+\alpha_3}$).

For the cases with power-law growth of the temporal scale factor $e^{6\beta+\sigma}$ in $t$, the integral defining the cosmic time $t'(t)$ grows as a power of $t$ (with the power raised by one) at large $t$. While, for the cases with exponential growth of the temporal scale factor in $t$, $t'(t)$ grows exponentially with $t$ at large $t$. Therefore, at late cosmic time $t'$ (which corresponds to large $t$, as discussed previously), the inversion $t(t')$ grows as a power of $t'$ or logarithmically in $t'$, respectively in these two types of cases. Consequently, in the second type of cases, the spatial scale factors that grow exponentially in $t$ also evolve as power laws in cosmic time $t'$, as in the first type. It will be interesting to study the spatial scale factors $e^{6\beta(t(t'))+\alpha_i(t(t'))}$ (e.g., those appear in appendix \ref{app:4DgHatEinProf}) in cosmic time $t'$ by numerically inverting the integral that defines $t'(t)$.

\subsection*{\textit{4. Homogeneous, anisotropic, and FLRW cosmologies}}

Below \eqref{eq:4DmetricAnsatz}, we have discussed the spatial scale factors $e^{6\beta(t(t'))+\alpha_i(t(t'))}$ of the 4D Einstein frame metric in cosmic time $t'$. In our constructions, generically the 4D Einstein frame metric is a Bianchi type I universe, i.e., spatially homogeneous yet anisotropic (for references, see \cite{ToledoSesma:2015wxj,Verma:2024lex}). However, for flux configurations A and B, our constructions allow all the spatial scale factor exponents $\alpha_i$ to be equal, so the Einstein frame metric takes a spatially flat FLRW form in those cases, as argued in section \ref{subsec:ExplicitSols} along with providing the corresponding choices of constant parameters. In contrast, for flux configuration C, only two of the spatial scale factors can be made equal. In all our constructions the Ricci scalar of the 4D Einstein frame metric is negative, $\h R^E<0$.\footnote{In the cases we obtain a 4D FLRW universe ($\alpha_i=\alpha$), one can check that the second derivative of the 3D spatial scale factor w.r.t. the cosmic time $t'$, computed as $e^{-3\beta-\sigma/2}\frac{d}{dt} [e^{-3\beta-\sigma/2}\frac{d}{dt} e^{3\beta+\alpha/2}]$, is negative.} Due to this, our backgrounds could be viable candidates for a kination epoch in the early universe. The latter also requires moduli fields in the 4D EFT whose kinetic energy dominates the energy density. Recently, \cite{Apers:2024dtn,Apers:2022cyl} studied perturbations of 10D Kasner backgrounds, relating them to rolling moduli in the 4D kination epoch by compactifying on a 6D torus. A study of the 4D EFT of the geometric moduli of the 6D internal metric of our constructed backgrounds could therefore be relevant in this context.

\subsection*{\textit{5. Time-dependent warped flux backgrounds}}

Here, by warped we refer to a nonzero $A(y)$ as in \eqref{eq:10DMetricDoubleWarp}, while we continue to refer to $\beta(x)$ as the 6D scale factor exponent. In this work, all the constructions work with $A(y)=0$. However, our quest is to construct viable time-dependent 4D spacetimes within the framework of 10D supergravity theory, regardless of other ingredients like $A(y)$. One time-dependent background with $A(y)\neq 0$, but $\beta(x)=0$, arises in the GKP setup \cite{Giddings:2001yu}. As also pointed out in \cite{Frey:2024jqy}, any 4D Ricci-flat metric (with conformally Ricci-flat internal manifold) solves type IIB supergravity equations in presence of D-branes and O-planes, and self-dual $\t F_5$ featuring $[4,1]$ and $[0,5]$ components and imaginary self-dual $G_3$ featuring harmonic $[0,3]$ components. This can be checked straightforwardly. Thus, for example, a 4D Kasner metric is a viable time-dependent 4D metric as it is Ricci-flat. Note that the GKP fluxes are of MN-type, which we did not consider in the current work. It would be important to solve the supergravity equations while keeping both $A(y),\beta(x)$ nontrivial, allowing both MN and non-MN type fluxes along with local sources, and searching for allowed time-dependent 4D spacetimes. We leave such a detailed scan for future work. Obtaining a time-dependent background that approaches a GKP background with a warped 4D Minkowski component at late times will be important, as such backgrounds are of particular interest in particle physics phenomenology (for reviews, see \cite{Agrawal:2022rqd,Maharana:2012tu,Cicoli:2023opf}). Moreover, effects of quantum corrections including higher-order curvature corrections (e.g., \cite{Dasgupta:2014pma, Dasgupta:2018rtp, Dasgupta:2019vjn}) on the allowed 4D spacetimes $\h g^E_{\mu\nu}(t)$, regardless of the sign of $\h R^E(t)$, will be important to study for early universe cosmologies.

\subsection*{\textit{6. Positive Ricci scalar of 4D Einstein frame metric}}

In absence of the 6D scale factor, non-MN types fluxes, and O-planes (which are objects with negative tension), \eqref{eq:MNRhatEInty} rules out $\h R^E\geq 0$ generically across any supergravity theories.\footnote{Note that in absence of the 6D scale factor, $\h R^E=\h R$.} GKP \cite{Giddings:2001yu} by inclusion of O-planes constructs 4D spacetimes with $\h R^E=0$ in type IIB supergravity. Our generalizations of the Maldacena--Nu\~nez analysis incorporating the 6D scale factor, non-MN types fluxes, do not rule out the existence of a positive $\h R^E$ solely by considering the traced Einstein equations (see, \eqref{eq:MNRhatEInty} and \eqref{eq:MNRtildeInty}). Even the contribution of the 6D scale factor to $\h R^E$ is noteworthy by itself. There are two approaches one can take to attempt drawing conclusive remarks: (i) by imposing energy conditions on top of the traced Einstein equations, along the lines of \cite{Russo:2019fnk, Das:2019vnx, Bernardo:2021zxo, Bernardo:2022ony, Faruk:2024usu}, or (ii) by conducting a comprehensive search in full generality, in which all components of the field strengths and both warp factors $A(y)$ and $\beta(x)$ are retained, and exploring all supergravity theories (even with branes included, as \eqref{eq:MNRhatEInty}, \eqref{eq:MNRtildeInty} hold for any contributions to the energy-momentum tensor). However, note that energy conditions are not absolute principles; there are known violations of all pointwise energy conditions by physically reasonable fields \cite{Kontou:2020bta, Curiel:2014zba, Visser:1999de, Barcelo:1999hq, Barcelo:2000zf}.

Notably, solution with a 4D component with $\h R^E>0$ to higher dimensional vacuum Einstein equations (i.e., with zero fluxes) was reported earlier, considering non-trivial time-dependent $\beta(t)$ and $A(y)=0$, however a compact hyperbolic internal space was taken \cite{Townsend:2003fx}. A negative scalar curvature of the internal space contributes positively to $\h R^E$, see \eqref{eq:10DRicciScalarDoubleWarp}. Our \eqref{eq:MNRhatEInty} does not directly see $\t R$, as in original Maldacena--Nu\~nez analysis. However, \eqref{eq:MNRtildeInty} does see it and feeds back into $\beta(x)$, which in turn affects \eqref{eq:MNRhatEInty}. It will be interesting to study if hyperbolic internal spaces are necessary in constructions with $\h R^E>0$ in presence of fluxes.

\subsection*{\textit{7. Completing the tabulation of solvable subcases}}

The constructions in section \ref{sec:TDepBackgIIBFlux} rely on the metric and flux ansatz \eqref{eq:CommonAnsatz}. Note that once we fix the internal components of the form fields in terms of known forms on a 6D Ricci-flat internal manifold, we are left with: (i) constraints on the noncompact components of the form fields and on the axiodilaton $\tau(x)$\footnote{Throughout this work, we ignore $y$-dependence of $\tau$.}, (ii) the noncompact components of the 10D Einstein field equations, and (iii) a compatibility condition on $\h R$ coming from the internal components of the 10D Einstein field equations. If we allow $n$ independent components in $\h g_{\mu\nu}$ (including any off-diagonal components), then (ii) yields the same number of independent equations. Together with (iii), this results in $n+1$ independent equations for $n+1$ unknown functions, namely the independent metric components and the scale-factor exponent $\beta$. (i) arise from the equations of motion and Bianchi identities for the form fields, as well as from the equation of motion for the axiodilaton, and these constraints also involve the metric components and $\beta$. Therefore, in order to avoid an overdetermined system, the constraints in (i) must be satisfied by appropriately choosing the independent noncompact components of the form fields and $\tau(x)$, without imposing additional restrictions on the metric components or on $\beta$.

\textbullet\ Now we discuss flux combinations within this specific ansatz \eqref{eq:CommonAnsatz} for which the supergravity equations become an overdetermined system:

As a subcase of \eqref{eq:CommonAnsatz}, turning off $\t F_5$ and keeping $\tau$ constant, while allowing only the $[1,2]$ components of $G_3$, the equation of motion for $\tau$ requires $\zeta_1$ to be null, i.e., $\zeta_\mu \zeta^{\h\mu} = 0$. As a result, we must allow both temporal and at least one spatial component for $\zeta_1$. Due to (i), not only are these components fixed, but also one independent component of $\h g_{\mu\nu}$ becomes determined in terms of the other independent metric components and $\beta$.\footnote{Note that we need to allow off-diagonal terms in $\h g_{\mu\nu}$ in this case, due to the $\zeta_\mu \zeta_\nu$ contribution to (ii).} Thus, (ii) and (iii) form an overdetermined system of equations.

Similarly, turning off $\t F_5$ and keeping $\tau$ constant, while allowing only the $[3,0]$ components of $G_3$, or both the $[1,2]$ and $[3,0]$ components, also leads to an overdetermined system of equations. In the former case, this arises from the nullness condition on $\vartheta_1$, while in the latter case this arises from a relation between $\zeta_\mu \zeta^{\h\mu}$ and $\vartheta_\mu \vartheta^{\h\mu}$, both originating from the equation of motion for $\tau$.

As a subcase of \eqref{eq:CommonAnsatz}, keeping $\tau$ constant, while turning on the $\t F_5$ and allowing only the $[1,2]$ or the $[3,0]$ components of $G_3$, or both the $[1,2]$ and $[3,0]$ components, leads to an overdetermined system of equations. This arises, respectively, from the nullness condition on $\zeta_1$, on $\vartheta_1$, and from a relation between $\zeta_\mu \zeta^{\h\mu}$ and $\vartheta_\mu \vartheta^{\h\mu}$, all originating from the equation of motion for $\tau$.

Now allowing time-dependence in $\tau$, while keeping the $\t F_5$ and allowing only the $[1,2]$ components of $G_3$ leads to an overdetermined system of equations. This arises from a constraint of vanishing of $\gamma_1\wedge\zeta_1$, on top of differential constraints on $\lambda_1$ and $\zeta_1$ individually.

\textbullet\ Now we discuss flux combinations within the ansatz \eqref{eq:CommonAnsatz} for which the supergravity equations become a determined system, however not illustrated in section \ref{subsec:SubCaseGenSolv}:

Turning off $\t F_5$, while allowing time-dependence in $\tau$, and including only the $[3,0]$ components of $G_3$, or both the $[1,2]$ and $[3,0]$ components, leads to a system in which the number of equations matches the number of unknowns. We have illustrated only the example with the $[1,2]$ components of $G_3$ in section \ref{sssec:ZeroF5TauT}.

Now turning on $\t F_5$, allowing time-dependence in $\tau$, and including both the $[1,2]$ and $[3,0]$ components of $G_3$, leads to a system in which the number of equations matches the number of unknowns. We have illustrated only the example with the $[3,0]$ components of $G_3$ in section \ref{sssec:SDF5NZG3TauT}.

\subsection*{\textit{8. Miscellaneous remarks}}

In the time-dependent backgrounds in \eqref{eq:0G3F5TauCEx} constructed with nontrivial self-dual $\tilde{F}_5$ featuring $[3,2]$ and $[1,4]$ components and vanishing $G_3$, the axiodilaton remains an unfixed constant.\footnote{The counting in the preceding discussion shows that this is the only case where $\tau$ can be kept constant. Moreover, the supergravity equations do not fix its value. This is the only case in which $\textrm{Im}\tau$ cannot be traded with $\beta(t)$ as the undetermined function.} Therefore, in this backgrounds the string coupling ($\textrm{Im}\tau=g_s^{-1}$) can be tuned to arbitrarily small values parametrically. This may be of particular importance for a worldsheet description of such backgrounds, as recently demonstrated by \cite{Cho:2023mhw} for warped toroidal flux compactification for the type IIB flux configurations that allow arbitrarily weak string coupling.

The 4D effective energy-momentum tensor that sources the 4D Einstein frame metric is given in \eqref{eq:4DEffEMTensorE}, written most generally to cover all our constructions. Note that it contains terms proportional to the metric $\h g^E_{\mu\nu}$ (such as the dynamical dark energy density). In addition to these terms, the $\vartheta_\mu\vartheta_\nu$ term contributes to the spatial components of the energy-momentum tensor. All other terms contribute strictly to the temporal component, i.e., the energy density. Also note that the metric dependence of the 4D effective energy-momentum tensor is such that it remains invariant under Weyl rescaling. However, the Einstein tensor transforms nontrivially under Weyl rescaling, contributing new terms to the effective energy-momentum tensor that sources the rescaled metric. Owing to this, our checks for energy conditions in 4D have been conducted directly on the Einstein frame energy-momentum tensor \eqref{eq:4DEffEMTensorE}.

Lastly, we would like to point out that it would be interesting to allow for off-diagonal components (i.e., $\mu m$ terms) in the 10D metric \eqref{eq:CommonAnsatz} in order to obtain a more general class of type IIB backgrounds. As long as the $x$- and $y$-dependences of such components factorize, integrating over the $y$-coordinates and studying the resulting 4D effective theory should not involve further conceptual subtleties. Moreover, if such components depend only on $t$, the 4D spacetime component may still admit FLRW isometries in some cases. The analysis of section \ref{sec:MNextension} could also be extended to incorporate such off-diagonal terms in \eqref{eq:10DMetricDoubleWarp}.

\acknowledgments

We thank Andrew R. Frey for reading an earlier version of the manuscript and for providing valuable comments. We also thank Evan McDonough and Keshav Dasgupta for interesting comments and correspondence on the earlier version. ARK is supported by the Czech Science Foundation GA\v{C}R grant "Dualities and higher derivatives" (GA23-06498S) and thanks the CERN theory group and Irene Valenzuela for the hospitality at CERN. RM is supported by the National Natural Science Foundation of China (NSFC) under Grants No. 12247103.

\appendix

\section{Conventions}
\label{app:Conven}

This appendix collects all the conventions used through out the paper. For the 10D (hereby 4D) spacetimes we use mostly plus signature: $(-+++\cdots)$. $x^\mu$ refer to the 4D spacetime coordinates, $y^m$ refer to the 6D internal space coordinates, while $X^M$ refer to the 10D spacetime coordinates i.e. $\{x^\mu,y^m\}$ collectively. Lowercase Greek indices $\mu,\nu,\dots$ run over the 4D spacetime coordinates, lowercase Latin indices $m,n,\dots$ run over the 6D internal coordinates, while uppercase Latin indices $M,N,\dots$ run over the 10D spacetime coordinates. On covariant derivatives or raised indices, hat indicates the use of the 4D spacetime metric $\h g_{\mu\nu}$ (or its inverse $\h g^{\mu\nu}$ as appropriate), while tilde denotes the use of the 6D internal space metric $\t g_{mn}$ (or its inverse $\t g^{mn}$ as appropriate). With these understandings, the d'Alembertian operators are denoted as:
\begin{align}
  \h\nabla^{\h 2}\equiv\h\nabla^{\h\mu}\h\nabla_{\mu}\,,\quad \t\nabla^{\t 2}\equiv\t\nabla^{\t m}\t\nabla_{m}\,.
\end{align}
Notably, there is one exception: when dealing with energy conditions in 4D in section \ref{subsec:EnergyConds4D}, a hat on raised indices refers to the contraction performed using the 4D Einstein frame metric $\hat g^E_{\mu\nu}$, not $\h g_{\mu\nu}$.

A $[i,j]$-form refers to a differential form in 10D with $i$ legs on 4D noncompact spacetime and $j$ legs on 6D internal space. The volume forms in 4D and 6D are defined as:
\begin{align}
  \h\epsilon\equiv \frac{1}{4!}\sqrt{-\h g}\epsilon_{\mu_1\cdots\mu_4}dx^{\mu_1}\wedge\cdots\wedge dx^{\mu_4}\,,\quad \t\epsilon\equiv \frac{1}{6!}\sqrt{\t g}\epsilon_{m_1\cdots m_6}dy^{m_1}\wedge\cdots\wedge dy^{m_6}\,,
\end{align}
using respective Levi-Civita symbols (with $\epsilon_{0123}=+1$) and metric determinants.

When writing the components of forms, we omit the their ranks, as is evident from the index counting. For example, the components of a $[0,2]$ form $\eta_2$ are written as $\eta_{mn}$.

For the Hodge dual of a $p$-form in $n$-dimensional space(time) with metric $g_{kl}$, we adopt the following definition:
\begin{align}
  \left(\star A_p\right)_{l_1\dots l_{n-p}}\equiv \frac{1}{p!}\sqrt{|g|}\epsilon_{l_1\cdots l_{n-p}k_1\cdots k_p}A^{k_1\cdots k_p}\,,
\end{align}
where the $n$-dimensional Levi-Civita symbol and the metric determinant $g$ are used, and indices are raised by the inverse metric $g^{kl}$. In the main text, the $\h\star$-operation denotes the use of the 4D metric $\h g_{\mu\nu}$, while the $\t\star$-operation denotes the use of the 6D metric $\t g_{mn}$.

Our definition of the dot product between two $p$-forms in $n$-dimensional space(time) includes a factor of $1/p!$:
\begin{align}
  A_p\cdot B_p\equiv \frac{1}{p!}A_{k_1\cdots k_p}B^{k_1\dots k_p}\,.
\end{align}
All the dot products that appear in the main text are between forms on the internal space and thus use the 6D inverse metric $\t g^{mn}$. To avoid confusion, contractions between forms on the 4D noncompact spacetime are written explicitly, for example, $\gamma_\mu\gamma^{\h\mu},\gamma_\mu\vartheta^{\h\mu}$, etc. Only in appendix \ref{app:SUGRAEOMs} do the dot products exclusively between forms on 10D spacetime appear, and they use the inverse metric $g^{MN}$.

Symmetrization and antisymmetrization of $p$ indices, respectively denoted by the first and second brackets, include a factor of $1/p!$. Lastly, in our explicit constructions, a dot on functions of time denotes their time derivative, e.g., $\dot\beta$.

\section{Type IIB supergravity EOMs}
\label{app:SUGRAEOMs}

In this appendix, we briefly collect the bosonic sector of the type IIB supergravity action in 10D Einstein frame, along with the corresponding equations of motion which are central to this paper. This sector contains the NS--NS fields $g_{MN}, B_{MN}, \Phi$, which denote the metric (with mostly plus signature), the Kalb--Ramond field, and the dilaton, respectively, as well as the R--R form fields $C_0$, $C_{MN}$, and $C_{MNPQ}$. For details, see \cite{Polchinski:1998rr,Johnson:2000ch}. We have:
\begin{align}
  \textrm{\it\color{NavyBlue} Action:}\quad S_{\textrm{IIB}}=&\frac1{2\kappa_{10}^2} \bigg[ \int d^{10}X \sqrt{-g}\ R_g+\frac12\int d\Phi\wedge\star d\Phi + \frac12\int e^{-\Phi} H_3\wedge \star H_3 \nn\\
  &+\frac1{2}\int e^{2\Phi} F_1\wedge \star F_1 +\frac12\int e^{\Phi} \tilde{F}_3\wedge \star \tilde{F}_3+ \frac{1}4\int \tilde{F}_5\wedge \star \tilde{F}_5\bigg] \nn\\
  &+\frac{1}{4\kappa_{10}^2}\int C_4\wedge H_3\wedge F_3\,. \label{eq:10DIIBSUGRAAction}
\end{align}
\begin{align}
  \textrm{\it\color{NavyBlue} Field strengths:}\quad &H_3=dB_2\,,\quad F_1=dC_0\,,\quad F_3=dC_2\,,\quad F_5=dC_4\,,\quad \tilde{F}_3=F_3-C_0 H_3\,, \nn\\
  &\tilde{F}_5=F_5-\frac12 C_2\wedge H_3+\frac12 B_2\wedge F_3\,.
\end{align}
\begin{align}
  \textrm{\it\color{NavyBlue} Bianchi identities:}\quad dF_1=0\,,\quad dH_3=0\,,\quad d\tilde{F}_3=H_3\wedge F_1\,,\quad d\tilde{F}_5=H_3\wedge F_3\,.
\end{align}
\begin{align}
  \textrm{\it\color{NavyBlue} EOMs for form fields:}\quad &d(\star e^{-\Phi} H_3)-F^{(1)}\wedge\star e^{\Phi}\tilde{F}_3+\tilde{F}_5\wedge \tilde{F}_3=0\,,\quad d(\star e^{\Phi}\tilde{F}_3)+H_3\wedge \tilde{F}_5=0\,, \nn\\
  &d\star d\Phi+\frac12 e^{-\Phi} H_3\wedge \star H_3-e^{2\Phi}F_1\wedge \star F_1 -\frac12 e^{\Phi} \tilde{F}_3\wedge \star \tilde{F}_3=0\,, \nn\\
  &d(\star e^{2\Phi}F_1)+H_3\wedge \star e^{\Phi}\tilde{F}_3=0\,.
\end{align}
\begin{align}
   \textrm{\it\color{NavyBlue} Self-duality:}\quad \star\tilde{F}_5=\tilde{F}_5\,.
\end{align}
The 10D self-duality constraint must be imposed when solving the equations of motion, as it arises from string theory\footnote{And, is also required for the closure of the 10D SUSY algebra \cite{Schwarz:1983qr}.}. This has already been used to simplify the equations of motion for the form fields shown above, while the equation of motion for the $C_4$ potential reduces to the Bianchi identity for $\tilde{F}_5$ and is hence not shown separately. These system of 10D equations of motion and Bianchi identities can be rewritten in terms of axio-dilaton $\tau$ and the 3-form flux $G_3$, as follows.
\begin{align}
  &\tau\equiv C_0+i e^{-\Phi}\,,\quad G_3\equiv F_3-\tau H_3=\tilde{F}_3-i e^{-\Phi}H_3\,.
\end{align}
\begin{align}
  \textrm{\it\color{NavyBlue} EOMs rewritten:}\quad &d\tilde{F}_5=\frac i2 \frac{G_3\wedge\bar{G}_3}{\textrm{Im}\tau}\,,\quad d\star G_3=-i \frac{d\tau\wedge\star \textrm{Re}G_3}{\textrm{Im}\tau}+i\tilde{F}_5\wedge G_3\,, \nn\\
  &d\star d\tau=-i\frac{d\tau\wedge\star d\tau}{\textrm{Im}\tau} -\frac{i}2 G_3\wedge\star G_3\,,\quad dG_3=d\tau\wedge\frac{G_3-\bar{G}_3}{2i\textrm{Im}\tau}\,. \label{eq:SUGRAEOMForms}
\end{align}
Now, we list the final set of 10D equations:
\begin{align}
  &\textrm{\it\color{NavyBlue} Einstein field equations:}\quad G_{MN}=T^1_{MN}+T^3_{MN}+T^5_{MN}\,,\quad G_{MN}\equiv R_{MN}-\frac12 g_{MN} R_g\,, \nn\\
  &\quad\quad\quad T^1_{MN}\equiv \frac1{2(\textrm{Im}\tau)^2}\partial_{(M}\tau\partial_{N)}\bar{\tau}-\frac1{4(\textrm{Im}\tau)^2} g_{MN}\ d\tau\cdot d\bar{\tau}\,,\nn\\
  &\quad\quad\quad T^3_{MN}\equiv \frac1{2\cdot 2!\textrm{Im}\tau} G{}_{(M}{}^{PQ} \bar{G}{}_{N)PQ} - \frac1{4\textrm{Im}\tau} g_{MN} G_3\cdot\bar{G}_3\,, \nn\\
  &\quad\quad\quad T^5_{MN}\equiv \frac{1}{4\cdot 4!} \tilde{F}{}_{MPQRS} \tilde{F}{{}_N}^{PQRS} -\frac{1}{8} g_{MN} \tilde{F}_5\cdot\tilde{F}_5=\frac{1}{4\cdot 4!} \tilde{F}{}_{MPQRS} \tilde{F}{{}_N}^{PQRS}\,, \label{eq:SUGRAEinsteinEq}
\end{align}
where the last equality uses the self-duality of $\tilde{F}_5$. The dot products appeared above use the 10D inverse metric $g^{MN}$, see appendix \ref{app:Conven}.

\section{Detailed terms in the EOMs}
\label{app:EOMTermDetail}

The metric and form field ansatz is given in \eqref{eq:CommonAnsatz} in the main text. In this appendix, we present in detail the resulting terms in the equations of motion and the Bianchi identities \eqref{eq:SUGRAEOMForms}--\eqref{eq:SUGRAEinsteinEq}. The terms are grouped according to their $[i,j]$ components, as indicated by the underbraces. Here, any dot products appearing in the final expressions are taken w.r.t. the 6D internal metric $\t g_{mn}$.

The terms that appear in the Bianchi identity of $\tilde F_5$ read as:
\begin{align}
  d\t F_5=\underbrace{\h d\h\star \gamma_1\wedge \lambda_2}_{[4,2]} \underbrace{-\h\star \gamma_1\wedge \t d\lambda_2}_{[3,3]} \underbrace{+\h d \left(e^{2\beta}\gamma_1\right)\wedge\t\star\lambda_2}_{[2,4]} \underbrace{-e^{2\beta}\gamma_1\wedge\t d\t\star\lambda_2}_{[1,5]}\,.
\end{align}
\begin{align}
  \frac i2 \frac{G_3\wedge\bar{G}_3}{\textrm{Im}\tau}=\underbrace{\frac{i}{2\textrm{Im}\tau} \zeta_1\wedge\h\star\vartheta_1\wedge \left(\bar{q}\eta_2 - q\bar\eta_2\right)}_{[4,2]}\,.
\end{align}

The terms that appear in the $G_3$ equation of motion read as:
\begin{align}
  d\star G_3=\underbrace{\h d\left(e^{2\beta}\h\star\zeta_1\right)\wedge \t\star\eta_2}_{[4,4]} \underbrace{- e^{2\beta}\h\star\zeta_1\wedge\t d\t\star\eta_2}_{[3,5]} \underbrace{+q\h d\left(e^{6\beta}\vartheta_1\right)\wedge\t\epsilon}_{[2,6]}\,.
\end{align}
\begin{align}
  i\t F_5\wedge G_3=\underbrace{i\h\star \gamma_1\wedge\zeta_1\wedge \lambda_2\wedge\eta_2 +i q e^{2\beta}\gamma_1\wedge\h\star\vartheta_1\wedge\t\star\lambda_2}_{[4,4]} \underbrace{+ie^{2\beta}\gamma_1\wedge\zeta_1\wedge\t\star\lambda_2\wedge\eta_2}_{[2,6]}\,.
\end{align}
\begin{align}
  -i \frac{d\tau\wedge\star \textrm{Re}G_3}{\textrm{Im}\tau} =\underbrace{-\frac{i}{\textrm{Im}\tau} e^{2\beta}\h d\tau\wedge\h\star\zeta_1\wedge \textrm{Re}\t\star\eta_2}_{[4,4]}  \underbrace{+\frac{i}{\textrm{Im}\tau}e^{2\beta} \h\star\zeta_1\wedge\t d\tau\wedge \textrm{Re}\t\star\eta_2}_{[3,5]} \underbrace{-\frac{i\textrm{Re}q}{\textrm{Im}\tau} e^{6\beta}\h d\tau\wedge\vartheta_1\wedge\t\epsilon}_{[2,6]}\,.
\end{align}

The terms that appear in the Bianchi identity of $G_3$ read as:
\begin{align}
  dG_3= \underbrace{q\h d\h\star\vartheta_1}_{[4,0]} \underbrace{+\h d\zeta_1\wedge\eta_2}_{[2,2]} \underbrace{-\zeta_1\wedge\t d\eta_2}_{[1,3]}\,.
\end{align}
\begin{align}
  d\tau\wedge\frac{\textrm{Im}G_3}{\textrm{Im}\tau}= \underbrace{\frac{\textrm{Im}q}{\textrm{Im}\tau}\h d\tau\wedge\h\star\vartheta_1}_{[4,0]} \underbrace{-\frac{\textrm{Im}q}{\textrm{Im}\tau}\h\star\vartheta_1\wedge\t d\tau}_{[3,1]} \underbrace{+\h d\tau\wedge\zeta_1\wedge\frac{\textrm{Im}\eta_2}{\textrm{Im}\tau}}_{[2,2]} \underbrace{-\zeta_1\wedge\t d\tau\wedge\frac{\textrm{Im}\eta_2}{\textrm{Im}\tau}}_{[1,3]}\,.
\end{align}

The terms (of type $[4,6]$) that appear in the $\tau$ equation of motion read as:
\begin{align}
  d\star d\tau=\h d\left(e^{6\beta}\h\star\h d\tau\right)\wedge \t\epsilon + e^{4\beta}\h\epsilon\wedge\t d\t\star\t d\tau\,.
\end{align}
\begin{align}
  \frac{-id\tau\wedge\star d\tau}{\textrm{Im}\tau}=-\frac{i}{\textrm{Im}\tau}\left(e^{6\beta}\h d\tau\wedge\h\star\h d\tau\wedge\t\epsilon + e^{4\beta}\h\epsilon\wedge\t d\tau\wedge\t\star\t d\tau\right)\,.
\end{align}
\begin{align}
  -\frac{i}{2}G_3\wedge\star G_3 = -\frac{i}{2}e^{2\beta}\zeta_1\wedge\h\star\zeta_1\wedge \eta_2\wedge\t\star\eta_2 +\frac{iq^2}{2} e^{6\beta}\vartheta_1\wedge\h\star\vartheta_1\wedge\t\epsilon\,.
\end{align}

The components of the energy-momentum tensor read as follows. The contributions from the 5-from flux are:
\begin{align}
  T^5_{\mu\nu}&=\frac{1}{4}e^{-4\beta}\lambda_2\cdot\lambda_2\left[2\gamma_\mu\gamma_\nu-\h g_{\mu\nu}\gamma_\sigma\gamma^{\h\sigma}\right]\,.
\end{align}
\begin{align}
  T^5_{mn}&=\frac{1}{4}e^{-2\beta}\gamma_\mu\gamma^{\h\mu}\left[\t g_{mn}\lambda_2\cdot\lambda_2-2\lambda_m{}^{\t p}\lambda_{np}\right]\,.
\end{align}
\begin{align}
  T^5_{\mu m}=0\,.
\end{align}
The contributions from the 3-form flux are:
\begin{align}
  T^3_{\mu\nu}=\frac{1}{4\textrm{Im}\tau}e^{-4\beta}\eta_2\cdot\bar\eta_2 \left[2\zeta_{\mu}\zeta_{\nu} - \h g_{\mu\nu} \zeta_\sigma\zeta^{\h\sigma}\right] +\frac{q\bar q}{4\textrm{Im}\tau} \left[2\vartheta_{\mu}\vartheta_{\nu} - \h g_{\mu\nu} \vartheta_\sigma\vartheta^{\h\sigma}\right]\,.
\end{align}
\begin{align}
  T^3_{mn}=\frac1{4\textrm{Im}\tau} e^{-2\beta} \zeta_\mu\zeta^{\h\mu} \left[2\eta{}_{(m}{}^{\t p} \bar\eta{}_{n)p} - \t g_{mn} \eta_2\cdot\bar\eta_2\right] +\frac{q\bar q}{4\textrm{Im}\tau}e^{2\beta}\vartheta_\sigma\vartheta^{\h\sigma}\t g_{mn}\,.
\end{align}
\begin{align}
  T^3_{\mu m}=0\,.
\end{align}
The contributions from the axiodilaton are:
\begin{align}
  T^1_{\mu\nu}=\frac1{2(\textrm{Im}\tau)^2}\partial_{(\mu}\tau\partial_{\nu)}\bar{\tau}-\frac1{4(\textrm{Im}\tau)^2} \h g_{\mu\nu}\partial_\sigma\tau\partial^{\h \sigma}\bar\tau -\frac1{4(\textrm{Im}\tau)^2} \h g_{\mu\nu} e^{-2\beta}\partial_m\tau\partial^{\t m}\bar\tau\,.
\end{align}
\begin{align}
  T^1_{mn}=\frac1{2(\textrm{Im}\tau)^2}\partial_{(m}\tau\partial_{n)}\bar{\tau}-\frac1{4(\textrm{Im}\tau)^2} \t g_{mn} \partial_l\tau\partial^{\t l}\bar\tau -\frac1{4(\textrm{Im}\tau)^2} \t g_{mn} e^{2\beta}\partial_\mu\tau\partial^{\h \mu}\bar\tau\,.
\end{align}
\begin{align}
  T^1_{\mu m}=\frac1{2(\textrm{Im}\tau)^2}\partial_{(\mu}\tau\partial_{m)}\bar{\tau}\,.
\end{align}

We now present the components of the Einstein tensor resulting from the 10D metric ansatz \eqref{eq:10DMetricDoubleWarp}. Setting $A(y)=0$ yields the components corresponding to the metric ansatz in \eqref{eq:CommonAnsatz}.
\begin{align}
  &G_{\mu\nu}=\hat{G}_{\mu \nu}-6\left(\hat{\nabla}_\mu \hat{\nabla}_\nu \beta+\partial_\mu \beta \partial_\nu \beta\right) +3\h g_{\mu\nu}\left(2\hat{\nabla}^{\h 2} \beta+7\partial_\sigma \beta \partial^{\h\sigma} \beta\right) \nn\\
  &\qquad\qquad +3\h g_{\mu\nu} e^{2A-2\beta}\left(\tilde{\nabla}^{\t 2} A+ 2\partial_m A \partial^{\t m} A\right) -\frac{1}{2}\h g_{\mu\nu}e^{2A-2\beta}\tilde{R}\,. \\
  &G_{mn}=\t G_{mn}-4 \left(\tilde{\nabla}_m \t \nabla_n A+\partial_m A \partial_n A\right) +2\t g_{mn}\left(2\tilde{\nabla}^{\t 2} A+5 \partial_l A \partial^{\t l} A\right) \nn\\
  &\qquad\qquad + 5e^{2\beta-2A}\t g_{mn}\left(\hat{\nabla}^{\h 2} \beta+3\partial_\mu \beta \partial^{\h\mu} \beta\right)-\frac{1}{2}\t g_{mn}e^{2\beta-2A}\hat{R}\,. \\
  &G_{\mu m}= 8\partial_\mu \beta \partial_m A\,.
\end{align}
We also provide the components of the Ricci tensor and the Ricci scalar, as relevant for the analysis in section \ref{sec:MNextension}:
\begin{align}
  &R_{\mu \nu}=\hat{R}_{\mu \nu}-6\left(\hat{\nabla}_\mu \hat{\nabla}_\nu \beta+\partial_\mu \beta \partial_\nu \beta\right)-\h g_{\mu \nu}e^{2A-2\beta}\left(\tilde{\nabla}^{\t 2} A+4 \partial_m A \partial^{\t m} A\right)\,. \\
  &R_{m n}=\tilde{R}_{m n}-4 \left(\tilde{\nabla}_m \t \nabla_n A+\partial_m A \partial_n A\right)-e^{2 \beta-2 A} \tilde{g}_{m n}\left(\hat{\nabla}^{\h 2} \beta+6 \partial_{\mu}\beta \partial^{\h\mu} \beta \right)\,. \\
  &R_{\mu m}=8\partial_\mu \beta \partial_m A\,.
\end{align}
\begin{align}
  R_g=e^{-2A}\hat{R} + e^{-2\beta}\tilde{R} -6e^{-2A}\left(2\hat{\nabla}^{\h 2} \beta+7\partial_\mu \beta \partial^{\h\mu} \beta\right)-4e^{-2\beta}\left(2\tilde{\nabla}^{\t 2} A+5 \partial_m A \partial^{\t m} A\right)\,. \label{eq:10DRicciScalarDoubleWarp}
\end{align}

\section{4D Einstein frame metric profiles}
\label{app:4DgHatEinProf}

In this appendix, we present the profiles of the metrics in the 4D Einstein frame, given by $\h g^{E}_{\mu\nu}=e^{6\beta} \h g_{\mu\nu}$ as in \eqref{eq:4DEinsteinFrameWeyl}. The functional forms of the corresponding scale factors are given in section~\ref{subsec:ExplicitSols}. Here, we consider representative choices for the unfixed functions and constant parameters to generate the resulting metric profiles. As presented below, across some cases the scale factor values match due to the specific choices of parameters; however, by varying the constants in each case, one can generate scale factors that differ among those cases. All the late-time behaviours are given up to positive proportionality constants.

\subsection{\texorpdfstring{$G_3= 0,\ \textrm{self-dual}\ \t F_5,\ \tau=\tau_0$}{z}}
\label{sapp:Zero3formConsT}

The scale factor exponents are given in \eqref{eq:0G3F5TauCEx}. Note that, due to the relation among the integration constants, no pair among $c_3, c_5, c_7$ can simultaneously vanish since $c_2^2 > 0$. For an admissible choice of $\beta$ (i.e., $0 < e^{8\beta} < 8 c_2^2$) such that $e^{8\beta}$ is monotonically decreasing in time, we need to choose $c_3, c_5, c_7 < 0$ to obtain growing scale factors from nonzero initial values.\footnote{Note that $\coth^{-1}[z] > 0$ and monotonically decreasing with $z$ for $z > 1$. See also footnote \ref{fn:AcothProp}.} To ensure a nonzero initial value of $e^{6\beta+\sigma}$, a nonvanishing initial $\dot\beta$ is also required.

\begin{figure}[!ht]
    \centering
    \includegraphics[width=1.05\linewidth]{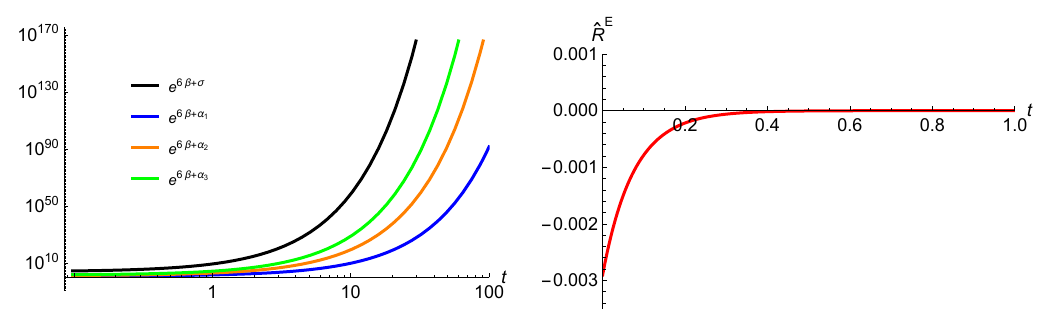}
    \rule{\linewidth}{0.5pt} 
    \includegraphics[width=1.05\linewidth]{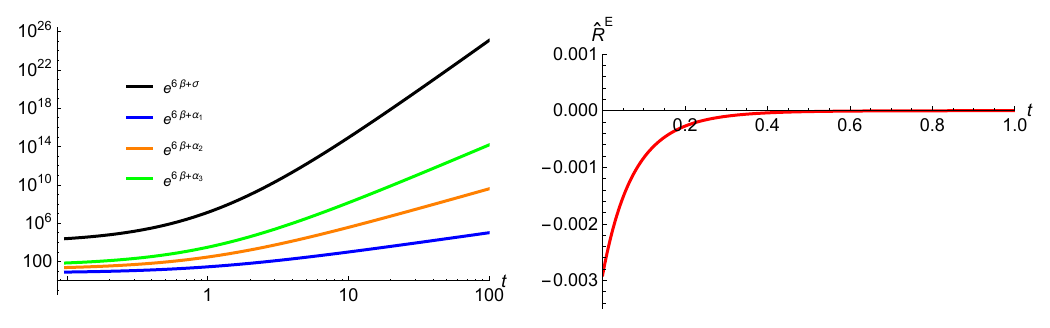}
    \rule{\linewidth}{0.5pt} 
    \caption{Profiles of the time-dependent scale factors and Ricci scalar associated with the 4D Einstein frame metric for the case with vanishing $G_3$, nontrivial self-dual $\tilde{F}_5$, and constant $\tau$. The quantities are plotted against time for the representative choice of parameters specified in the text, with $\beta=-t$ for the top row and $\beta=-\ln[t+1]$ for the bottom row. In the left panels, both axes are shown on a logarithmic scale. The initial values of the plotted quantities are all nonzero, as given in the text.}
    \label{fig:Zero3formConsT}
\end{figure}

Figure \ref{fig:Zero3formConsT} (top row) shows the scale factor profiles for the following consistent choice of the parameters and $\beta(t)$: $k=1,c_1=1,c_2=1,c_4=c_6=c_8=1,c_5=2 c_3,c_7=3 c_3,c_3=-8 \sqrt{\frac{6}{11}},\beta=-t$. The corresponding initial values of the relevant quantities are: $e^{6 \beta +\sigma }\approx 9448.27,\ e^{6 \beta+\alpha _1 }\approx 6.605,\ e^{6 \beta+\alpha _2}\approx 16.049,\ e^{6 \beta+\alpha _3}\approx 38.996,\ \h R^E=-0.003,\ \gamma_t \approx 1.512$. The scale factors grow exponentially with $t$ at late times: $e^{6 \beta +\sigma }\sim e^{24 \sqrt{\frac{3}{11}} t},\ e^{6 \beta +\alpha_1}\sim e^{4 \sqrt{\frac{3}{11}} t},\ e^{6 \beta +\alpha_2}\sim e^{8 \sqrt{\frac{3}{11}} t},\ e^{6 \beta +\alpha_3}\sim e^{12 \sqrt{\frac{3}{11}} t}$.\footnote{Up to proportionality constants.} Meanwhile, $\h R^E$ and $\gamma_t$ decay to zero as: $\h R^E\sim - e^{-24 \sqrt{\frac{3}{11}} t},\ \gamma_t\sim e^{-6 t}$.

With the same choice of parameters, if we instead take $\beta=-\ln [t+1]$,\footnote{More generally, one can choose $\beta=-\ln [t+a]$ with any $a>1/2^{3/8}$, which produces similar late-time profiles but with different initial values.} the scale factors grow as powers of $t$ at late times: $e^{6 \beta +\sigma }\sim t^{24 \sqrt{\frac{3}{11}}-2},\ e^{6 \beta +\alpha_1}\sim t^{4 \sqrt{\frac{3}{11}}},\ e^{6 \beta +\alpha_2}\sim t^{8 \sqrt{\frac{3}{11}}},\ e^{6 \beta +\alpha_3}\sim t^{12 \sqrt{\frac{3}{11}}}$, while starting from the same initial values of all the relevant quantities as given above since the initial values of $\beta,\dot\beta$ coincide with the previous case. In this case, $\h R^E$ and $\gamma_t$ decay to zero as: $\h R^E\sim -t^{-24 \sqrt{\frac{3}{11}}},\ \gamma_t\sim t^{-7}$. See, figure \ref{fig:Zero3formConsT} (bottom row).

\subsection{\texorpdfstring{$G_3= 0,\ \textrm{self-dual}\ \t F_5,\ \tau(t)$}{z}}
\label{sapp:Zero3formTauT}

\textbf{For the class of $\textrm{Re}\tau=\textrm{constant}$}, the scale factor exponents are given in \eqref{eq:0G3F5ImTauTEx}. Note that, due to the relation among the integration constants, no pair among $c_5, c_7,c_9$ can simultaneously vanish since $c_2^2,c_3^2 > 0$. $c_4 > 0$ keeps $\text{Im}\tau > 0$, and by choosing $c_4$ sufficiently large, one can achieve a large value of $\text{Im}\tau$. For an admissible choice of $\beta$ (i.e., $0 < e^{8\beta} < 8 c_2^2$) such that $e^{8\beta}$ is monotonically decreasing in time, we need to choose $c_5, c_7, c_9 < 0$ to obtain growing scale factors from nonzero initial values, while $c_3<0$ for growing $\textrm{Im}\tau$. To ensure a nonzero initial value of $e^{6\beta+\sigma}$, a nonvanishing initial $\dot\beta$ is also required.

\begin{figure}[!ht]
    \centering
    \includegraphics[width=1.05\linewidth]{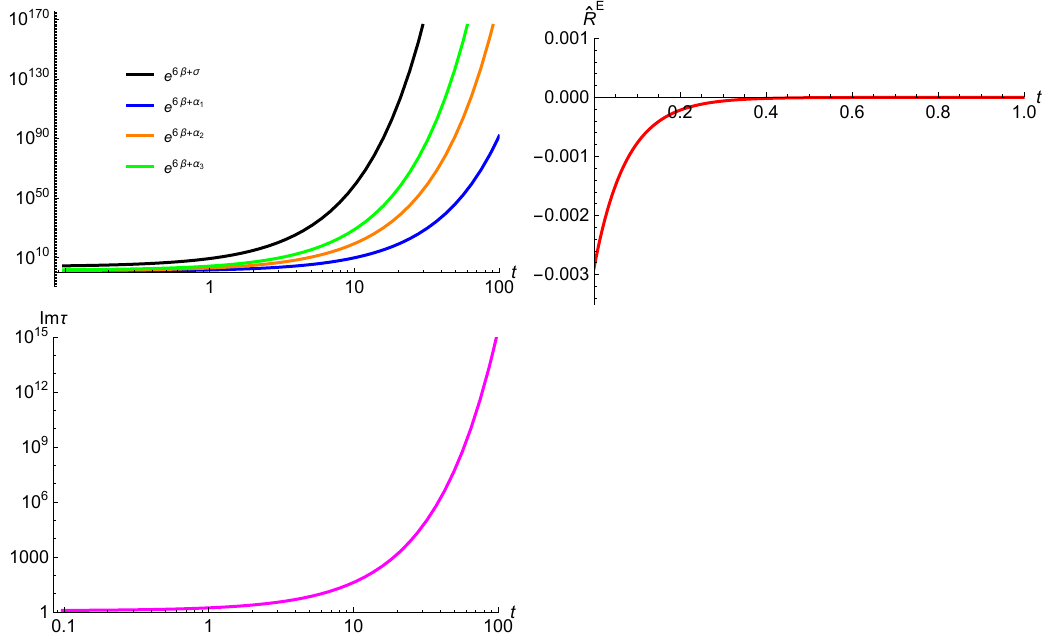}
    \caption{Profiles of the time-dependent scale factors and Ricci scalar associated with the 4D Einstein frame metric, and the imaginary part of axiodilaton, for the case with vanishing $G_3$, nontrivial self-dual $\tilde{F}_5$, and constant $\textrm{Re}\tau$. The quantities are plotted against time for the representative choice of parameters specified in the text, with $\beta=-t$. In the left panels, both axes are shown on a logarithmic scale. The initial values of the plotted quantities are all nonzero, as given in the text.}
    \label{fig:Zero3formImTauT}
\end{figure}

Figure \ref{fig:Zero3formImTauT} shows the profiles of the scale factors and $\textrm{Im}\tau$ for the following consistent choice of the parameters and $\beta(t)$: $k=1,c_1=1,c_2=1,c_3=-1,c_4=1, c_6=c_8=c_{10}=1,c_7=2 c_5,c_9=3 c_5,c_5=-\sqrt{35},\beta=-t$. The corresponding initial values of the relevant quantities are: $e^{6 \beta +\sigma }\approx 9513.99,\ e^{6 \beta+\alpha_1 }\approx 6.613,\ e^{6 \beta+\alpha_2 }\approx 16.086,\ e^{6 \beta+\alpha_3 }\approx 39.131,\ \h R^E\approx -0.003,\ \textrm{Im}\tau\approx 1.162,\ \gamma _t\approx 1.512$. The scale factors and $\textrm{Im}\tau$ grow exponentially with $t$ at late times: $e^{6 \beta +\sigma }\sim e^{3 \sqrt{\frac{35}{2}} t},\ e^{6 \beta +\alpha_1}\sim e^{\frac{1}{4} \sqrt{70} t},\ e^{6 \beta +\alpha_2}\sim e^{\sqrt{\frac{35}{2}} t},\ e^{6 \beta +\alpha_3}\sim e^{\frac{3}{2} \sqrt{\frac{35}{2}} t},\ \textrm{Im}\tau\sim e^{\frac{t}{2 \sqrt{2}}}$. Meanwhile, $\h R^E$ and $\gamma_t$ decay to zero as: $\h R^E\sim - e^{-3 \sqrt{\frac{35}{2}} t},\ \gamma_t\sim e^{-6 t}$.

With the same choice of parameters, if we instead take $\beta=-\ln [t+1]$, the scale factors and $\textrm{Im}\tau$ grow as powers of $t$ at late times: $e^{6 \beta +\sigma }\sim t^{3 \sqrt{\frac{35}{2}}-2},\ e^{6 \beta +\alpha_1}\sim t^{\frac{\sqrt{\frac{35}{2}}}{2}},\ e^{6 \beta +\alpha_2}\sim t^{\sqrt{\frac{35}{2}}},\ e^{6 \beta +\alpha_3}\sim t^{\frac{3 \sqrt{\frac{35}{2}}}{2}},\ \textrm{Im}\tau\sim t^{\frac{1}{2 \sqrt{2}}}$, while starting from the same initial values of all the relevant quantities as given above since the initial values of $\beta,\dot\beta$ coincide with the previous case. In this case, $\h R^E$ and $\gamma_t$ decay to zero as: $\h R^E\sim -t^{-3 \sqrt{\frac{35}{2}}},\ \gamma_t\sim t^{-7}$.

\textbf{For the class of time-dependent $\textrm{Re}\tau,\textrm{Im}\tau$}, the scale factor exponents are given in \eqref{eq:0G3F5ReImTauTEx}. Note that, due to the relation among the integration constants, no pair among $c_7, c_9,c_{11}$ can simultaneously vanish since $c_2^2,c_3^2,c_5^2 > 0$. $c_3 > 0$ keeps $\text{Im}\tau > 0$.\footnote{Note that for real $z$, $0<\textrm{sech}[z]\leq 1$, while $-1<\tanh[z]<1$.} Notably, only $\textrm{Re}\tau$ and $\textrm{Im}\tau$ vary with the sign of $c_5$, while the remaining functions depend solely on its magnitude, since the relation among the constants involve only $c_5^2$. For an admissible choice of $\beta$ (i.e., $0 < e^{8\beta} < 8 c_2^2$) such that $e^{8\beta}$ is monotonically decreasing in time, we need to choose $c_7, c_9 ,c_{11}< 0$ to obtain growing scale factors from nonzero initial values. To ensure a nonzero initial value of $e^{6\beta+\sigma}$, a nonvanishing initial $\dot\beta$ is also required.

\begin{figure}[!ht]
    \centering
    \includegraphics[width=1.05\linewidth]{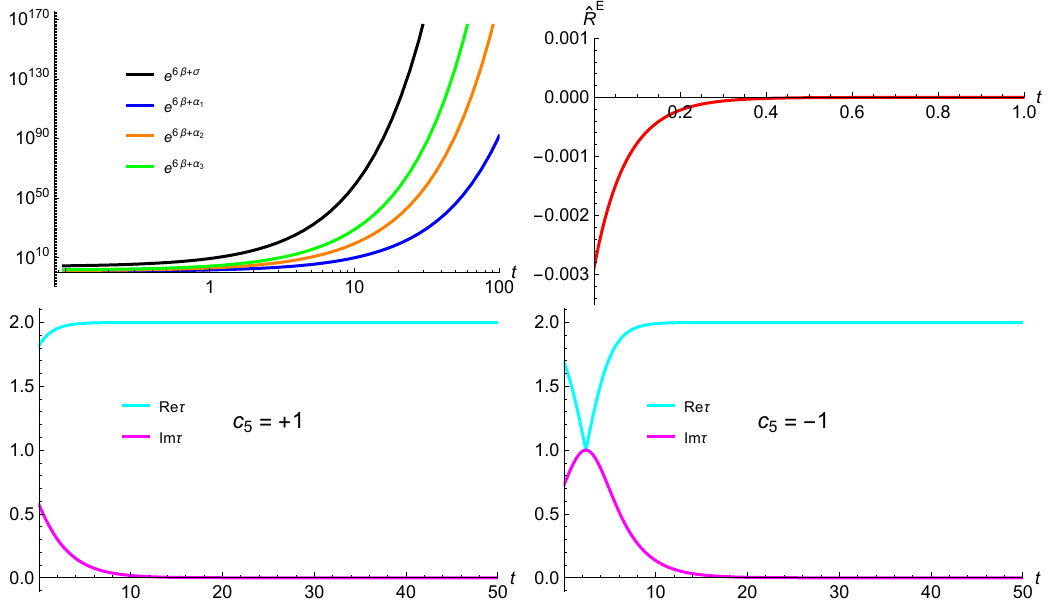}
    \caption{Profiles of the time-dependent scale factors and Ricci scalar associated with the 4D Einstein frame metric, and the real and imaginary parts of axiodilaton, for the case with vanishing $G_3$ and nontrivial self-dual $\tilde{F}_5$. The quantities are plotted against time for the representative choice of parameters specified in the text, with $\beta=-t$. $\textrm{Re}\tau=\textrm{Re}\tau_{+}$ is displayed here only. In the top left panel, both axes are shown on a logarithmic scale. The initial values of the plotted quantities are all nonzero, as given in the text. The two plots in top row do not depend on the sign of $c_5$.}
    \label{fig:Zero3formReImTauT}
\end{figure}

Figure \ref{fig:Zero3formReImTauT} shows the profiles of the scale factors and $\textrm{Re}\tau=\textrm{Re}\tau_{+},\ \textrm{Im}\tau$ for the following consistent choice of the parameters and $\beta(t)$: $k=1,c_1=1,c_2=1,c_3=1,c_4=1,c_5^2=1,c_6=1,c_8=c_{10}=c_{12}=1,c_9=2 c_7,c_{11}=3 c_7,c_7=-\sqrt{35},\beta=-t$. The corresponding initial values of the relevant quantities are: $e^{6 \beta +\sigma }\approx 9513.99,\ e^{6 \beta+\alpha _1 }\approx 6.613,\ e^{6 \beta+\alpha _2 }\approx 16.086,\ e^{6 \beta+\alpha _3 }\approx 39.131,\ \h R^E\approx -0.003,\ \gamma _t\approx 1.512$, while $\textrm{Re}\tau\approx 1.818,\ \textrm{Im}\tau\approx 0.575$ for $c_5 = +1$, and $\textrm{Re}\tau\approx 1.691,\ \textrm{Im}\tau\approx 0.723$ for $c_5=-1$. The scale factors grow exponentially with $t$ at late times: $e^{6 \beta +\sigma }\sim e^{3 \sqrt{\frac{35}{2}} t},\ e^{6 \beta +\alpha_1}\sim e^{\frac{1}{4} \sqrt{70} t},\ e^{6 \beta +\alpha_2}\sim e^{\sqrt{\frac{35}{2}} t},\ e^{6 \beta +\alpha_3}\sim e^{\frac{3}{2} \sqrt{\frac{35}{2}} t}$. Meanwhile, $\textrm{Re}\tau$ approaches a constant value as: $\textrm{Re}\tau\sim 2-2 e^{-\frac{t}{\sqrt{2}}}$, and $\h R^E$, $\gamma_t$ and $\textrm{Im}\tau$ decay to zero as: $\h R^E\sim - e^{-3 \sqrt{\frac{35}{2}} t},\ \gamma_t\sim e^{-6 t},\ \textrm{Im}\tau\sim e^{-\frac{t}{2 \sqrt{2}}}$, for both the cases $c_5=\pm 1$.\footnote{The profile of $\textrm{Re}\tau_{-}$ is not shown, but it exhibits a decay to zero at late times, with $\textrm{Re}\tau_{-}\sim e^{-\frac{t}{\sqrt{2}}}$.}

\subsection{\texorpdfstring{$\t F_5=0,\ G_3,\tau\neq 0,\ \tau(t)$}{z}}
\label{sapp:ZeroF5TauT}

\textbf{For the class of $\textrm{Re}\tau=\textrm{constant}$}, the scale factor exponents are given in \eqref{eq:0F5G3ImTauTEx}. Note that, due to the relation among the integration constants, no pair among $c_4, c_6,c_8$ can simultaneously vanish since $c_3^2 > 0$. $c_2 > 0$ keeps $\text{Im}\tau > 0$, and by choosing $c_2$ sufficiently large, one can achieve a large value of $\text{Im}\tau$. For an admissible choice of $\beta$ (i.e., $0 < e^{16\beta} < 16 c_3^2$) such that $e^{16\beta}$ is monotonically decreasing in time, we need to choose $c_4, c_6, c_8 < 0$ to obtain growing scale factors from nonzero initial values, while $\textrm{Im}\tau$ grows as $e^{16\beta}$ decreases. To ensure a nonzero initial value of $e^{6\beta+\sigma}$, a nonvanishing initial $\dot\beta$ is also required.

\begin{figure}[!ht]
    \centering
    \includegraphics[width=1.05\linewidth]{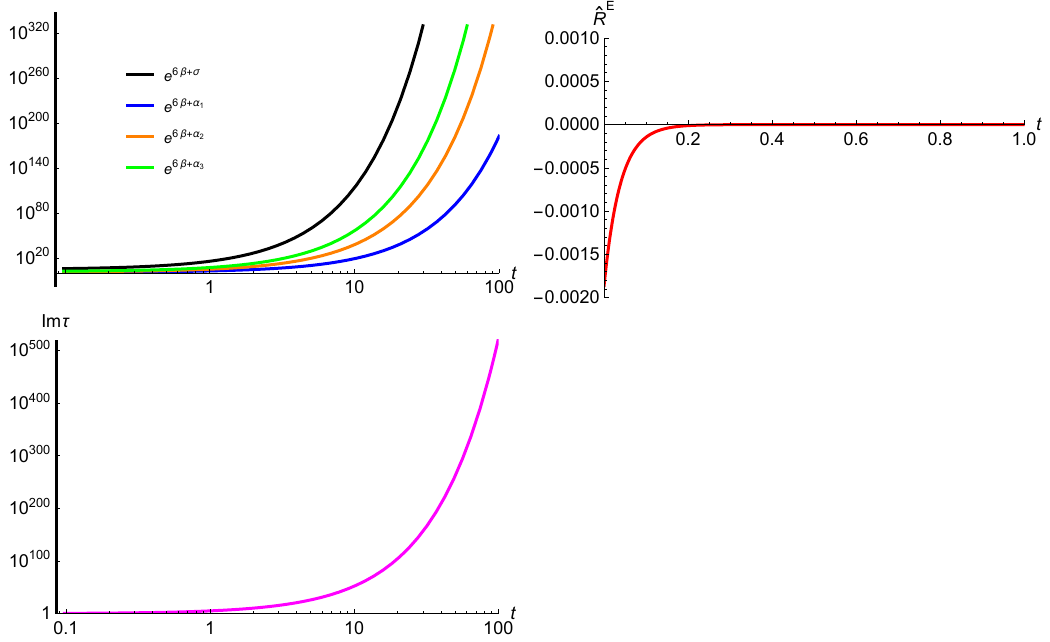}
    \caption{Profiles of the time-dependent scale factors and Ricci scalar associated with the 4D Einstein frame metric, and the imaginary part of axiodilaton, for the case with vanishing $\t F_5$, nontrivial $G_3$ flux, and constant $\textrm{Re}\tau$. The quantities are plotted against time for the representative choice of parameters specified in the text, with $\beta=-t$. In the left panels, both axes are shown on a logarithmic scale. The initial values of the plotted quantities are all nonzero, as given in the text.}
    \label{fig:ZeroF5ImTauT}
\end{figure}

Figure \ref{fig:ZeroF5ImTauT} shows the profiles of the scale factors and $\textrm{Im}\tau$ for the following consistent choice of the parameters and $\beta(t)$: $k=1,c_1=1,c_2=1,c_3=1,c_5=c_7=c_9=1,c_6=2 c_4,c_8=3 c_4,c_4=-32 \sqrt{\frac{3}{11}},\beta=-t$. The corresponding initial values of the relevant quantities are: $e^{6 \beta +\sigma }\approx 55072.296,\ e^{6 \beta+\alpha _1 }\approx 7.985,\ e^{6 \beta+\alpha _2 }\approx 23.457,\ e^{6 \beta+\alpha _3 }\approx 68.909,\ \h R^E\approx -0.002,\ \textrm{Im}\tau= 1,\ \zeta _t\approx 2.066$. The scale factors and $\textrm{Im}\tau$ grow exponentially with $t$ at late times: $e^{6 \beta +\sigma }\sim e^{48 \sqrt{\frac{3}{11}} t},\ e^{6 \beta +\alpha_1}\sim e^{8 \sqrt{\frac{3}{11}} t},\ e^{6 \beta +\alpha_2}\sim e^{16 \sqrt{\frac{3}{11}} t},\ e^{6 \beta +\alpha_3}\sim e^{24 \sqrt{\frac{3}{11}} t},\ \textrm{Im}\tau\sim e^{12 t}$. Meanwhile, $\h R^E$ and $\zeta_t$ decay to zero as: $\h R^E\sim - e^{-48 \sqrt{\frac{3}{11}} t},\ \zeta_t\sim e^{-4 t}$.

With the same choice of parameters, if we instead take $\beta=-\ln [t+1]$, the scale factors and $\textrm{Im}\tau$ grow as powers of $t$ at late times: $e^{6 \beta +\sigma }\sim t^{48 \sqrt{\frac{3}{11}}-2},\ e^{6 \beta +\alpha_1}\sim t^{8 \sqrt{\frac{3}{11}}},\ e^{6 \beta +\alpha_2}\sim t^{16 \sqrt{\frac{3}{11}}},\ e^{6 \beta +\alpha_3}\sim t^{24 \sqrt{\frac{3}{11}}},\ \textrm{Im}\tau\sim t^{12}$, while starting from the same initial values of all the relevant quantities as given above since the initial values of $\beta,\dot\beta$ coincide with the previous case. In this case, $\h R^E$ and $\zeta_t$ decay to zero as: $\h R^E\sim -t^{-48 \sqrt{\frac{3}{11}}},\ \zeta_t\sim t^{-5}$.

\textbf{For the class of time-dependent $\textrm{Re}\tau,\textrm{Im}\tau$}, the scale factor exponents are given in \eqref{eq:0F5G3ReImTauTEx}. Note that, due to the relation among the integration constants, no pair among $c_5, c_7, c_9$ can simultaneously vanish since $c_4^2 > 0$. $\text{Im}\tau > 0$ always. Notably, by altering the sign of $c_2$, $\textrm{Re}\tau$ can be made to increase or decrease in time. For an admissible choice of $\beta$ (i.e., $0 < 9e^{\frac{8\beta}{3}} < 8 c_4^2$) such that $e^{\frac{8\beta}{3}}$ is monotonically decreasing in time, we need to choose $c_5, c_7, c_9< 0$ to obtain growing scale factors from nonzero initial values. To ensure a nonzero initial value of $e^{6\beta+\sigma}$, a nonvanishing initial $\dot\beta$ is also required.

\begin{figure}[!ht]
    \centering
    \includegraphics[width=1.05\linewidth]{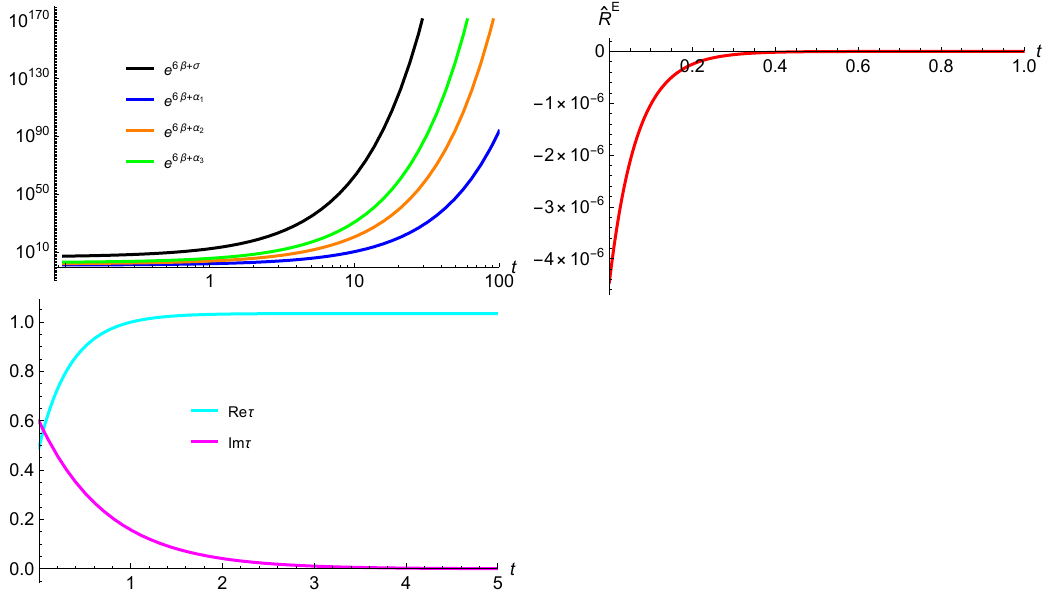}
    \caption{Profiles of the time-dependent scale factors and Ricci scalar associated with the 4D Einstein frame metric, and the real and imaginary parts of axiodilaton, for the case with vanishing $\t F_5$ and nontrivial $G_3$. The quantities are plotted against time for the representative choice of parameters specified in the text, with $\beta=-t$. In the top left panel, both axes are shown on a logarithmic scale. The initial values of the plotted quantities are all nonzero, as given in the text.}
    \label{fig:ZeroF5ReImTauT}
\end{figure}

Figure \ref{fig:ZeroF5ReImTauT} shows the profiles of the scale factors and $\textrm{Re}\tau,\ \textrm{Im}\tau$ for the following consistent choice of the parameters and $\beta(t)$: $k=1,c_1=1,c_2=1,c_3=1,c_4=2,c_6=c_8=c_{10}=1,c_7=2 c_5,c_9=3 c_5,c_5=-\frac{32}{3} \sqrt{\frac{14}{11}},\beta=-t$. The corresponding initial values of the relevant quantities are: $e^{6 \beta +\sigma }\approx 7.78029\times 10^6,\ e^{6 \beta+\alpha _1 }\approx 19.916,\ e^{6 \beta+\alpha _2 }\approx 145.915,\ e^{6 \beta+\alpha _3 }\approx 1069.06,\ \h R^E\approx -4.45073\times 10^{-6},\ \textrm{Re}\tau\approx 0.491,\ \textrm{Im}\tau\approx 0.6,\ \zeta _t\approx 1.583$. The scale factors grow exponentially with $t$ at late times: $e^{6 \beta +\sigma }\sim e^{16 \sqrt{\frac{7}{11}} t},\ e^{6 \beta +\alpha_1}\sim e^{\frac{8}{3} \sqrt{\frac{7}{11}} t},\ e^{6 \beta +\alpha_2}\sim e^{\frac{16}{3} \sqrt{\frac{7}{11}} t},\ e^{6 \beta +\alpha_3}\sim e^{8 \sqrt{\frac{7}{11}} t}$. Meanwhile, $\textrm{Re}\tau$ approaches a constant value as: $\textrm{Re}\tau\sim 1.03513-\frac{9}{8 \sqrt{5}} e^{-\frac{8 t}{3}}$, and $\h R^E$, $\textrm{Im}\tau$ and $\zeta_t$ decay to zero as: $\h R^E\sim - e^{-16 \sqrt{\frac{7}{11}} t},\ \textrm{Im}\tau\sim e^{-\frac{4 t}{3}},\ \zeta_t\sim e^{-4 t}$.

\subsection{\texorpdfstring{$\textrm{Self-dual}\ \t F_5,\ G_3,\tau\neq 0,\ \tau(t)$}{x}}
\label{sapp:AllFluxTauT}

\textbf{For the class of $\textrm{Re}\tau=\textrm{constant}$}, the scale factor exponents are given in \eqref{eq:AllFluxImTauTEx}. Note that, due to the first relation among the integration constants, we have $c_3 > 0$ which keeps $\text{Im}\tau > 0$, and by choosing $c_3$ sufficiently large one can achieve a large value of $\text{Im}\tau$. For an admissible choice of $\beta$ (i.e., $0 < e^{8\beta} < c_6^2/c_8^2$) such that $e^{8\beta}$ is monotonically decreasing in time, we obtain growing scale factors from nonzero initial values, while $\textrm{Im}\tau$ grows as $e^{8\beta}$ decreases, for any $c_6>0$. To ensure a nonzero initial value of $e^{6\beta+\sigma}$, a nonvanishing initial $\dot\beta$ is also required.

\begin{figure}[!ht]
    \centering
    \includegraphics[width=1.05\linewidth]{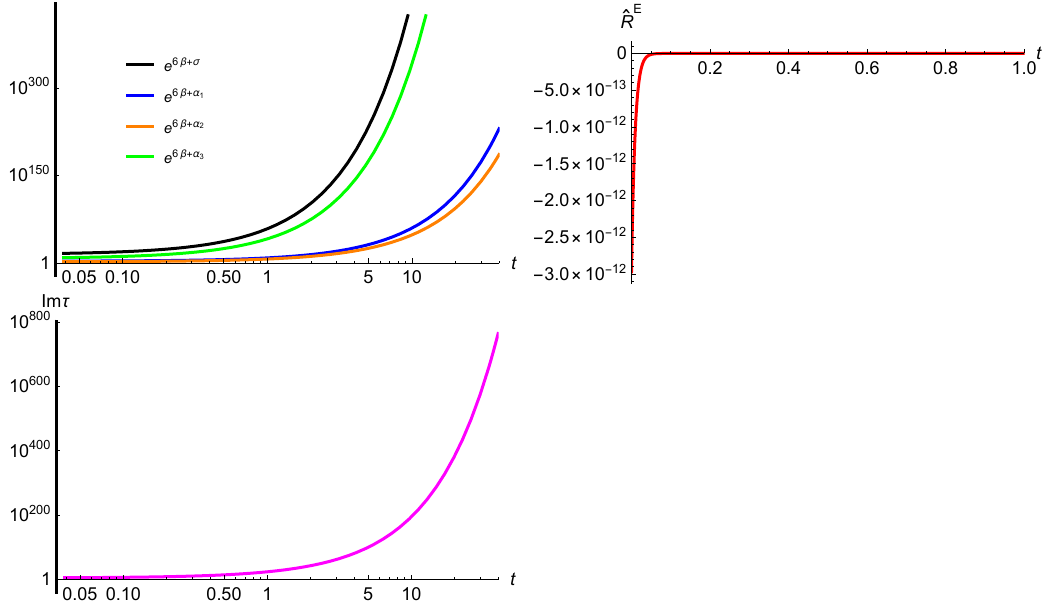}
    \caption{Profiles of the time-dependent scale factors and Ricci scalar associated with the 4D Einstein frame metric, and the imaginary part of axiodilaton, for the case with nontrivial $\t F_5$, $G_3$ fluxes, and constant $\textrm{Re}\tau$. The quantities are plotted against time for the representative choice of parameters specified in the text, with $\beta=-t$. In the left panels, both axes are shown on a logarithmic scale. The initial values of the plotted quantities are all nonzero, as given in the text.}
    \label{fig:AllFluxImTauT}
\end{figure}

Figure \ref{fig:AllFluxImTauT} shows the profiles of the scale factors and $\textrm{Im}\tau$ for the following consistent choice of the parameters and $\beta(t)$: $k=1,r=1,c_1=\sqrt{\frac{3}{14}},c_2=1,c_3=\frac{14}{5},c_4=1,c_5=1,c_6=2,c_7=2,c_8=1,c_9=1,\beta=-t$. The corresponding initial values of the relevant quantities are: $e^{6 \beta +\sigma }\approx 3.92566\times 10^{14},\ e^{6 \beta+\alpha _1 }\approx 694.334,\ e^{6 \beta+\alpha _2 }\approx 88.619,\ e^{6 \beta+\alpha _3 }\approx 4.2723\times 10^7,\ \h R^E\approx -2.96809\times 10^{-12},\ \textrm{Im}\tau= 172287,\ \gamma _t\approx 5.657,\ \vartheta_3=1$. The scale factors, $\textrm{Im}\tau$ and $\vartheta_3$ grow exponentially with $t$ at late times: $e^{6 \beta +\sigma }\sim e^{4 \left(10-\sqrt{7}+3 \sqrt{35}\right) t},\ e^{6 \beta +\alpha_1}\sim e^{4 \left(\sqrt{5}-1\right) \sqrt{7} t},\ e^{6 \beta +\alpha_2}\sim e^{4 \sqrt{7} t},\ e^{6 \beta +\alpha_3}\sim e^{\left(40-4 \sqrt{7}+8 \sqrt{35}\right) t},\ \textrm{Im}\tau\sim e^{4 \left(\sqrt{35}+5\right) t},\ \vartheta_3\sim e^{6 t}$. Meanwhile, $\h R^E$ and $\gamma_t$ decay to zero as: $\h R^E\sim - e^{-4 \left(10-\sqrt{7}+3 \sqrt{35}\right) t},\ \gamma_t\sim e^{-6 t}$.

With the same choice of parameters, if we instead take $\beta=-\ln [t+1]$, the scale factors, $\textrm{Im}\tau$ and $\vartheta_3$ grow as powers of $t$ at late times: $e^{6 \beta +\sigma }\sim t^{38-4 \sqrt{7}+12 \sqrt{35}},\ e^{6 \beta +\alpha_1}\sim t^{4 \sqrt{7} \left(\sqrt{5}-1\right)},\ e^{6 \beta +\alpha_2}\sim t^{4 \sqrt{7}},\ e^{6 \beta +\alpha_3}\sim t^{40-4 \sqrt{7}+8 \sqrt{35}},\ \textrm{Im}\tau\sim t^{4 \left(\sqrt{35}+5\right)},\ \vartheta_3\sim t^6$, while starting from the same initial values of all the relevant quantities as given above since the initial values of $\beta,\dot\beta$ coincide with the previous case. In this case, $\h R^E$ and $\gamma_t$ decay to zero as: $\h R^E\sim -t^{-4 \left(10-\sqrt{7}+3 \sqrt{35}\right)},\ \gamma_t\sim t^{-7}$.

\textbf{For the class of time-dependent $\textrm{Re}\tau,\textrm{Im}\tau$}, the scale factor exponents are given in \eqref{eq:AllFluxReImTauTEx}. For monotonically increasing $\beta$ with $\beta>0$ for $t>0$, we need to choose $c_5$: $-\frac{2}{3}<c_5<2$, to obtain growing scale factors from nonzero initial values. To ensure a nonzero initial value of $e^{6\beta+\sigma}$, a nonvanishing initial $\dot\beta$ is also required. The sign of $c_9$ does affect these growths, however, the second relation among the integration constants together with the above condition on $c_5$ fixes the sign of $c_9$: $c_9<0$. For $\textrm{Im}\tau>0$, we need to have: $c_1k/c_6>0$, which is guaranteed by the first relation among the integration constants. Finally, $c_7$ can be adjusted to offset any negative contribution arising from the lower limit of the integral in the expression for $\textrm{Re}\tau$.

\begin{figure}[!ht]
    \centering
    \includegraphics[width=1.05\linewidth]{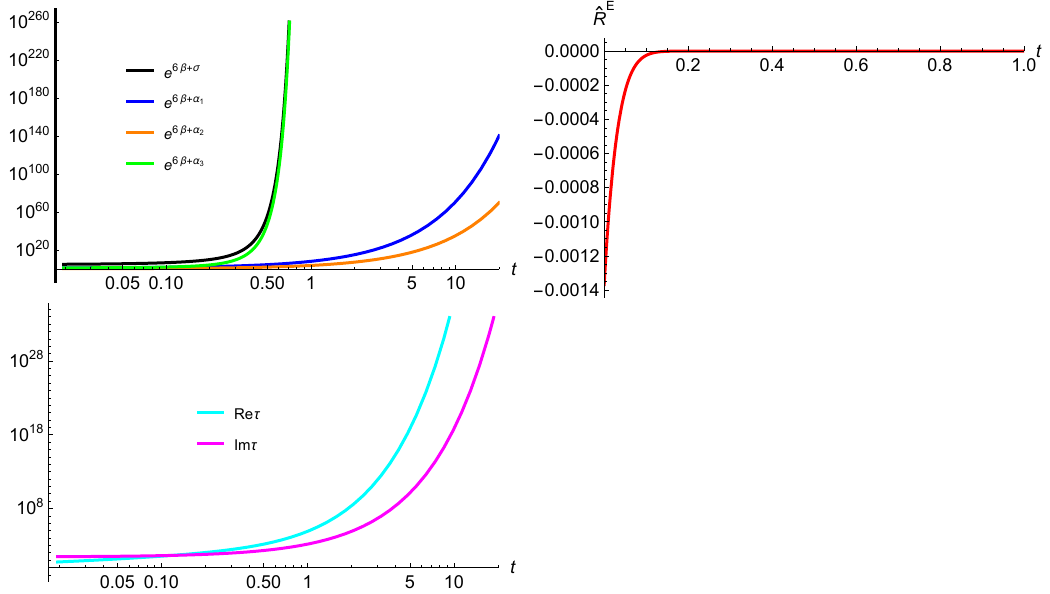}
    \caption{Profiles of the time-dependent scale factors and Ricci scalar associated with the 4D Einstein frame metric, and the real and imaginary parts of axiodilaton, for the case with nontrivial $\t F_5$, $G_3$ fluxes. The quantities are plotted against time for the representative choice of parameters specified in the text, with $\beta=+t$. In the left panels, both axes are shown on a logarithmic scale. The initial values of the plotted quantities are all nonzero, as given in the text.}
    \label{fig:AllFluxReImTauT}
\end{figure}

Figure \ref{fig:AllFluxReImTauT} shows the profiles of the scale factors and $\textrm{Re}\tau,\ \textrm{Im}\tau$ for the following consistent choice of the parameters and $\beta(t)$: $k=1,r=1,c_1=1,c_2=1,c_3=1,c_4=1,c_5=1,c_8=1,c_{10}=1,c_6=\frac{2}{e^4},c_9=-\frac{26}{3},c_7=1-e^4/2+e^{12}/2,\beta=+t$. The corresponding initial values of the relevant quantities are: $e^{6 \beta +\sigma }=64 e^7,e^{6 \beta+\alpha _1 }=e^3,e^{6\beta+\alpha _2}=e,e^{6\beta+\alpha _3}=e^3,\h R^E=-\frac{3}{2} e^{-7},\textrm{Re}\tau =1,\textrm{Im}\tau =e^4/2,\gamma _t=8,\vartheta_3=1$. The scale factors, $\textrm{Re}\tau$, $\textrm{Im}\tau$ and $\gamma_t$ grow with $t$ at late times as: $e^{6 \beta +\sigma }\sim e^{\frac{64 t}{3}+2 e^{8 t}},\ e^{6 \beta +\alpha_1}\sim e^{16 t},\ e^{6 \beta +\alpha_2}\sim e^{8 t},\ e^{6 \beta +\alpha_3}\sim e^{2 e^{8 t}-\frac{8 t}{3}},\ \textrm{Re}\tau\sim e^{8 t},\ \textrm{Im}\tau\sim e^{4 t},\ \gamma_t\sim e^{6 t}$. Meanwhile, $\h R^E$ and $\vartheta_3$ decay to zero as: $\h R^E\sim - e^{-\frac{40 t}{3}-2 e^{8 t}},\ \vartheta_t\sim e^{-6 t}$. Note that some of the growth and decay behaviours are double-exponential.

With the same choice of parameters (except $c_7$, for which we now take $c_7=\frac{(1+e)^8-1}{2 e^4}+1$), if we instead take $\beta=+\ln [t+e^{-1}]$, the scale factors, $\textrm{Re}\tau$, $\textrm{Im}\tau$ and $\gamma_t$ grow with $t$ at late times as: $e^{6 \beta +\sigma }\sim e^{2 t^8} t^{58/3},\ e^{6 \beta +\alpha_1}\sim t^{16},\ e^{6 \beta +\alpha_2}\sim t^8,\ e^{6 \beta +\alpha_3}\sim t^{-8/3}e^{2 t^8},\ \textrm{Re}\tau\sim t^8,\ \textrm{Im}\tau\sim t^4,\ \gamma_t\sim t^5$. Meanwhile, $\h R^E$ and $\vartheta_3$ decay to zero as: $\h R^E\sim -t^{-40/3}e^{-2 t^8},\ \vartheta_t\sim t^{-6}$. Note that only some of the growth and decay behaviours follow power laws in $t$. The corresponding initial values of the relevant quantities are: $e^{6 \beta +\sigma }=64 e^{\frac{2}{e^8}-\frac{43}{3}},e^{6\beta+\alpha _1}=e^{-13},e^{6\beta+\alpha _2}=e^{-7},e^{6\beta+\alpha _3}=e^{\frac{11}{3}+\frac{2}{e^8}},R^{\text{E}}\approx -6.20078\times 10^6,\text{Re$\tau $}=1,\text{Im$\tau $}=1/2,\gamma _t=8e^{-5},\vartheta _3=e^6$.

\bibliography{biblio}
\bibliographystyle{JHEP}

\end{document}